\documentclass[preprint,10pt]{elsarticle}
\pdfoutput=1


\makeatletter
\def\ps@pprintTitle{%
   \let\@oddhead\@empty
   \let\@evenhead\@empty
   \def\@oddfoot{\reset@font\hfil\thepage\hfil}
   \let\@evenfoot\@oddfoot
}
\makeatother

\usepackage{titlesec}
\usepackage{xcolor}
\usepackage{enumitem}
\usepackage{hyperref}
\hypersetup{colorlinks=true,urlcolor=black}
\usepackage[T1]{fontenc}\usepackage{lmodern}

\usepackage{mathtools,amsfonts}
\usepackage{caption}
\usepackage{subcaption}
\usepackage{booktabs}
\usepackage{pifont}
\usepackage{placeins}
\usepackage{multirow}

\definecolor{green}{rgb}{0,0.5,0}

\usepackage{tikz}
\usetikzlibrary{calc}
\usepackage{tkz-euclide}
\usepackage{calc}
\usepackage{tikz,pgfplots}
\pgfplotsset{compat=1.12}
\usetikzlibrary{arrows.meta,3d,matrix,positioning,decorations.markings}

\newcommand{\pz}{\phantom{0}}
\newcommand{\ntriangles}{n_t}

\newcommand{\nbasis}{n_b}
\newcommand{\srcidx}{j}
\newcommand{\testidx}{i}

\makeatletter
\tikzoption{canvas is xy plane at z}[]{%
  \def\tikz@plane@origin{\pgfpointxyz{0}{0}{#1}}%
  \def\tikz@plane@x{\pgfpointxyz{1}{0}{#1}}%
  \def\tikz@plane@y{\pgfpointxyz{0}{1}{#1}}%
  \tikz@canvas@is@plane
}
\makeatother

\makeatletter
\tikzoption{canvas is plane}[]{\@setOxy#1}
\def\@setOxy O(#1,#2,#3)x(#4,#5,#6)y(#7,#8,#9)%
  {\def\tikz@plane@origin{\pgfpointxyz{#1}{#2}{#3}}%
   \def\tikz@plane@x{\pgfpointxyz{#4}{#5}{#6}}%
   \def\tikz@plane@y{\pgfpointxyz{#7}{#8}{#9}}%
   \tikz@canvas@is@plane
  }
\makeatother 

\definecolor{orange}{RGB}{217,95,2}
\definecolor{green}{RGB}{102,166,30}
\definecolor{purple}{RGB}{231,41,138}
\newcommand{\reviewerOne}[1]{#1}
\newcommand{\reviewerBoth}[1]{#1}
\newcommand{\reviewerTwo}[1]{#1}
\usepackage[margin=1in]{geometry}
\biboptions{sort&compress}
\begin{document}

\begin{frontmatter}
\title{Code-Verification Techniques for the Method-of-Moments Implementation of the Electric-Field Integral Equation}

\author[freno]{Brian A.\ Freno}
\ead{bafreno@sandia.gov}
\author[freno]{Neil R.\ Matula}
\author[freno]{Justin I.\ Owen}
\author[freno]{William A.\ Johnson}

\address[freno]{Sandia National Laboratories, Albuquerque, NM 87185}

\begin{abstract}
The method-of-moments implementation of the electric-field integral equation yields many code-verification challenges due to the \reviewerOne{various} sources of numerical error and their possible interactions.  Matters are further complicated by singular integrals, which arise from the presence of a Green's function.  In this paper, we provide approaches to separately assess the numerical errors arising from the use of basis functions to approximate the solution and the use of quadrature to approximate the integration.  Through these approaches, we are able to verify the code and compare the error from different quadrature options.
\end{abstract}


\end{frontmatter}

\section{Introduction}

In computational electromagnetics, the method-of-moments (MoM) implementation of the electric-field integral equation (EFIE) is frequently used to model electromagnetic scattering and radiation problems.  Through this approach, the surface of the electromagnetic scatterer is discretized using planar or curvilinear mesh elements, and four-dimensional integrals are evaluated over two-dimensional source and test elements.  However, the presence of a Green's function in these equations yields singularities when the test and source elements share one or more edges or vertices and near-singularities when they are otherwise close.  Many approaches have been developed to address the singularity and near-singularity for the inner, source-element integral~\cite{graglia_1993,wilton_1984,rao_1982,khayat_2005,fink_2008,khayat_2008,vipiana_2011,vipiana_2012,botha_2013,rivero_2019}, as well as for the outer, test-element integral~\cite{vipiana_2013,polimeridis_2013,wilton_2017,rivero_2019b,freno_em}.

As with any computational physics code, an important step towards establishing the credibility of the results of the MoM implementation of the EFIE is performing code verification~\cite{roache_1998,salari_2000,oberkampf_2010}.  Through code verification, the correctness of the implementation of the numerical methods is assessed.  The discretization of differential, integral, or integro-differential equations incurs some truncation error, and thus the approximate solutions produced from the discretized equations will incur an associated discretization error.  If the discretization error tends to zero as the discretization is refined, the consistency of the code is verified~\cite{roache_1998}.  This may be taken a step further by examining not only consistency, but the rate at which the error decreases as the discretization is refined, thereby verifying the order of accuracy of the discretization scheme.  The correctness of the numerical-method implementation may then be verified by comparing the expected and observed orders of accuracy obtained from numerous test cases with known solutions. 

In general, exact solutions are limited and may not sufficiently exercise the capabilities of the code.  Therefore, manufactured solutions~\cite{roache_2001} are a popular alternative, permitting the construction of problems of arbitrary complexity with known solutions.  Through the method of manufactured solutions (MMS), a solution is manufactured and substituted directly into the governing equations to yield a residual term, which is added as a source term to coerce the solution to the manufactured solution. 

However, integral equations yield an additional challenge.  While analytical differentiation is straightforward, analytical integration is not always possible.  Therefore, the residual source term arising from the manufactured solution may not be representable in closed form, and its implementation may be accompanied by numerical techniques that carry their own uncertainties.  Furthermore, in many applications, such as the MoM implementation of the EFIE, singular integrals appear, which can further complicate the numerical evaluation of the source term.  Therefore, many of the benefits associated with MMS are lost when applied to integral equations in a straightforward manner.

Code verification has been performed on computational physics codes associated with several physics disciplines, including fluid dynamics~\cite{roy_2004,bond_2007,veluri_2010,oliver_2012,eca_2016,freno_2021}, solid mechanics~\cite{chamberland_2010}, fluid--structure interaction~\cite{etienne_2012}, heat transfer in fluid--solid interaction~\cite{veeraragavan_2016}, \reviewerTwo{multiphase flows~\cite{brady_2012,lovato_2021}}, radiation hydrodynamics~\cite{mcclarren_2008}, electrodynamics~\cite{ellis_2009}, and ablation~\cite{amar_2008,amar_2009,amar_2011,freno_ablation}.  
For integral equations in computational electromagnetics, code-verification activities that employ manufactured solutions have been limited to the works of Marchand et al.~\cite{marchand_2013,marchand_2014}, through which the MMS source term is computed using additional quadrature points, and Freno et al.~\cite{freno_em_mms_2020}, through which the Green's function is additionally manufactured, permitting the numerical-integration error to be eliminated and the solution-discretization error isolated.

This work continues the efforts of Reference~\cite{freno_em_mms_2020} by assessing the numerical errors from the basis functions and numerical integration together and separately for a manufactured Green's function, as well as the actual Green's function.  We accomplish this by first measuring the error norms of the MoM implementation of the EFIE using the actual Green's function and considering the contributions from the magnetic vector potential and electric scalar potential separately and together.  
\reviewerOne{Next, we isolate the numerical-integration error using two complementary approaches:
\begin{enumerate}
\item \textbf{Solution-discretization error cancellation.} Instead of computing the MMS source term as a function of the manufactured surface current, we compute it as a function of the basis-function representation of the manufactured surface current.  
Due to the presence of the basis functions on both sides of the equation, we cancel the solution-discretization error.

\item \textbf{Solution-discretization error elimination.} Instead of projecting the governing equations onto the basis functions, we project them onto the manufactured surface current.  Additionally, we do not represent the manufactured surface current in terms of the basis functions.  
By eliminating the presence of the basis functions, we eliminate the solution-discretization error.

\end{enumerate}
}

This paper is organized as follows.  In Section~\ref{sec:efie}, we describe the MoM implementation of the EFIE.  In Section~\ref{sec:mms}, we describe the challenges of using MMS with the MoM implementation of the EFIE, and we describe our approach to mitigating them.  In Section~\ref{sec:results}, we demonstrate the effectiveness of our approach.  In Section~\ref{sec:conclusions}, we summarize this work.

\section{The Method-of-Moments Implementation of the EFIE}
\label{sec:efie}

In time-harmonic form, the scattered electric field $\mathbf{E}^\mathcal{S}$ can be computed from the surface current by
\begin{align}
\mathbf{E}^\mathcal{S} = -\left(j\omega\mathbf{A}+\nabla\Phi\right), 
\label{eq:Es}
\end{align}
where the magnetic vector potential $\mathbf{A}$ is defined by
\begin{align}
\mathbf{A}(\mathbf{x})= \mu \int_{S'} \mathbf{J} (\mathbf{x}')G_k(\mathbf{x},\mathbf{x}')dS',
\label{eq:A}
\end{align}
and, by employing the Lorenz gauge condition and the continuity equation, the electric scalar potential $\Phi$ is defined by 
\begin{align}
\Phi(\mathbf{x})=  \frac{j}{\epsilon\omega} \int_{S'} \nabla'\cdot\mathbf{J}(\mathbf{x}')G_k(\mathbf{x},\mathbf{x}')dS'.
\label{eq:Phi}
\end{align}
In~\eqref{eq:A} and~\eqref{eq:Phi}, the integration domain is the surface $S$ of a perfectly conducting scatterer.  Additionally, $\mathbf{J}$ is the surface current, $\mu$ and $\epsilon$ are the permeability and permittivity of surrounding medium, and $G_k$ is the Green's function 
\begin{align}
G_k(\mathbf{x},\mathbf{x}') = \frac{e^{-jkR}}{4\pi R},
\label{eq:G}
\end{align}
where $R=|\mathbf{x}-\mathbf{x}'|$, and $k=\omega\sqrt{\mu\epsilon}$ is the wave number.  If $S$ is open, the component of $\mathbf{J}$ normal to the boundary of $S$ must vanish on the boundary of $S$ to reflect that the total current is zero.

The total electric field $\mathbf{E}$ is the sum of the incident electric field $\mathbf{E}^\mathcal{I}$ (which induces $\mathbf{J}$) and $\mathbf{E}^\mathcal{S}$.  On $S$, the tangential component of $\mathbf{E}$ is zero, such that
\begin{align}
\mathbf{E}_t^\mathcal{S}=-\mathbf{E}_t^\mathcal{I},
\label{eq:tan_BC}
\end{align}
where the subscript $t$ denotes the tangential component.
Substituting~\eqref{eq:Es} into~\eqref{eq:tan_BC}, we can compute $\mathbf{J}$ from $\mathbf{E}^\mathcal{I}$:
\begin{align}
\mathbf{E}_t^\mathcal{I} =\left(j\omega\mathbf{A} + \nabla\Phi\right)_t.
\label{eq:tan}
\end{align}

Inserting~\eqref{eq:A} and~\eqref{eq:Phi} into~\eqref{eq:tan}, projecting~\eqref{eq:tan} onto an appropriate space $\mathbb{V}$ containing vector fields that are tangential to $S$ with no components normal to the boundary of $S$, and integrating by parts yields the variational form: find $\mathbf{J}\in\mathbb{V}$, such that
\begin{align}
\int_S \mathbf{E}^\mathcal{I}\cdot \bar{\mathbf{v}} dS =
j\omega\mu\int_S \bar{\mathbf{v}}(\mathbf{x})\cdot \int_{S'} \mathbf{J}(\mathbf{x}')G_k(\mathbf{x},\mathbf{x}')dS' dS - \frac{j}{\epsilon\omega}\int_S \nabla\cdot\bar{\mathbf{v}}(\mathbf{x}) \int_{S'} \nabla'\cdot \mathbf{J}(\mathbf{x}')G_k(\mathbf{x},\mathbf{x}')dS' dS 
\label{eq:variational}
\end{align} 
for all $\mathbf{v}\in\mathbb{V}$, \reviewerOne{where the bar notation denotes complex conjugation}.  It can be shown that the appropriate test and solution space is
$\mathbb{V} = H_\text{div}^{-1/2}(S)$~\cite{christiansen_2004}. \reviewerOne{The space $H_\text{div}^{-1/2}(S)$ is the space of functions $\mathbf{v}$ with domain $S$, such that both $\mathbf{v}$ and $\nabla\cdot\mathbf{v}$ are in $H^{-1/2}(S)$, where $H^{-1/2}(S)$ is the $-1/2$ order dual space to the $1/2$ order space $H^{1/2}(S)$.}
We can write~\eqref{eq:variational} more succinctly in terms of a sesquilinear form and inner product: find $\mathbf{J}\in\mathbb{V}$, such that
\begin{align}
a(\mathbf{J},\mathbf{v}) = \left(\mathbf{E}^\mathcal{I}, \mathbf{v}\right) \qquad \forall \mathbf{v}\in\mathbb{V},
\label{eq:var_sesquilinear}
\end{align}
where the sesquilinear form and inner product are defined by
\begin{align}
a(\mathbf{u},\mathbf{v}) &{}= a^\mathbf{A}(\mathbf{u},\mathbf{v}) + a^\Phi(\mathbf{u},\mathbf{v}), \label{eq:a}
\\ \nonumber
a^\mathbf{A}(\mathbf{u},\mathbf{v}) &{}= j\omega\mu \int_S \bar{\mathbf{v}}(\mathbf{x})\cdot\int_{S'} \mathbf{u}(\mathbf{x}')G_k(\mathbf{x},\mathbf{x}')dS'dS,
\\ \nonumber
a^\Phi(\mathbf{u},\mathbf{v}) &{}= -\frac{j}{\epsilon\omega} \int_S \nabla\cdot\bar{\mathbf{v}}(\mathbf{x})\int_{S'} \nabla'\cdot\mathbf{u}(\mathbf{x}')G_k(\mathbf{x},\mathbf{x}')dS' dS,
\\ \nonumber
(\mathbf{u},\mathbf{v})  &{}= \int_S \mathbf{u}(\mathbf{x})\cdot \bar{\mathbf{v}}(\mathbf{x}) dS.
\end{align}

To solve the variational problem~\eqref{eq:var_sesquilinear}, we discretize $S$ with a mesh composed of triangular elements and approximate $\mathbf{J}$ with $\mathbf{J}_h$ using a Galerkin method with Rao--Wilton--Glisson (RWG) basis functions $\boldsymbol{\Lambda}_{\srcidx}(\mathbf{x})$~\cite{rao_1982}:
\begin{align}
\mathbf{J}_h(\mathbf{x}) =\sum_{\srcidx=1}^{\nbasis} J_{\srcidx} \boldsymbol{\Lambda}_{\srcidx}(\mathbf{x}),
\label{eq:J_h}
\end{align}
where $\nbasis$ is the total number of basis functions.
\reviewerOne{%
The RWG basis functions are defined for a triangle pair by
%
%
\begin{align*}
\boldsymbol{\Lambda}_{\srcidx}(\mathbf{x}) = \left\{
\begin{matrix}
\displaystyle\frac{\ell_{\srcidx}}{2A_{\srcidx}^+}\boldsymbol{\rho}_{\srcidx}^+, & \text{for }\mathbf{x}\in T_{\srcidx}^+ \\[1em]
\displaystyle\frac{\ell_{\srcidx}}{2A_{\srcidx}^-}\boldsymbol{\rho}_{\srcidx}^-, & \text{for }\mathbf{x}\in T_{\srcidx}^- \\[1em]
\mathbf{0}, & \text{otherwise}
\end{matrix}
\right.,
\end{align*}
where $\ell_{\srcidx}$ is the length of the edge shared by the triangle pair, and $A_{\srcidx}^+$ and $A_{\srcidx}^-$ are the areas of the triangles $T_{\srcidx}^+$ and $T_{\srcidx}^-$ associated with basis function $\srcidx$.  $\boldsymbol{\rho}_{\srcidx}^+$ denotes the vector from the vertex of $T_{\srcidx}^+$ opposite the shared edge to $\mathbf{x}$, and $\boldsymbol{\rho}_{\srcidx}^-$ denotes the vector to the vertex of $T_{\srcidx}^-$ opposite the shared edge from $\mathbf{x}$.
}

These basis functions ensure that $\mathbf{J}_h$ is tangential to $S$ and has no component normal to the outer boundary of the triangle pair associated with basis function $\srcidx$.  
Additionally, along the shared edge of the triangle pair, the component of $\boldsymbol{\Lambda}_{\srcidx}(\mathbf{x})$ normal to that edge is unity.  Therefore, for a triangle edge shared by only two triangles, the component of $\mathbf{J}_h$ normal to that edge is $J_\srcidx$.  The solution is considered most accurate at the midpoint of the edge~\cite[pp.\ 155--156]{warnick_2008}.

Defining $\mathbb{V}_h$ to be the span of RWG basis functions associated with the mesh on $S$, the Galerkin approximation of the original problem is now: find $\mathbf{J}_h\in\mathbb{V}_h$, such that
\begin{align}
a(\mathbf{J}_h,\boldsymbol{\Lambda}_{\testidx}) = \left(\mathbf{E}^\mathcal{I}, \boldsymbol{\Lambda}_{\testidx}\right)
\label{eq:proj_disc}
\end{align}
for $i=1,\hdots,\nbasis$.
Letting $\mathbf{J}^h$ denote the vector of coefficients used to construct $\mathbf{J}_h$~\eqref{eq:J_h}, \eqref{eq:proj_disc} can be written in matrix form as
\begin{align}
\mathbf{Z}\mathbf{J}^h = \mathbf{V},
\label{eq:zjv}
\end{align}
where $J_{\srcidx}^h = J_{\srcidx}$, $V_{\testidx} =\left(\mathbf{E}^\mathcal{I}, \boldsymbol{\Lambda}_{\testidx}\right)$, and
\begin{align}
Z_{\testidx,\srcidx} &{}= a(\boldsymbol{\Lambda}_{\srcidx},\boldsymbol{\Lambda}_{\testidx}). \label{eq:matrix_elements}
\end{align}

\section{Manufactured Solutions}
\label{sec:mms}

For each test basis function, we can write Equation~\eqref{eq:var_sesquilinear} in terms of its residual vector,
%
$\mathbf{r}(\mathbf{J})$,
%
where
\begin{align*}
r_{\testidx}(\mathbf{J}) = a(\mathbf{J},\boldsymbol{\Lambda}_{\testidx}) -\left(\mathbf{E}^\mathcal{I}, \boldsymbol{\Lambda}_{\testidx}\right) = 0.
\end{align*}
Similarly, we can define the vector of residuals for the discretized problem~\eqref{eq:zjv} to be
\begin{align}
\mathbf{r}_h(\mathbf{J}_h) = \mathbf{0},
\label{eq:res_disc}
\end{align}
where $\mathbf{r}_h(\mathbf{J}_h) = \mathbf{Z}\mathbf{J}^h - \mathbf{V}$.

The method of manufactured solutions modifies~\eqref{eq:res_disc} to be
\begin{align}
\mathbf{r}_h(\mathbf{J}_h) = \mathbf{r}(\mathbf{J}_\text{MS}),
\label{eq:mms}
\end{align}
where $\mathbf{J}_\text{MS}$ is the manufactured solution, and $\mathbf{r}(\mathbf{J}_\text{MS})$ is computed exactly.

Algebraically, we can then solve~\eqref{eq:mms} from
\begin{align}
\mathbf{Z}\mathbf{J}^h = \mathbf{V}_{\text{MS}}(\mathbf{J}_\text{MS}),
\label{eq:zjv_mmsp}
\end{align}
where 
\begin{align}
V_{\text{MS}_{\testidx}}(\mathbf{u})= a(\mathbf{u},\boldsymbol{\Lambda}_{\testidx}).
\label{eq:V_mms}
\end{align}

However, as described in the introduction, not only is it impossible to analytically evaluate the integrals in $V_{\text{MS}_{\testidx}}$~\eqref{eq:V_mms}, but the singular nature of the Green's function~\eqref{eq:G} when $R\to 0$ complicates their accurate approximation, potentially contaminating convergence studies.  In Reference~\cite{freno_em_mms_2020}, this problem is overcome by manufacturing a Green's function, such that~\eqref{eq:V_mms} can be computed exactly.  
Alternatively, in this paper, we integrate~\eqref{eq:V_mms} approximately but with sufficient accuracy using the deterministic globally adaptive subdivision quadrature approach~\cite{berntsen_1991a,berntsen_1991b} in the Cuhre subroutine of the \texttt{CUBA} library~\cite{hahn_2005}.  \reviewerBoth{In Section~\ref{sec:results}, we describe the tolerance criteria.}
We denote this approximation to the sesquilinear form~\eqref{eq:a} in~\eqref{eq:V_mms} as $a'(\cdot,\cdot)$.   
To expedite convergence for singular and near-singular integrals, we perform the radial--angular transformation~\cite{khayat_2008}.\looseness=-1

In this work, we seek to verify the expected convergence rate of the basis functions with the actual Green's function and to assess the convergence properties of the numerical-integration schemes.

\subsection{Solution-Discretization Error} 
\label{sec:part1}
We begin by verifying the convergence rates of the solution as the mesh is refined.  The primary focus is on the error incurred by representing the solution in terms of the basis functions; however, in practice, because the actual Green's function is involved, additional error is incurred from the numerical integration of~\eqref{eq:matrix_elements}, through which the sesquilinear form~\eqref{eq:a} is approximately integrated using quadrature.  
We denote this approximation to the sesquilinear form~\eqref{eq:a} in~\eqref{eq:V_mms} as $a_h(\cdot,\cdot)$.  
For the EFIE, we expect that~\cite{christiansen_2004,marchand_2013,marchand_2014,nedelec_2001}
\begin{align*}
\left\|\mathbf{e}\right\|_{H_\text{div}^{-1/2}(S)} \le C h^{3/2},
\end{align*}
where $\mathbf{e}(\mathbf{x})=\mathbf{J}_h(\mathbf{x})-\mathbf{J}_\text{MS}(\mathbf{x})$, and we define the norms to be
\begin{align}
\left\|\mathbf{e}\right\|_{H_\text{div}^{-1/2}(S)}^2 &{}= 
\omega\mu \left\|\mathbf{e}\right\|_{H^{-1/2}(S)}^2 + \frac{1}{\epsilon\omega}\left\|\mathbf{\nabla\cdot e}\right\|_{H^{-1/2}(S)}^2, \label{eq:hdiv_norm}
\\
\left\|\mathbf{e}\right\|_{H^{-1/2}(S)}^2 &{}= \int_S \bar{\mathbf{e}}(\mathbf{x})\cdot\int_{S'} \mathbf{e}(\mathbf{x}')G_0(\mathbf{x},\mathbf{x}')dS'dS. \label{eq:norm_integral}
\end{align}

As a result, numerical-integration error is incurred not only when computing $\mathbf{Z}$, but also when computing the norm~\eqref{eq:norm_integral}.  Therefore, the usefulness of this approach alone in the context of code verification is limited.  The solution-discretization error can be verified without contamination from numerical-integration error using the approach of Reference~\cite{freno_em_mms_2020}, whereas this assessment can confirm the singularities are integrated suitably.

\subsection{Numerical-Integration Error} 
A complementary assessment to that of Section~\ref{sec:part1} is to isolate the numerical-integration error without the contamination from the solution-discretization error.  We accomplish this in two ways: 1) canceling the solution-discretization error, and 2) eliminating the solution-discretization error.

\subsubsection{Solution-Discretization Error Cancellation} 
\label{sec:part2}

To cancel the solution-discretization error, we modify the right-hand side of~\eqref{eq:mms} to be
\begin{align}
\mathbf{r}_h(\mathbf{J}_h) = \mathbf{r}(\mathbf{J}_{h_\text{MS}}),
\label{eq:part2}
\end{align}
where $\mathbf{J}_{h_\text{MS}}$ is the basis-function representation of $\mathbf{J}_\text{MS}$, obtained from~\eqref{eq:J_h} by setting the coefficients $J_\srcidx$ equal to the normal component of $\mathbf{J}_\text{MS}$ at the midpoint of each edge associated with $\Lambda_\srcidx(\mathbf{x})$.  
We can then solve~\eqref{eq:part2} from
\begin{align}
\mathbf{Z}\mathbf{J}^h = \mathbf{V}_{\text{MS}}(\mathbf{J}_{h_\text{MS}}),
\label{eq:part2_problem}
\end{align}
where $\mathbf{V}_{\text{MS}}(\mathbf{J}_{h_\text{MS}})$ is computed using the definition in~\eqref{eq:V_mms}.  \reviewerOne{Due to the presence of the basis functions on both sides of~\eqref{eq:part2_problem}, we cancel the solution-discretization error.}

For a manufactured Green's function $G_\text{MS}$~\cite{freno_em_mms_2020}, $\mathbf{V}_{\text{MS}}(\mathbf{J}_{h_\text{MS}})$ in~\eqref{eq:part2_problem} can be computed exactly.  For the actual Green's function $G_k$~\eqref{eq:G}, $\mathbf{V}_{\text{MS}}(\mathbf{J}_{h_\text{MS}})$ can be computed with sufficient accuracy using the \texttt{CUBA} library.  \reviewerBoth{In Section~\ref{sec:results}, we describe the tolerance criteria.}  If done successfully, the solution-discretization error is canceled on each side, leaving only the numerical-integration error in the computation of $\mathbf{Z}$.

For this approach, we consider the $L^\infty$-norm 
\begin{align}
\left\|\mathbf{e}_n\right\|_\infty = \max_j|e_{n_\srcidx}|,
\label{eq:linf_norm}
\end{align}
where $\mathbf{e}_n=\mathbf{J}^h-\mathbf{J}_n$, and $J_{n_\srcidx}$ is the component of $\mathbf{J}_\text{MS}$ that flows normal to the edge associated with basis function $\srcidx$.  The use of~\eqref{eq:linf_norm} avoids contamination of the norm computation from integration \reviewerTwo{as in Section~\ref{sec:part1}, and the $L^\infty$-norm is the most sensitive of the $L^p$-norms}.

\subsubsection{Solution-Discretization Error Elimination} 
\label{sec:part3}
The approach of Section~\ref{sec:part2} enables us to assess how the numerical integration performs for the integrands arising from the approximated solution, which is limited to the quality of the basis functions.  Alternatively, instead of assessing the performance of the numerical integration relative to limited-order basis functions, we can remove the basis-function restriction.

Instead of~\eqref{eq:proj_disc}, we consider
\begin{align*}
a(\mathbf{J}_\text{MS},\mathbf{J}_\text{MS}) = (\mathbf{E}^\mathcal{I},\mathbf{J}_\text{MS}) .
\end{align*}
We then compute the relative error $|I_h - I|/|I|$,
where
\begin{align}
I &{}= a(\mathbf{J}_\text{MS},\mathbf{J}_\text{MS}),
\label{eq:I}
\\ \nonumber
I_h &{}= a_h(\mathbf{J}_\text{MS},\mathbf{J}_\text{MS}).
\end{align}
$I_h$ denotes the approximation to $I$ obtained by discretizing the domain into triangles and integrating over each triangle using quadrature.  \reviewerOne{By eliminating the presence of the basis functions, we eliminate the solution-discretization error.}

For a manufactured Green's function $G_\text{MS}$, $I$ can be computed exactly.  For the actual Green's function $G_k$, $I$ can be approximated by $I'=a'(\mathbf{J}_\text{MS},\mathbf{J}_\text{MS})$, which is computed with sufficient accuracy using the \texttt{CUBA} library. \reviewerBoth{In Section~\ref{sec:results}, we describe the tolerance criteria.} This approach provides an idea of how the numerical integration would perform without error from basis functions as it approximates the sesquilinear form.

\section{Numerical Examples} 
\label{sec:results}

In this section, we demonstrate the effectiveness of the approaches described in Section~\ref{sec:mms}.
We consider two unit-square flat plates, as shown in Figure~\ref{fig:two_squares}, with one rotated out of the plane of the other by an angle $\theta$.

\begin{figure}[!t]
\centering
\begin{subfigure}[b]{\textwidth}
\centering
\includegraphics[scale=.28,clip=true,trim=0in 0in 0in 0in]{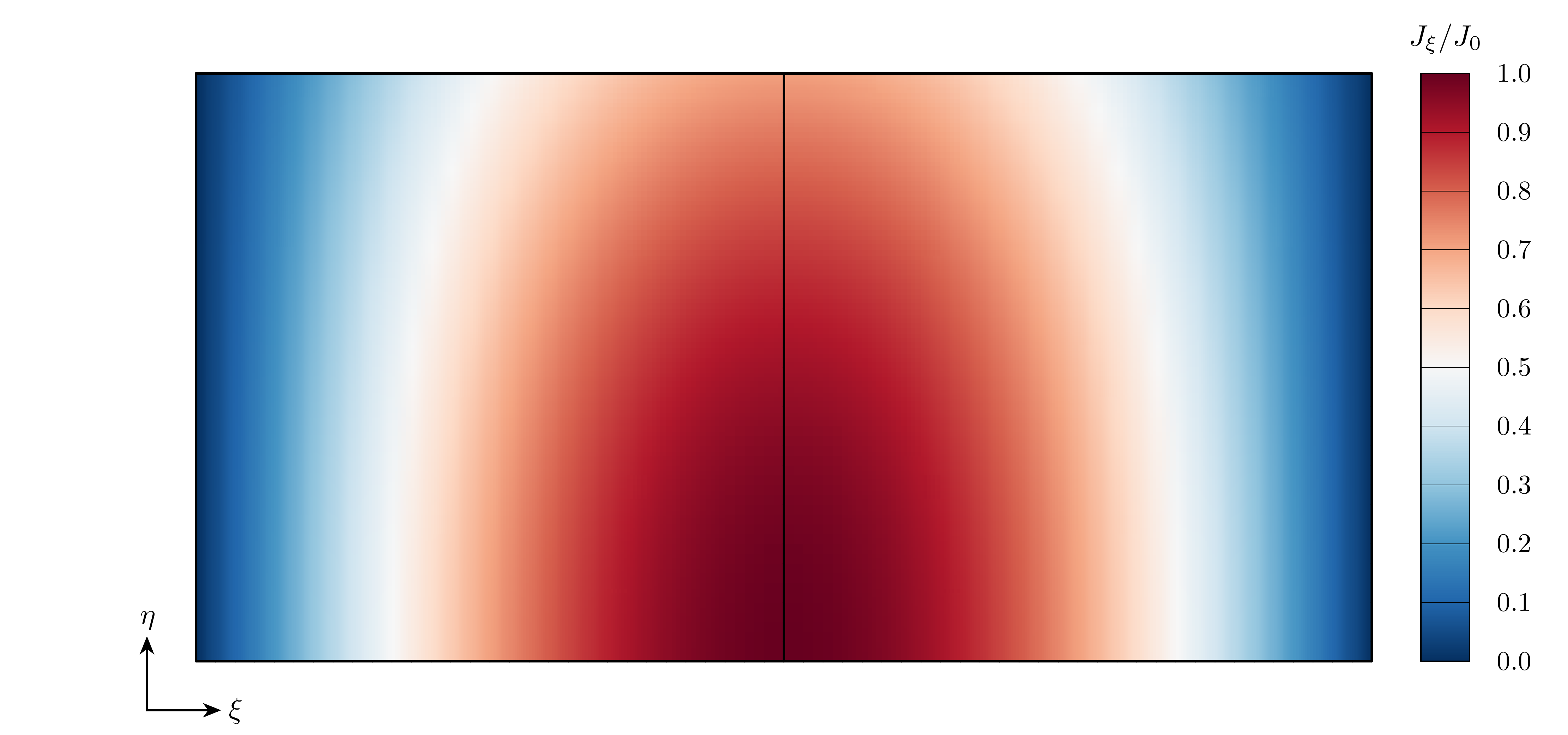}
\caption{\strut$J_\xi$}
\label{fig:Jxi}
\end{subfigure}

\begin{subfigure}[b]{\textwidth}
\centering
\includegraphics[scale=.28,clip=true,trim=0in 0in 0in 0in]{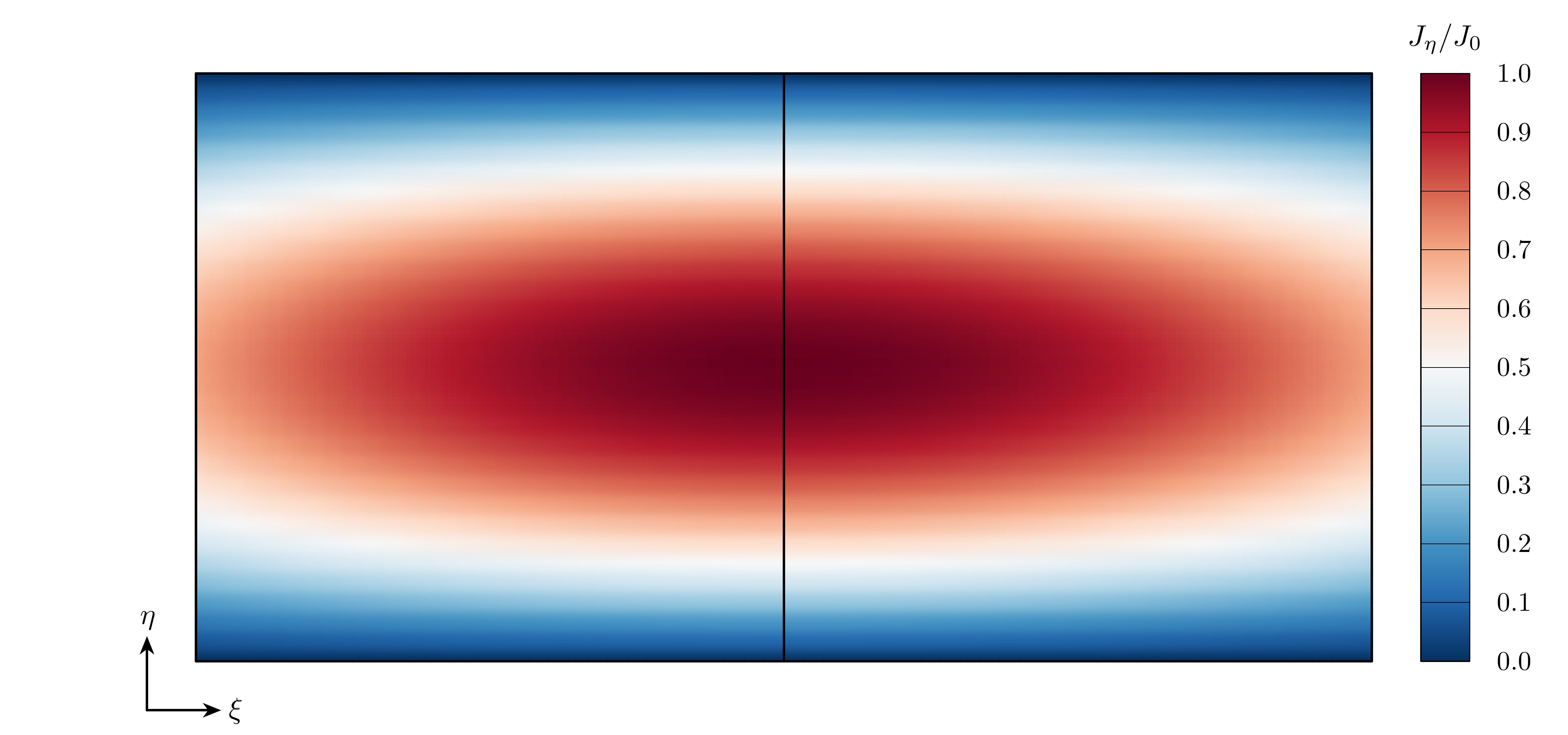}
\caption{\strut$J_\eta$}
\label{fig:Jeta}
\end{subfigure}
\caption{\strut Manufactured surface current $\mathbf{J}_\text{MS}$.}
\vskip-\dp\strutbox
\label{fig:J_MS}
\end{figure}

We manufacture the surface current on the plate $\mathbf{J}_\text{MS}(\mathbf{x}) = \{J_\xi(\boldsymbol{\xi}),\,J_\eta(\boldsymbol{\xi})\}$ using sinusoidal functions:
\begin{align}
J_\xi (\boldsymbol{\xi}) &{}= J_0 \cos(\pi \xi/2)\cos(\pi \eta/4), \label{eq:Jxi}\\
J_\eta(\boldsymbol{\xi}) &{}= J_0 \cos(\pi \xi/4)\sin(\pi \eta), \label{eq:Jeta}
\end{align}
where $J_0=1$ A/m and the plate-fixed coordinate system $\boldsymbol{\xi}(\mathbf{x};\theta)$ is given by
\begin{align*}
\xi(\mathbf{x};\theta) &{}= \frac{1}{L_0}\left\{\begin{matrix}
x, & \text{for } x\le 0 \text{ m} \\
x\cos\theta + z\sin\theta, & \text{for } x>0 \text{ m}
\end{matrix}\right.,
\\
\eta(\mathbf{x}) &{}= y/L_0,
\end{align*}
where $L_0=1$ m.
At the edges of the domain, the normal component of $\mathbf{J}_\text{MS}(\mathbf{x})$ is zero, satisfying the boundary conditions.  Figures~\ref{fig:Jxi} and~\ref{fig:Jeta} provide plots of~\eqref{eq:Jxi} and~\eqref{eq:Jeta}.

\newsavebox{\twosquares}%
\sbox{\twosquares}{%
\newcommand*\projA{-20}
\newcommand*\projB{ 35}
%
\definecolor{darkred} {RGB}{227,26,28}%
\definecolor{darkorange} {RGB}{255,127,0}%
\definecolor{lightred} {RGB}{243.8,163.4,164.2}%
\definecolor{darkblue}  {RGB}{ 31,120,180}%
\definecolor{lightblue}  {RGB}{165.4,201,225}
\definecolor{darkgreen} {RGB}{ 51,160, 44}%
\definecolor{lightgreen}{RGB}{173.4,217,170.6}
\def\xangle{100}
\def\yangle{60}
\begin{tikzpicture}[scale=0.6,
x={({ sin(\xangle)*1cm},{ cos(\xangle)*1cm})}, 
y={({.9*sin(\yangle)*1cm},{.9*cos(\yangle)*1cm})}, 
z={(0cm,1cm)}
    ]   

   \def\Deltaz{0};    
   \def\Deltay{0};    
   \def\h{6};         
   \def\xl{-6};       
   \def\xr{0};        
   \def\meas{1};      
   \def\mw{.25};      
   \def\thetaval{45}; 
   \def\arcrad{1};    
   \def\ep{.1}        
   \def\ymax{7}      
   
   \useasboundingbox (-11cm,-3cm) rectangle (7cm,9cm);
   
   \begin{scope}[canvas is yz plane at x=0,transform shape]
      \draw[line cap=round,gray] (\arcrad,0) arc (0:\thetaval:\arcrad);
      \node[scale=2] at ({1.5*\arcrad*cos(\thetaval/2)},{1.5*\arcrad*sin(\thetaval/2)}) {$\theta$};
      \pgflowlevelsynccm
      
   \end{scope}

   \begin{scope}[canvas is plane={O(0,0,0)x(1,0,0)y(0,{cos(\thetaval)},{sin(\thetaval)})}]
      \draw[line cap=round,dashed,gray] (0,0) -- (0,7-\ep);
      
      \pgflowlevelsynccm
      \draw[-{Stealth[scale=2.0,fill=gray,gray]},line cap=round,dash pattern=on 0cm off 100cm,text=gray] (0,0) to (0,7) node[above,scale=2] {$\xi$};

      \pgflowlevelsynccm
      \draw[-{Stealth[scale=2.0,fill=gray,gray]},line cap=round,dash pattern=on 0cm off 100cm,darkgreen] (0,0) to (0,7); 

   \end{scope}
   
   \begin{scope}[canvas is plane={O(0,0,0)x(1,0,0)y(0,{cos(\thetaval)},{sin(\thetaval)})}]
      \draw[thick,draw=darkgreen,fill=lightgreen,fill opacity=.5,draw opacity=1, line join=round] (\xr,0) -- (\xr,\h) -- (\xl,\h) -- (\xl,0) -- cycle;
      
      \node[transform shape,scale=2,text=darkgreen,anchor=west] at (\xr,\h) {$(1,0)$};
      \node[transform shape,scale=2,text=darkgreen,anchor=east] at (\xl,\h) {$(1,1)$};
      \draw[draw=darkgreen,fill=darkgreen] (\xr,\h) circle (.1);
      \draw[draw=darkgreen,fill=darkgreen] (\xl,\h) circle (.1);
   \end{scope}

   \begin{scope}[canvas is xy plane at z=0]
      \draw[thick,draw=darkblue,fill=lightblue,fill opacity=.5,draw opacity=1, line join=round] (\xr,-\h) -- (\xr,0) -- (\xl,0) -- (\xl,-\h) -- cycle;    

      \node[transform shape,scale=2,text=darkblue,anchor=west] at (\xr,-\h) {$(-1,0)$};
      \node[transform shape,scale=2,text=darkblue,anchor=east] at (\xl,-\h) {$(-1,1)$};
      \draw[draw=darkblue,fill=darkblue] (\xr,-\h) circle (.1);
      \draw[draw=darkblue,fill=darkblue] (\xl,-\h) circle (.1);   
   \end{scope} 
   
   \begin{scope}[canvas is xy plane at z=0,transform shape]
   \draw[line cap=round] (0,0) -- (-7+\ep, 0);
   \draw[line cap=round] (0,0) -- (0,\ymax-\ep);
   \pgflowlevelsynccm
   \draw[-{Stealth[scale=2.0]},line cap=round] (0,0) to (-7, 0) node[left,scale=2,dash pattern=on 0cm off 100cm] {$y=\eta$};
   \draw[-{Stealth[scale=2.0]},line cap=round] (0,0) to (0,\ymax) node[above,scale=2,dash pattern=on 0cm off 100cm] {$x$};
   \end{scope} 
  
   \begin{scope}[canvas is xz plane at y=0]
   \draw[line cap=round] (0,0) -- (0,8-\ep);
   \pgflowlevelsynccm
   \draw[-{Stealth[scale=2.0]},line cap=round,dash pattern=on 0cm off 100cm] (0,0) to (0,8) node[above,scale=2] {$z$};
   \end{scope}

\end{tikzpicture}%
}%
\begin{figure}[!b]
\centering
\usebox{\twosquares}%

\caption{Computational domain consisting of two unit-square plates.  Coordinates are expressed in the plate-fixed coordinate system $\boldsymbol{\xi}(\mathbf{x};\theta)$.}
\label{fig:two_squares}
\end{figure}

\begin{figure}[!t]
\centering
\begin{subfigure}[b]{\textwidth}
\centering
\includegraphics[scale=.28,clip=true,trim=0in 0in 0in 0in]{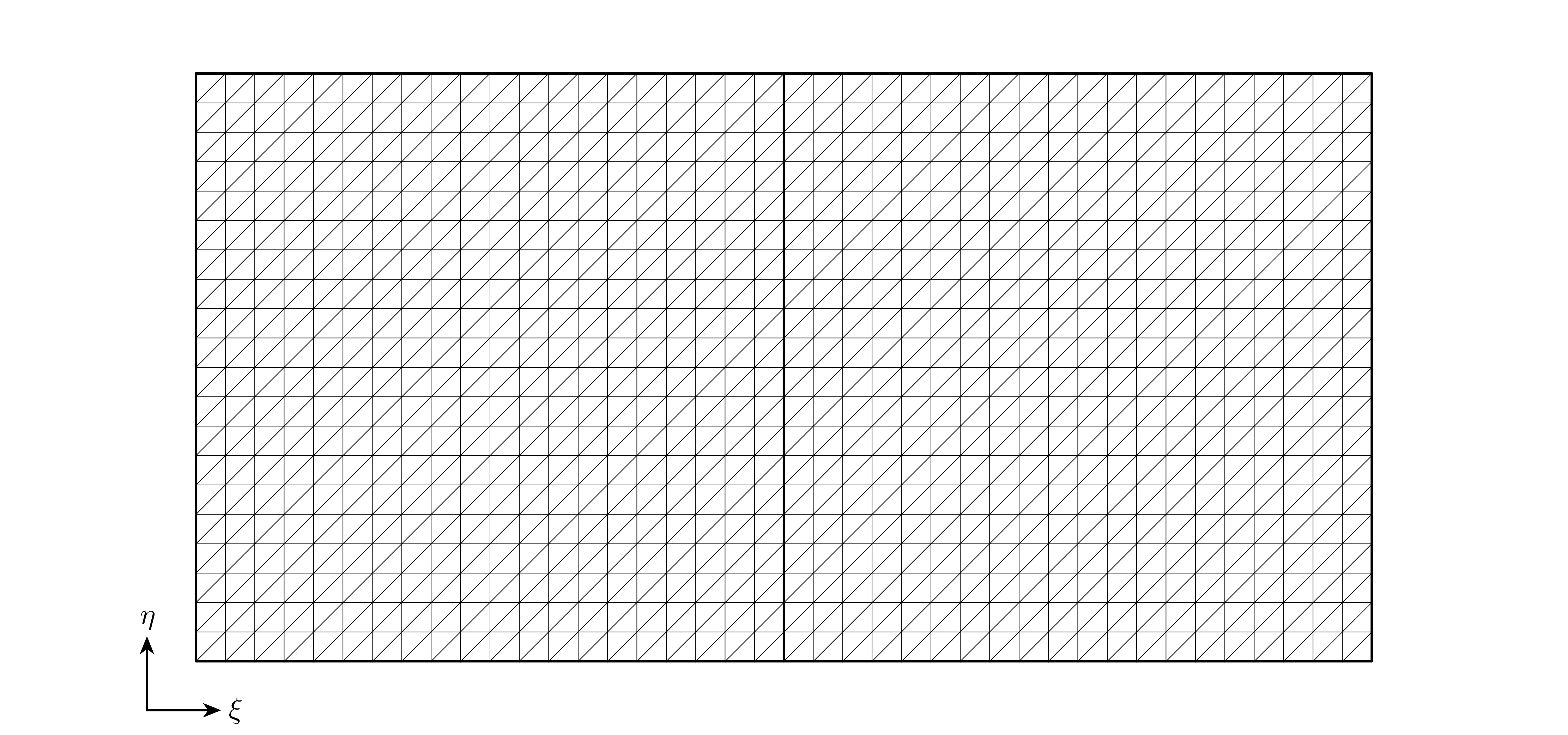}
\caption{\strut Uniform}
\label{fig:uniform}
\end{subfigure}

\begin{subfigure}[b]{\textwidth}
\centering
\includegraphics[scale=.28,clip=true,trim=0in 0in 0in 0in]{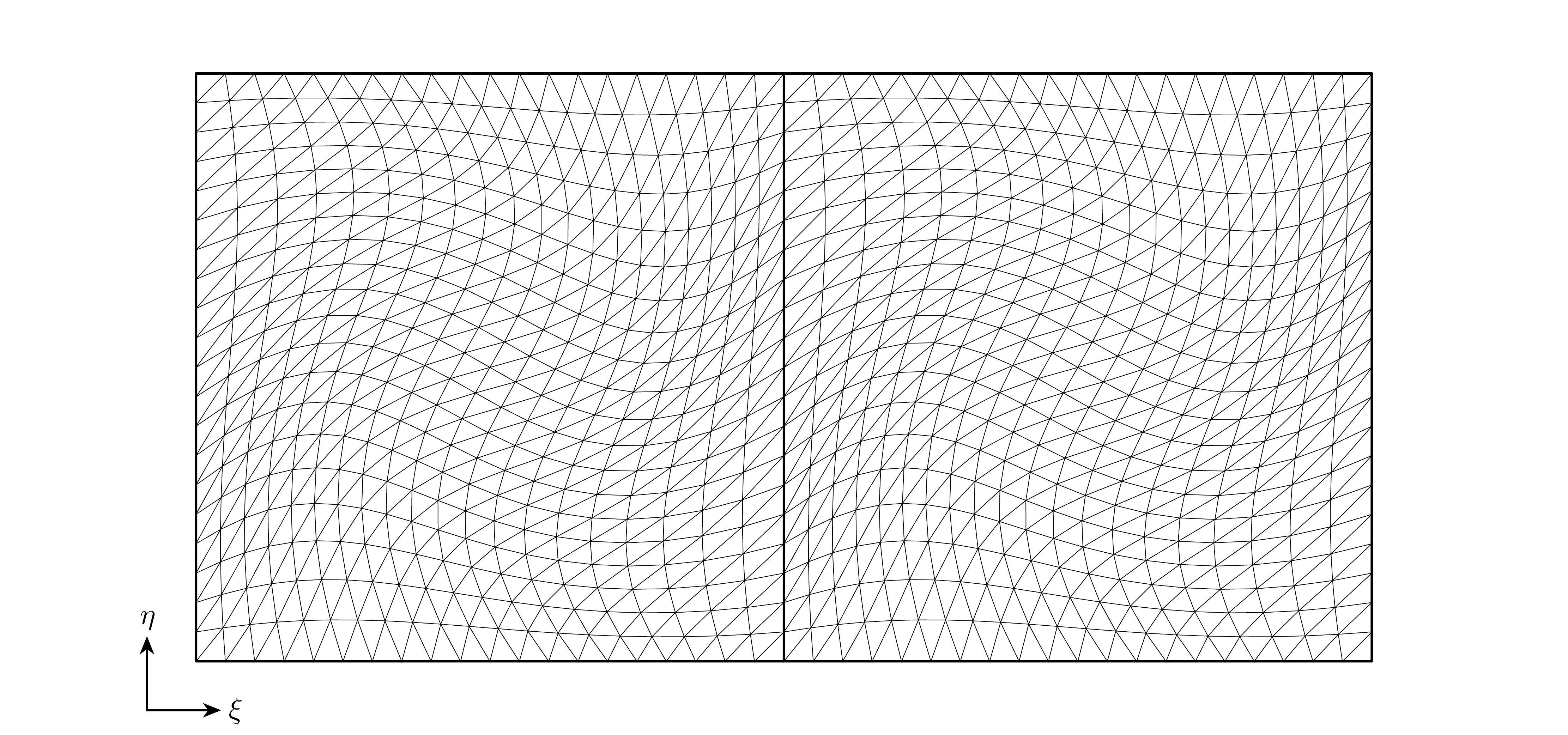}
\caption{\strut Twisted}
\label{fig:twisted}
\end{subfigure}
\caption{\strut Two different types of meshes, shown with $\ntriangles=1600$.}
\vskip-\dp\strutbox
\label{fig:mesh}
\end{figure}

We consider two types of meshes: a uniform mesh and a twisted mesh, examples of which are shown in Figures~\ref{fig:uniform} and~\ref{fig:twisted} with the total number of triangles $\ntriangles=1600$.  The twisted mesh is obtained by transforming the uniform mesh, using the transformation provided in Reference~\cite{freno_2021}.

As in Reference~\cite{freno_em_mms_2020}, we account for potential disparities in the magnitudes of the contributions to $\mathbf{Z}$~\eqref{eq:matrix_elements} from $\mathbf{A}$ and $\Phi$:  
\begin{align*}
\mathbf{Z} = \mathbf{Z}^\mathbf{A} + \mathbf{Z}^\Phi,
\end{align*}
where
%
$Z_{\testidx,\srcidx}^\mathbf{A} = a^\mathbf{A}(\boldsymbol{\Lambda}_{\srcidx},\boldsymbol{\Lambda}_{\testidx})$
and
$Z_{\testidx,\srcidx}^\Phi =  a^\Phi(\boldsymbol{\Lambda}_{\srcidx},\boldsymbol{\Lambda}_{\testidx})$.
%
\reviewerOne{%
To ensure the errors of one are not overshadowed by those of the other, we consider the contributions $\mathbf{Z}^\mathbf{A}$ and $\mathbf{Z}^\Phi$ together and separately, with $\epsilon=1$ F/m and $\mu=1$ H/m.  
$\mathbf{Z}^\mathbf{A}$ contains the factor $\omega\mu=k^2/(\epsilon\omega)$, whereas $\mathbf{Z}^\Phi$ contains the factor $1/(\epsilon\omega)$.
When we consider the contributions separately, with the exception of $G_k$, we are effectively taking the limits as $k\to\infty$ for $\mathbf{Z}^\mathbf{A}$ and $k\to 0$ for $\mathbf{Z}^\Phi$.  However, for the considered term, we set $k=1$ m$^{-1}$.
When we consider the contributions together, we set $k=1$ m$^{-1}$ for $\mathbf{Z}_\text{MS}$}.    We adjust $V_{\text{MS}_{\testidx}}$~\eqref{eq:V_mms} accordingly.  We similarly consider the contributions to $I$~\eqref{eq:I}:
\begin{align*}
I = I^\mathbf{A} + I^\Phi,
\end{align*}
where
%
$I^\mathbf{A} = a^\mathbf{A}(\mathbf{J}_\text{MS}, \mathbf{J}_\text{MS})$
and
$I^\Phi = a^\Phi(\mathbf{J}_\text{MS},\mathbf{J}_\text{MS})$.

We consider two Green's functions: $G_k$~\eqref{eq:G} and $G_\text{MS}=1-R^2/R_m^2$, where $R_m=\max_{\mathbf{x},\mathbf{x}'\in S} R$ is the maximum possible distance between two points on $S$.  For these choices, $\mathbf{Z}$, $\mathbf{Z}^\mathbf{A}$, and $\mathbf{Z}^\Phi$ are practically singular for $G_\text{MS}$, and $\mathbf{Z}^\Phi$ is practically singular for $G_k$.  For these cases, we use the optimization approach in Reference~\cite{freno_em_mms_2020} to obtain a unique solution.


\reviewerOne{%
We evaluate the integrals in $\mathbf{Z}$ and $I$ using quadrature.  For $G_k$, we consider the eight quadrature combinations listed in Table~\ref{tab:G_quad_combinations}.
When the integrals do not contain singularities, we employ polynomial triangle quadrature rules~\cite{lyness_1975,dunavant_1985}, for which the numbers of points are listed in Table~\ref{tab:G_quad_combinations}.
However, for $G_k$, the integrals contain singularities when the test and source triangles share one or more edges or vertices and near-singularities when they are otherwise close.  Therefore, various approaches have been developed to address these (near-)singularities in both the source and test integrals.  For (near-)singularities in the source integral, we employ the radial--angular transformation~\cite{khayat_2008}, for which the numbers of points in the radial and transverse directions are listed in Table~\ref{tab:G_quad_combinations}.  For (near-)singularities in the test integral, we either use polynomial triangle quadrature rules or the Approach 1, 1D rules of Reference~\cite{freno_quad}.  If applicable, in Table~\ref{tab:G_quad_combinations}, we list the number of points for the Approach 1, 1D rules in the `(Near-)Singular' row.
}


\begin{table}
\centering
\reviewerOne{%
\begin{tabular}{c r c c c c c c c c}
\toprule
& Combination                      & Q1     & Q2     & Q3     & Q4     & Q5     & Q6     & Q7     & Q8     \\ \midrule
\multirow{2}{*}{\begin{tabular}{@{} c @{}}Test \\ Points \end{tabular}}   & Polynomial Rules & \pz7   & \pz7   & 16     & 16     & \pz7   & \pz7   & 16     & 16     \\
                                                                          & (Near-)Singular  & ---    & ---    & ---    & ---    & \pz7   & \pz7   & 16     & 16     \\ \midrule
\multirow{3}{*}{\begin{tabular}{@{} c @{}}Source \\ Points \end{tabular}} & Polynomial Rules & \pz7   & 16     & \pz7   & 16     & \pz7   & 16     & \pz7   & 16     \\
                                                                          & Radial           & \pz3   & \pz6   & \pz3   & \pz6   & \pz3   & \pz6   & \pz3   & \pz6   \\
                                                                          & Transverse       & 12     & 24     & 12     & 24     & 12     & 24     & 12     & 24     \\
\bottomrule
\end{tabular}}
\caption{Quadrature combinations for $G_k$.}
\vskip-\dp\strutbox
\label{tab:G_quad_combinations}
\end{table}


\subsection{Solution-Discretization Error} 

We begin by performing the assessment described in Section~\ref{sec:part1}, measuring the error norm $\left\|\mathbf{e}\right\|_{H_\text{div}^{-1/2}(S)}$~\eqref{eq:hdiv_norm}, adjusted accordingly for $\mathbf{Z}$, $\mathbf{Z}^\mathbf{A}$, and $\mathbf{Z}^\Phi$.  We consider the uniform and twisted meshes and four $\theta$ values.  We integrate the matrix elements~\eqref{eq:matrix_elements} and norm integrals~\eqref{eq:norm_integral} using Quadrature Combination~Q8 in Table~\ref{tab:G_quad_combinations}.  The relative convergence tolerance $\epsilon$ for the adaptive integration of $\mathbf{V}_{\text{MS}}(\mathbf{J}_\text{MS})$~\eqref{eq:zjv_mmsp} \reviewerBoth{via the \texttt{CUBA} library} is initialized to $\epsilon=10^{-3}$ and reduced by a factor of ten until the relative changes in $\left\|\mathbf{e}\right\|_{H_\text{div}^{-1/2}(S)}$~\eqref{eq:hdiv_norm} and $\left\|\mathbf{e}_n\right\|_\infty$~\eqref{eq:linf_norm} are both less than $10^{-3}$.  

\begin{figure}
\centering
\begin{subfigure}[b]{.49\textwidth}
\includegraphics[scale=.64,clip=true,trim=2.3in 0in 2.8in 0in]{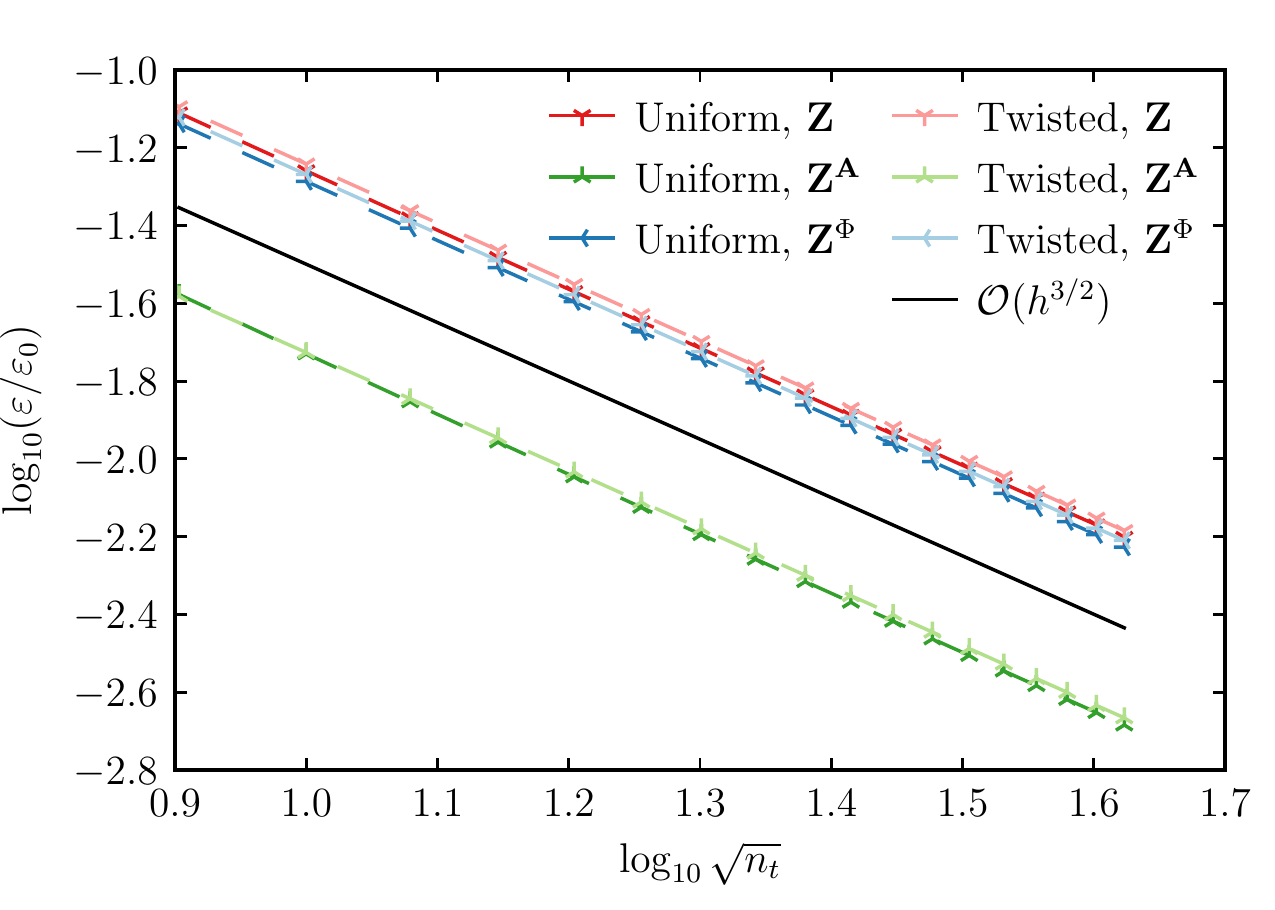}

\caption{$\theta=0^\circ$}
\end{subfigure}
\hspace{0.25em}
\begin{subfigure}[b]{.49\textwidth}
\includegraphics[scale=.64,clip=true,trim=2.3in 0in 2.8in 0in]{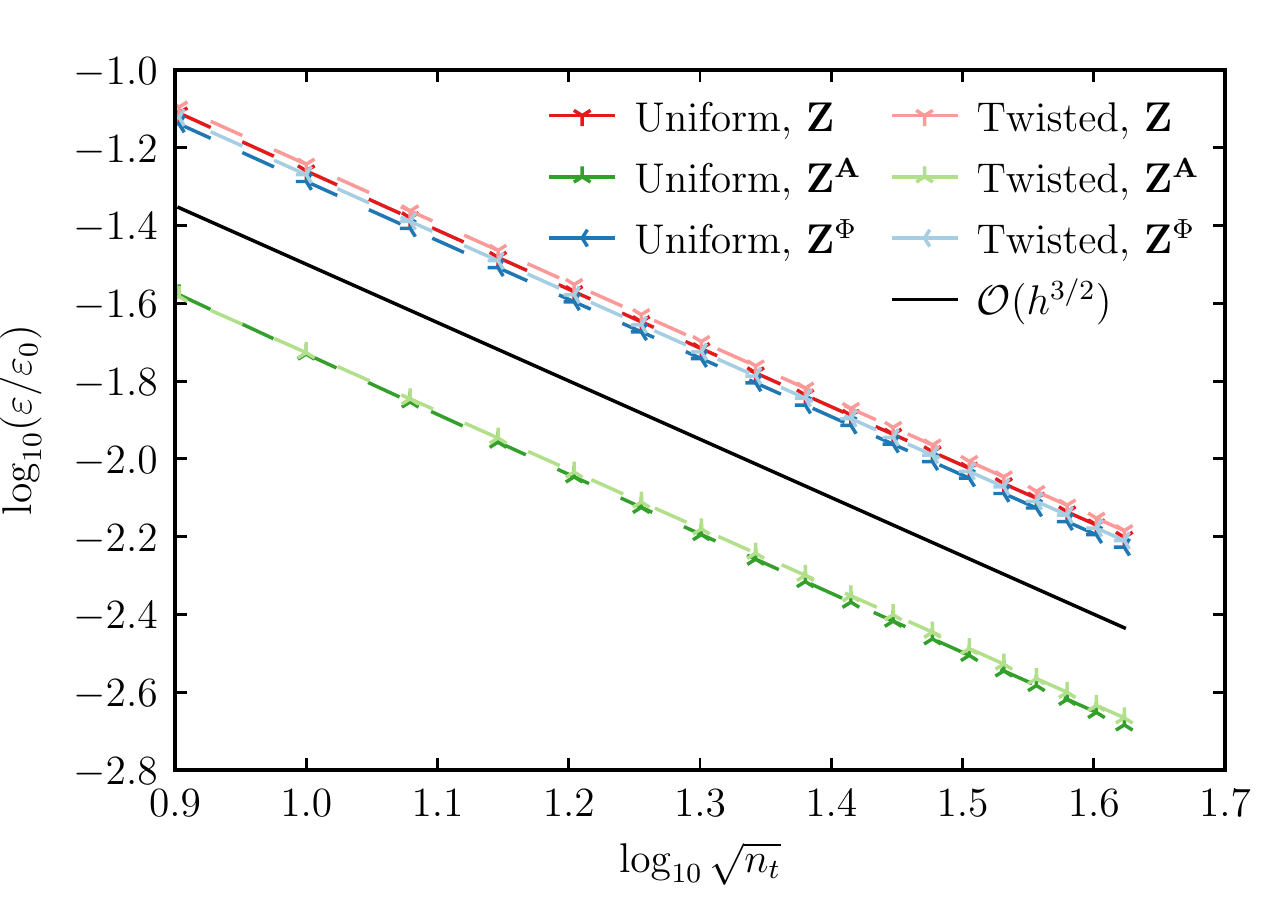}

\caption{$\theta=45^\circ$}
\end{subfigure}
\\
\begin{subfigure}[b]{.49\textwidth}
\includegraphics[scale=.64,clip=true,trim=2.3in 0in 2.8in 0in]{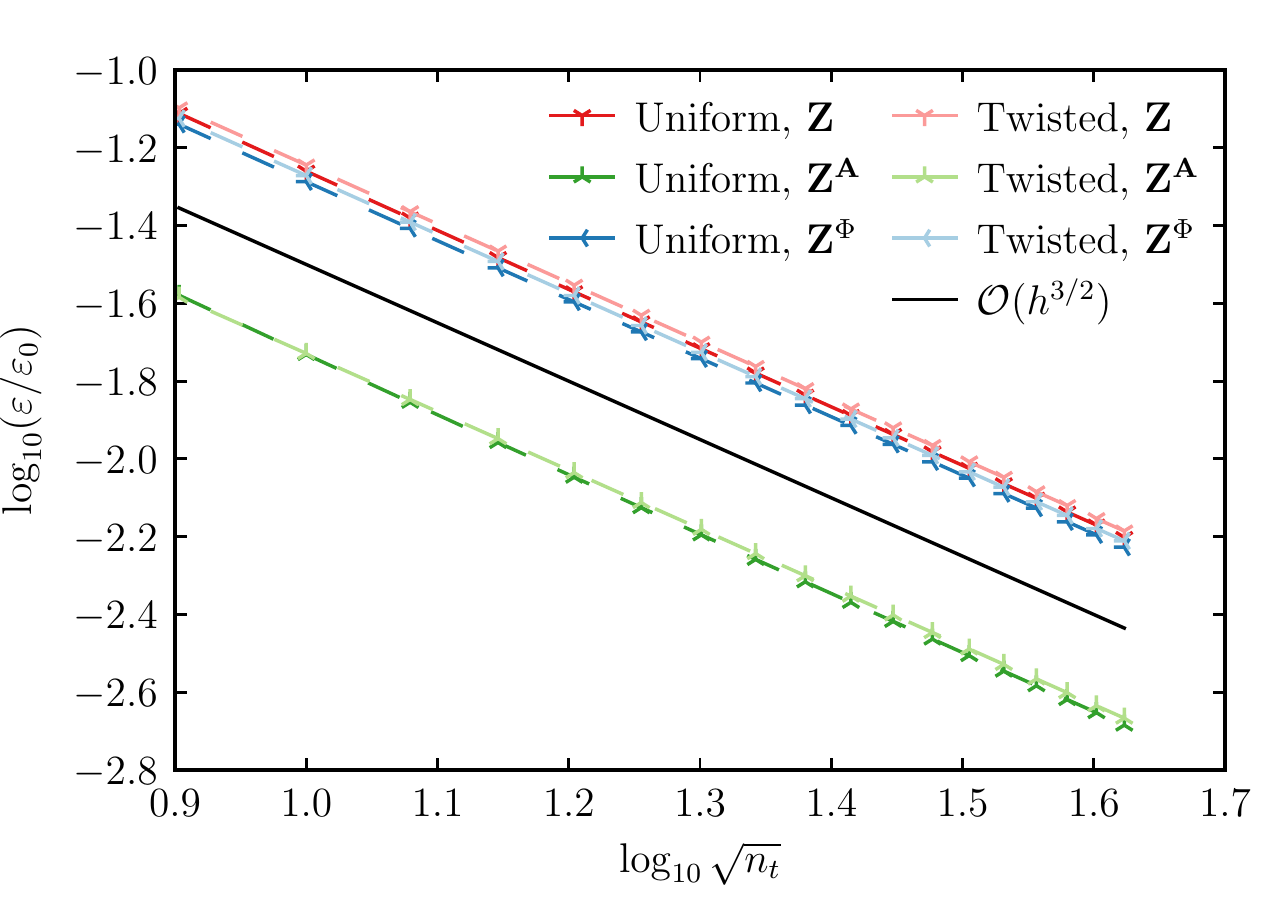}

\caption{$\theta=90^\circ$}
\end{subfigure}
\hspace{0.25em}
\begin{subfigure}[b]{.49\textwidth}
\includegraphics[scale=.64,clip=true,trim=2.3in 0in 2.8in 0in]{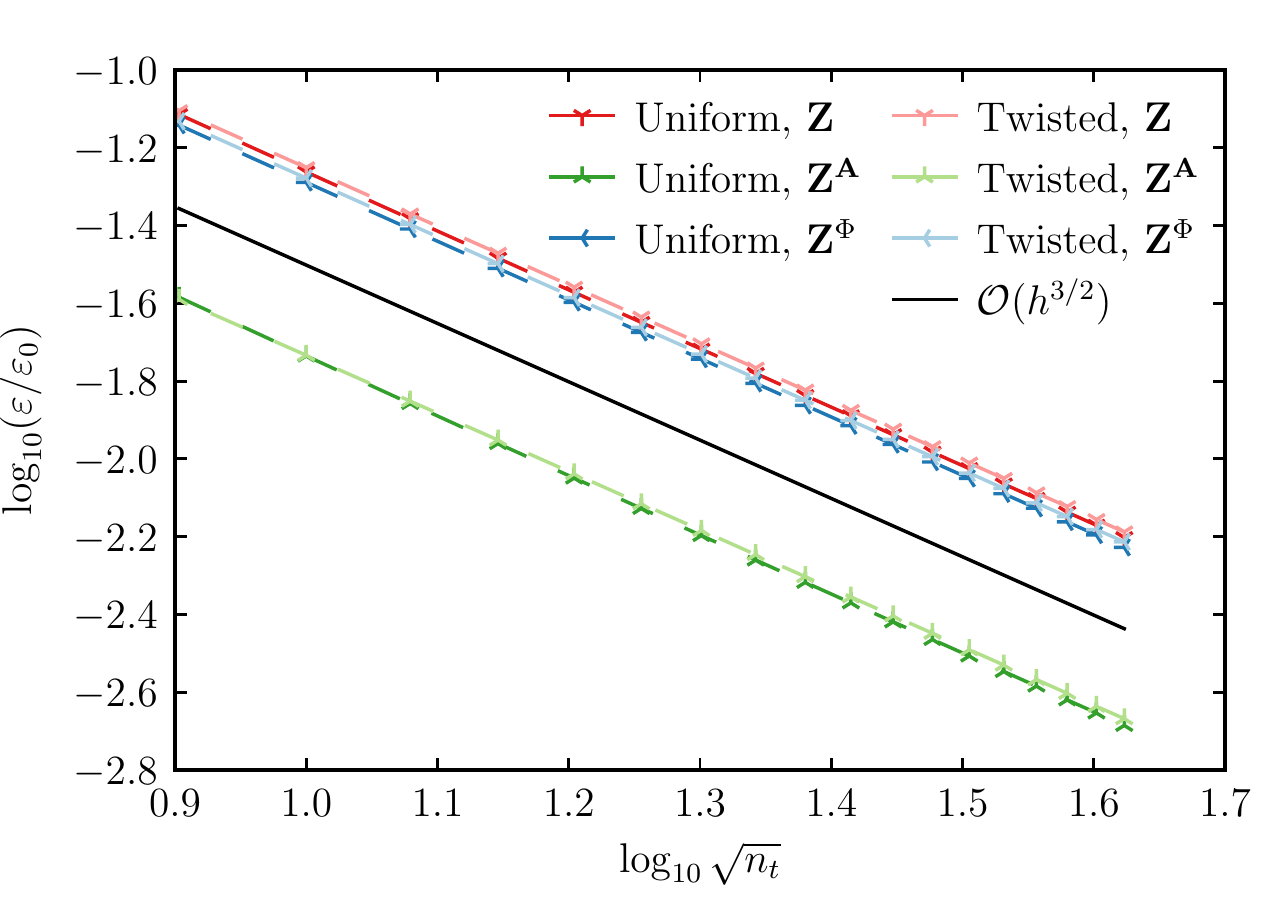}

\caption{$\theta=135^\circ$}
\end{subfigure}
\caption{Solution-discretization error: $\varepsilon=\left\|\mathbf{e}\right\|_{H_\text{div}^{-1/2}(S)}$~\eqref{eq:hdiv_norm}, with $G_k$ and Q8.}
\vskip-\dp\strutbox
\label{fig:part1}
\end{figure}

Figure~\ref{fig:part1} shows the error norm $\varepsilon=\left\|\mathbf{e}\right\|_{H_\text{div}^{-1/2}(S)}$~\eqref{eq:hdiv_norm}, nondimensionalized by the constant $\varepsilon_0=1$ W$^{1/2}$.  Each of the simulations achieves the expected $\mathcal{O}(h^{3/2})$ rate.

\begin{figure}
\centering
\begin{subfigure}[b]{.49\textwidth}
\includegraphics[scale=.64,clip=true,trim=2.3in 0in 2.8in 0in]{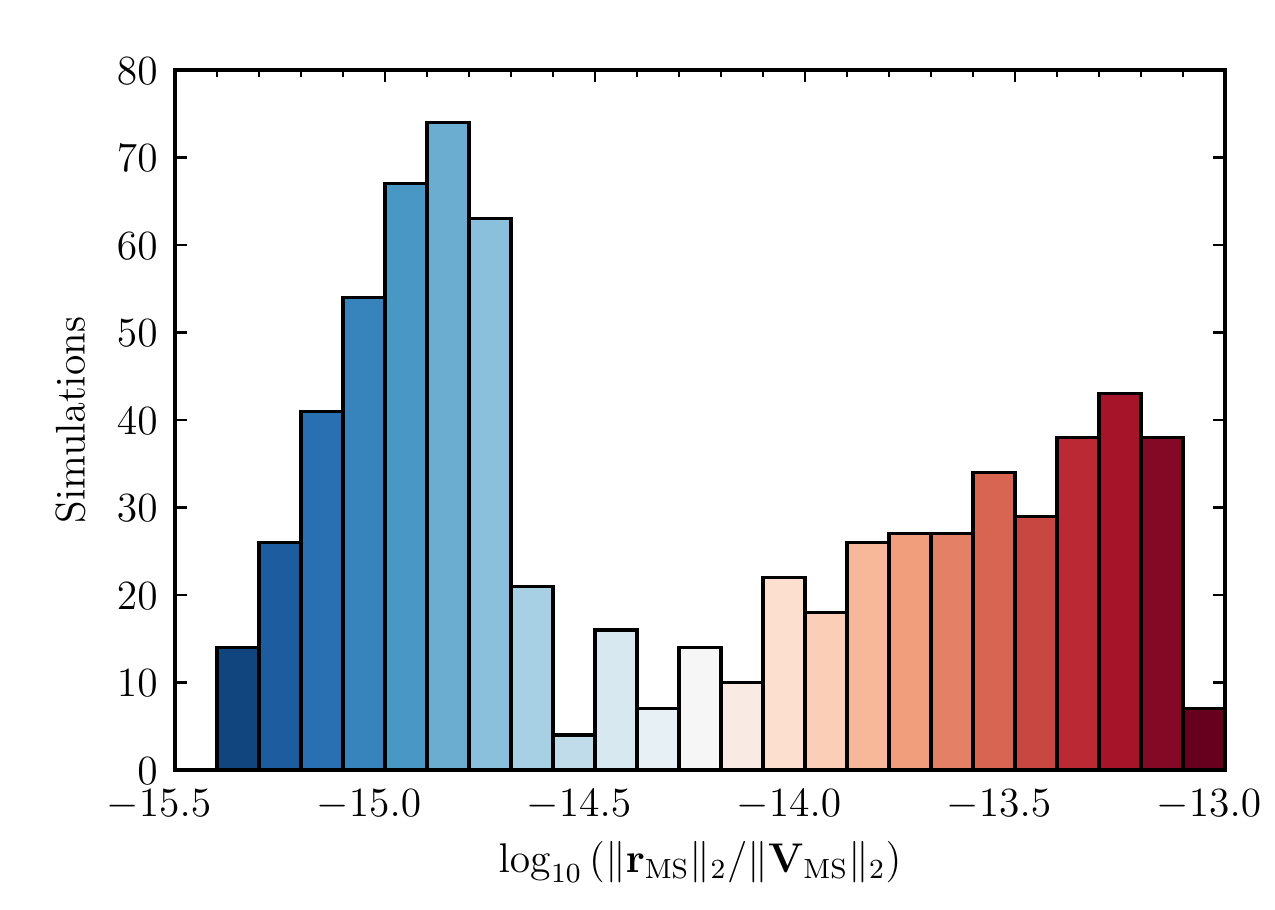}

\caption{$\mathbf{Z}$ and $\mathbf{Z}^\mathbf{A}$}
\end{subfigure}
\hspace{0.25em}
\begin{subfigure}[b]{.49\textwidth}
\includegraphics[scale=.64,clip=true,trim=2.3in 0in 2.8in 0in]{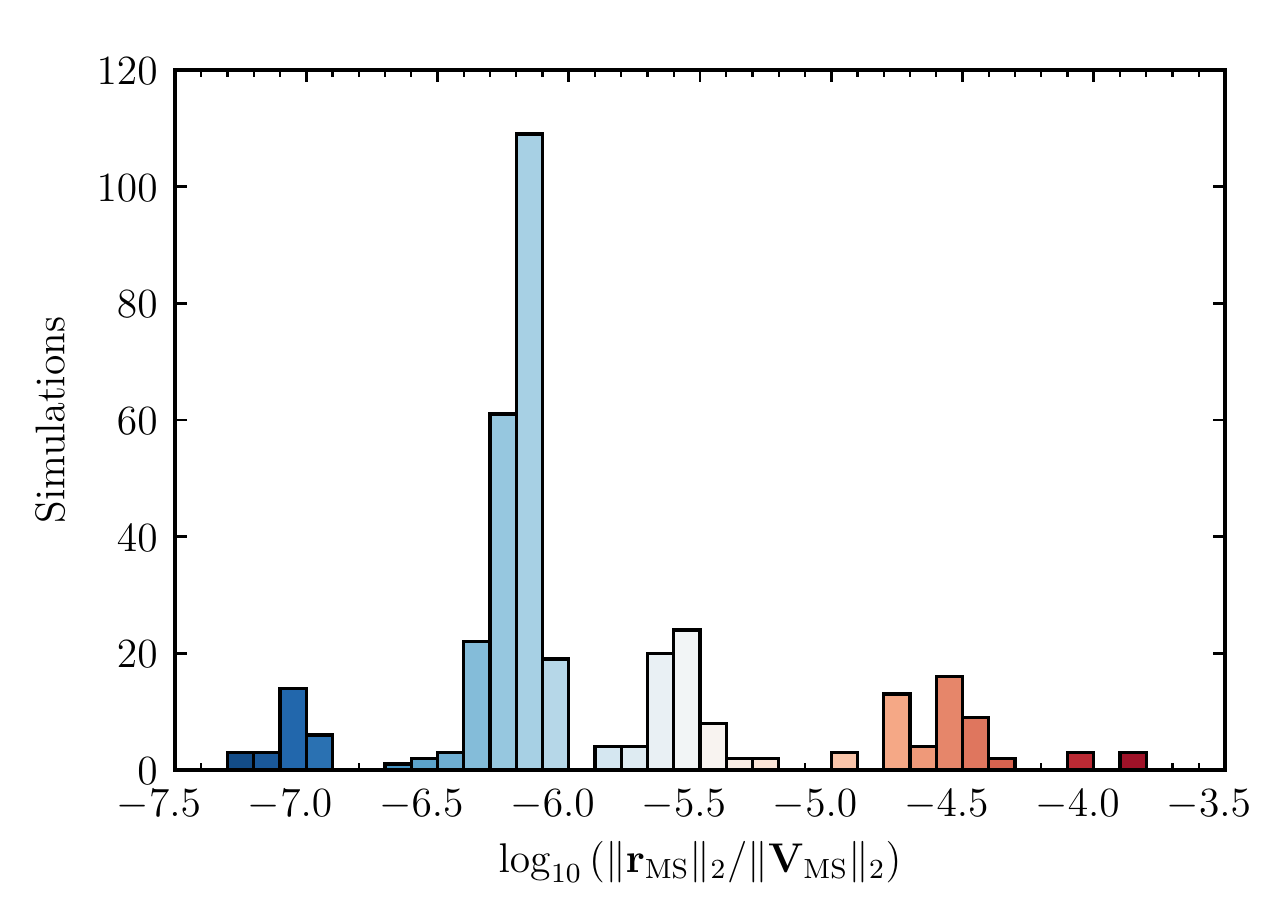}

\caption{$\mathbf{Z}^\Phi$}
\end{subfigure}
\caption{Solution-discretization error: $\mathbf{r}_\text{MS}=\mathbf{Z}\mathbf{J}^h - \mathbf{V}_{\text{MS}}(\mathbf{J}_\text{MS})$~\eqref{eq:zjv_mmsp} for the simulations in Figure~\ref{fig:part1}.}
\vskip-\dp\strutbox
\label{fig:part1_res}
\end{figure}

Because $\mathbf{Z}^\Phi$ is practically singular, we perform the optimization approach in Reference~\cite{freno_em_mms_2020} to obtain a unique solution.  
\reviewerOne{However, the integrals in Reference~\cite{freno_em_mms_2020} were computed exactly.  Therefore, the redundant equations that caused the matrix to be singular were computed consistently on both sides of the system of equations, such that the residual was zero.  On the other hand, in this study, the nonzero numerical-integration errors yield discrepancies between redundant equations, such that the $L^2$-norms of the residuals for $\mathbf{Z}^\Phi$ in Figure~\ref{fig:part1_res} are noticeably nonzero for the simulations when compared to residual norms for $\mathbf{Z}$ and $\mathbf{Z}^\mathbf{A}$, which are nonsingular.  Nonetheless, given the lack of a suitable alternative for addressing the singular matrices in this study, we employ the approach of Reference~\cite{freno_em_mms_2020} outside of its intended scope and view the residuals as a measure of the appropriateness of doing so.}

\subsection{Numerical-Integration Error} 

\begin{figure}[!t]
\centering
\begin{subfigure}[b]{.49\textwidth}
\includegraphics[scale=.64,clip=true,trim=2.3in 0in 2.8in 0in]{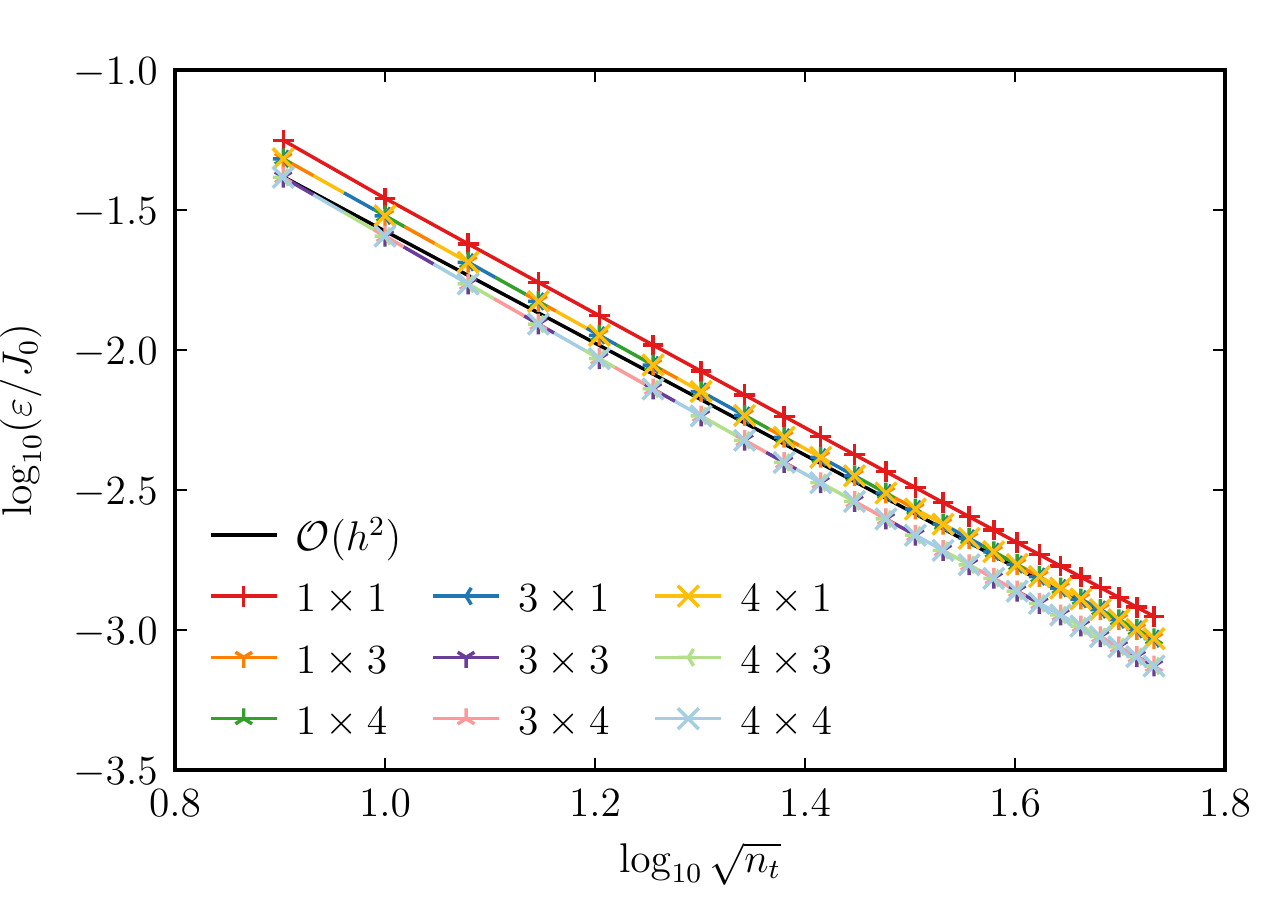}

\caption{$\mathbf{Z}^\mathbf{A}$}
\end{subfigure}
\hspace{0.25em}
\begin{subfigure}[b]{.49\textwidth}
\includegraphics[scale=.64,clip=true,trim=2.3in 0in 2.8in 0in]{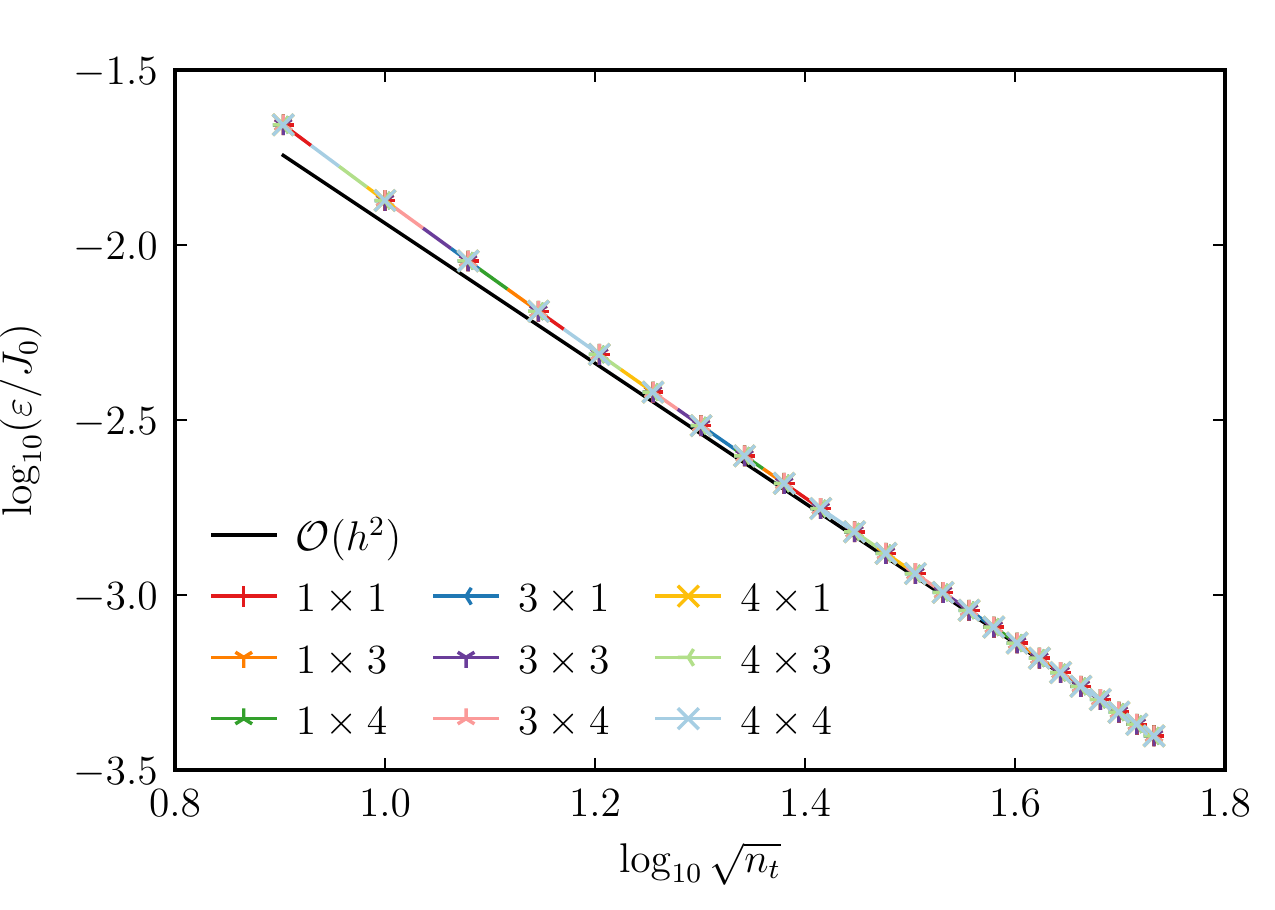}

\caption{$\mathbf{Z}^\Phi$}
\end{subfigure}
\caption{Solution-discretization error: $\varepsilon=\left\|\mathbf{e}_n\right\|_\infty$, with $G_\text{MS}$, $\theta=0^\circ$, and the uniform mesh.}
\vskip-\dp\strutbox
\label{fig:part4}
\end{figure}

\begin{figure}[!t]
\centering
\begin{subfigure}[b]{.49\textwidth}
\includegraphics[scale=.64,clip=true,trim=2.3in 0in 2.8in 0in]{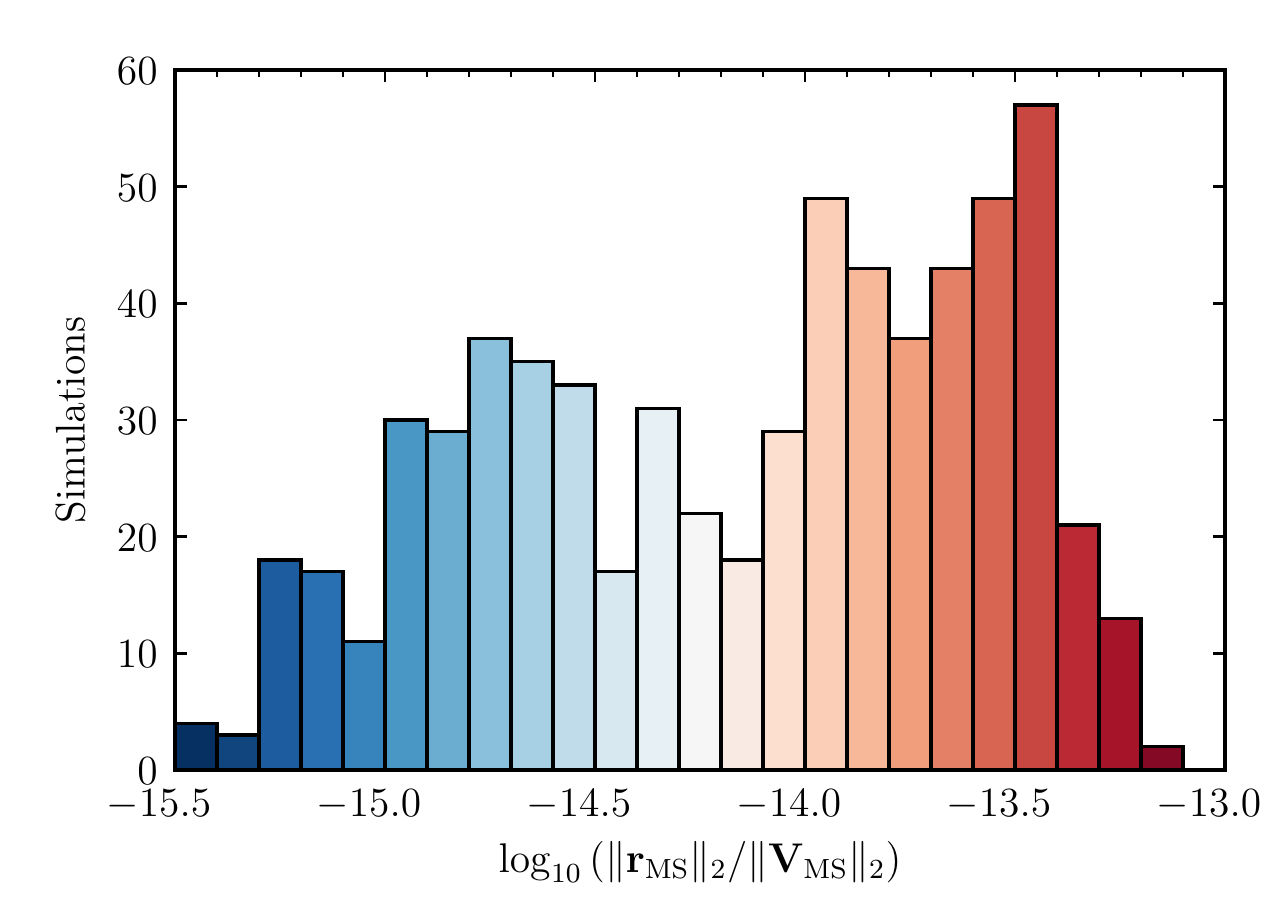}
\end{subfigure}
\caption{Solution-discretization error: $\mathbf{r}_\text{MS}=\mathbf{Z}\mathbf{J}^h - \mathbf{V}_{\text{MS}}(\mathbf{J}_\text{MS})$~\eqref{eq:zjv_mmsp} for the simulations in Figure~\ref{fig:part4}.}
\vskip-\dp\strutbox
\label{fig:part4_res}
\end{figure}

\begin{table}[!t]
\centering
\begin{tabular}{l c c c c c c c c}
\toprule
Number of points         & 1     & 3     & 4     &     6 &     7 &    12 &    13 &   16 \\ \midrule
Maximum integrand degree & 1     & 2     & 3     &     4 &     5 &  \pz6 &  \pz7 & \pz8 \\
Convergence rate         & $\mathcal{O}(h^2)$ & $\mathcal{O}(h^4)$ & $\mathcal{O}(h^4)$ & $\mathcal{O}(h^6)$ & $\mathcal{O}(h^6)$ & $\mathcal{O}(h^8)$ & $\mathcal{O}(h^8)$ & $\mathcal{O}(h^{10})$ \\
\bottomrule
\end{tabular}
\caption{Polynomial triangle quadrature properties.}
\vskip-\dp\strutbox
\label{tab:dunavant_properties}
\end{table}

To provide motivation for this section, we begin with Figure~\ref{fig:part4}, which shows the error norm $\varepsilon=\left\|\mathbf{e}_n\right\|_\infty$~\eqref{eq:linf_norm}, nondimensionalized by the constant $J_0=1$ A/m, for $G_\text{MS}$, $\theta=0^\circ$, and the uniform mesh, with multiple $n_\text{test} \times n_\text{source}$ polynomial triangle quadrature point combinations. 
As stated previously, $G_\text{MS}$ yields practically singular matrices, such that we perform the optimization approach in Reference~\cite{freno_em_mms_2020} to obtain a unique solution.  Once again, with a generally nonzero numerical-integration error, we are not assured nonzero residuals; however, Figure~\ref{fig:part4_res} shows the residuals are \reviewerTwo{less than $10^{-13}$} for this case.

For multiple polynomial triangle quadrature point amounts, Table~\ref{tab:dunavant_properties} lists the maximum integrand polynomial degree the points can integrate exactly, as well as the convergence rates of the errors for inexact integrations of nonsingular integrands.  For nonsingular integrands, the slowest expected quadrature convergence rate is $\mathcal{O}(h^2)$.
For $G_\text{MS}$, the integrands of $\mathbf{Z}$, $\mathbf{Z}^\mathbf{A}$, and $\mathbf{Z}^\Phi$ consist of polynomials of the same degree in the test integration variables as in the source integration variables.  Therefore, the smaller amount of quadrature points between those used for the test and source domains limits the accuracy.  The integrands of $\mathbf{Z}$ and $\mathbf{Z}^\mathbf{A}$ contain polynomials of degree three, such that they are exactly integrated with $4\times 4$ points.  The integrand of $\mathbf{Z}^\Phi$ contains polynomials of degree two, such that it is exactly integrated with $3\times 3$ points.

\begin{figure}
\centering
\begin{subfigure}[b]{.49\textwidth}
\includegraphics[scale=.64,clip=true,trim=2.3in 0in 2.8in 0in]{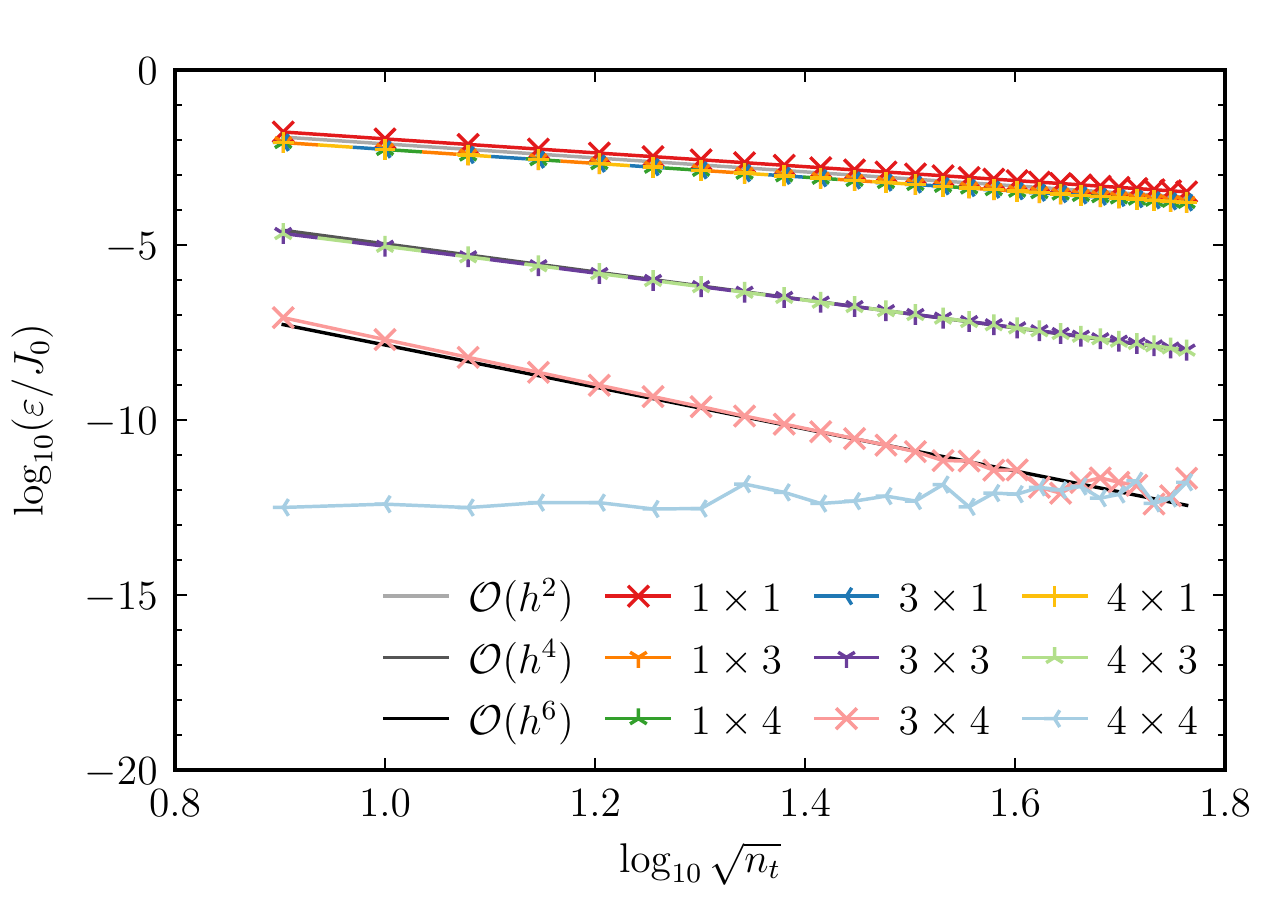}
\caption{$\mathbf{Z}$}
\end{subfigure}
\hspace{0.25em}
\begin{subfigure}[b]{.49\textwidth}
\includegraphics[scale=.64,clip=true,trim=2.3in 0in 2.8in 0in]{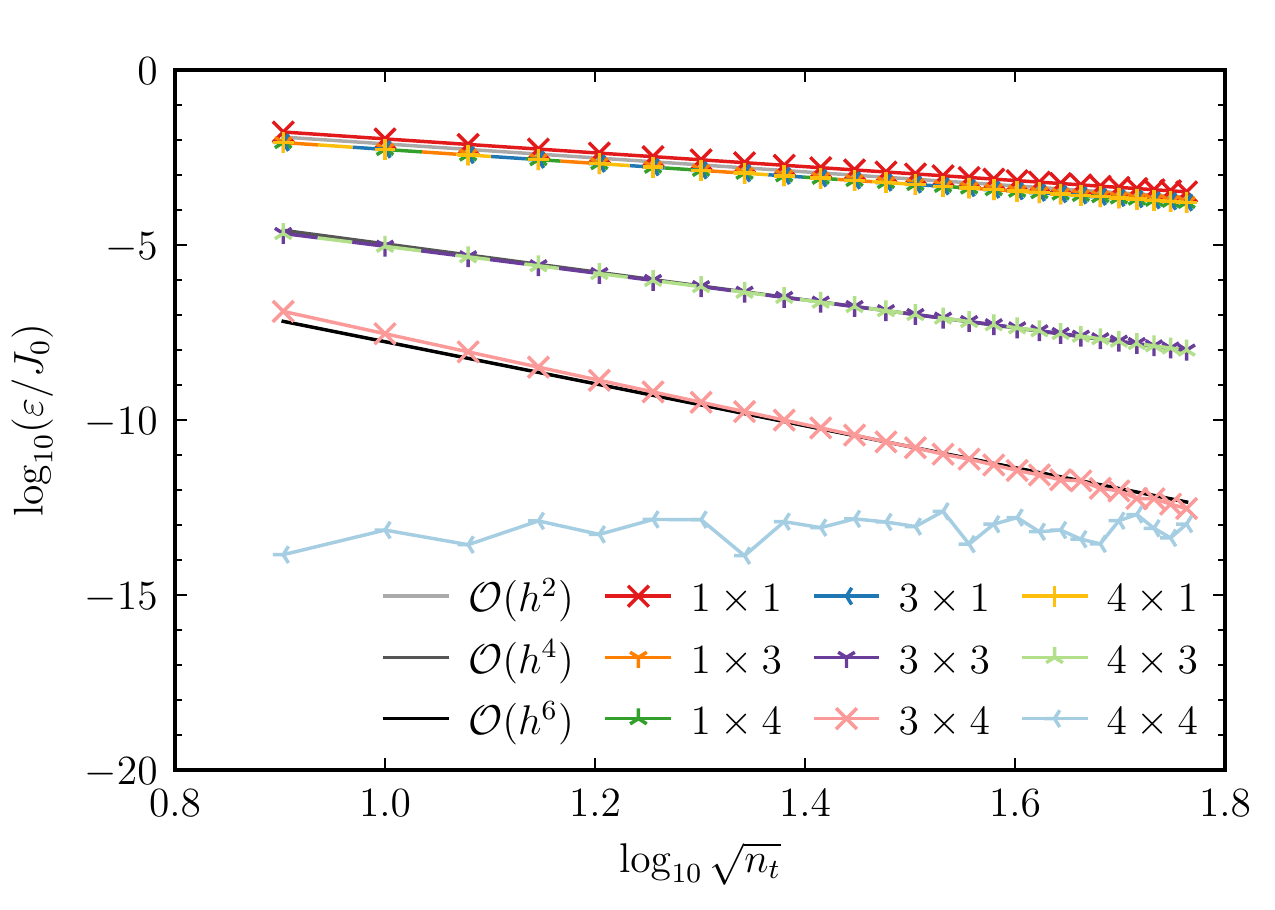}
\caption{$\mathbf{Z}^\mathbf{A}$}
\end{subfigure}
\\
\begin{subfigure}[b]{.49\textwidth}%
\includegraphics[scale=.64,clip=true,trim=2.3in 0in 2.8in 0in]{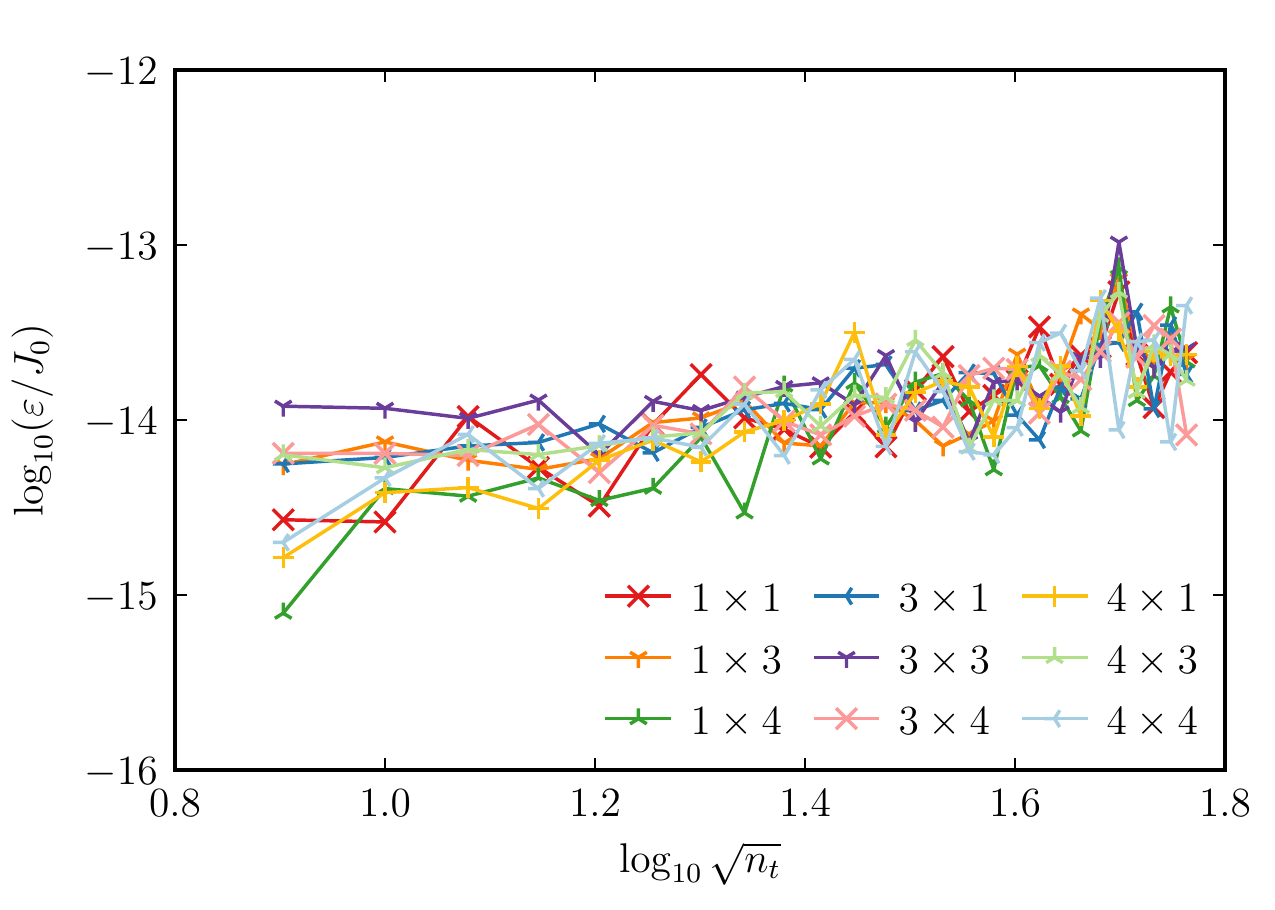}
\caption{$\mathbf{Z}^\Phi$}
\end{subfigure}
\caption{Solution-discretization error cancellation: $\varepsilon=\left\|\mathbf{e}_n\right\|_\infty$, with $G_\text{MS}$, $\theta=0^\circ$, and the uniform mesh.}
\vskip-\dp\strutbox
\label{fig:part2}
\end{figure}

\begin{figure}
\centering
\begin{subfigure}[b]{.49\textwidth}
\includegraphics[scale=.64,clip=true,trim=2.3in 0in 2.8in 0in]{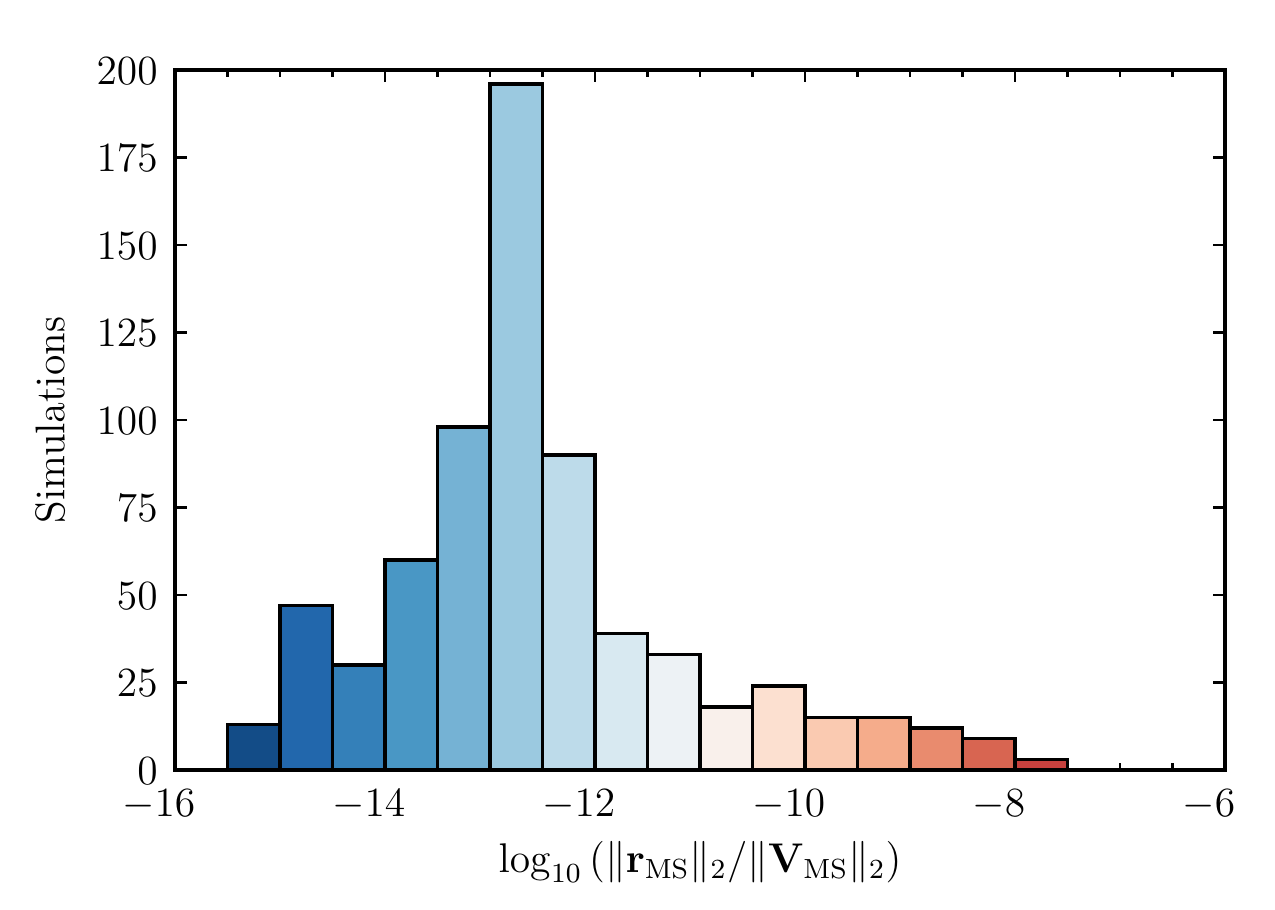}
\end{subfigure}
\caption{Solution-discretization error cancellation: $\mathbf{r}_\text{MS}=\mathbf{Z}\mathbf{J}^h - \mathbf{V}_{\text{MS}}(\mathbf{J}_{h_\text{MS}})$~\eqref{eq:zjv_mmsp} for the simulations in Figure~\ref{fig:part2}.}
\vskip-\dp\strutbox
\label{fig:part2_res}
\end{figure}

\begin{figure}
\centering
\begin{subfigure}[b]{.49\textwidth}
\includegraphics[scale=.64,clip=true,trim=2.3in 0in 2.8in 0in]{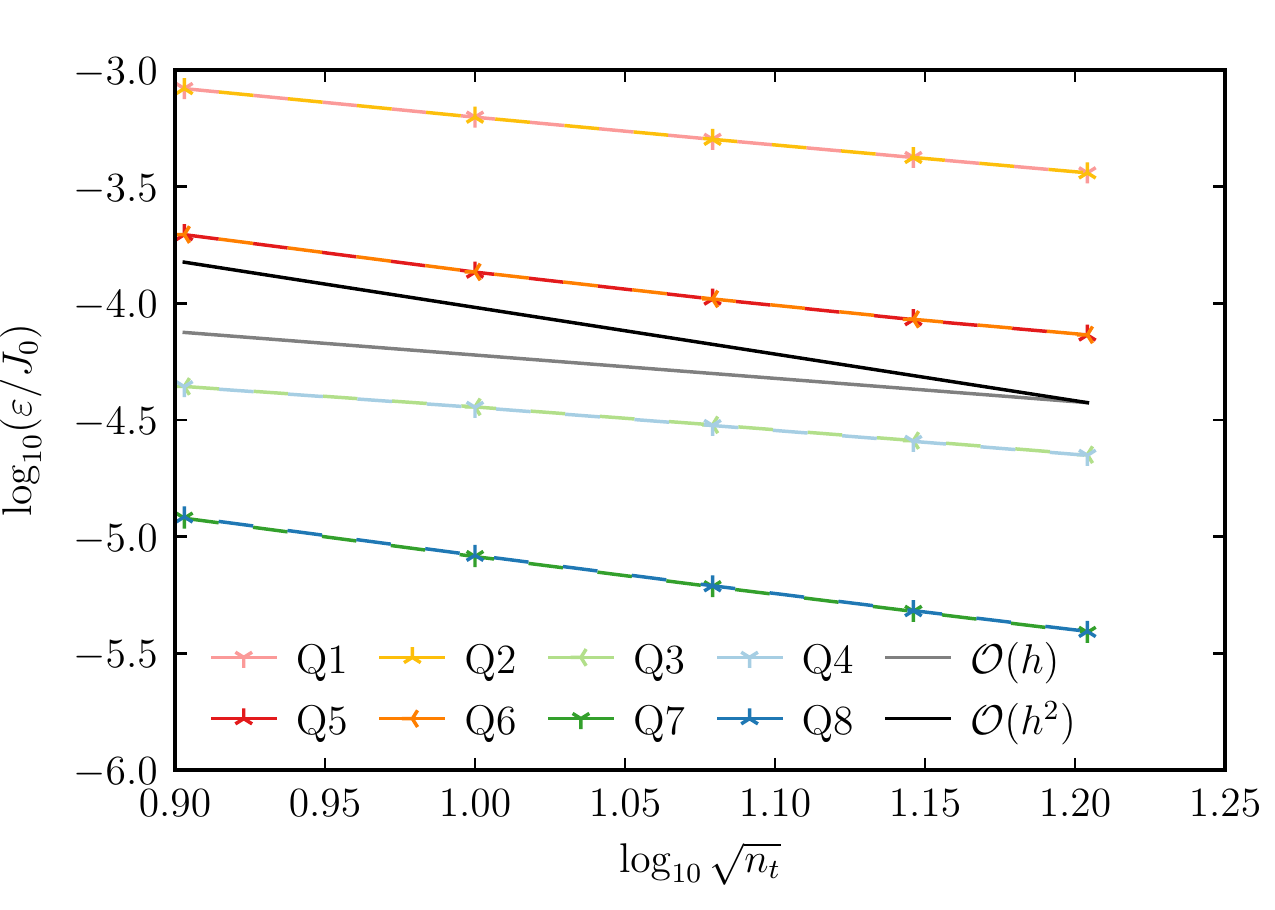}
\caption{Uniform mesh, $\theta=0^\circ$}
\end{subfigure}
\hspace{0.25em}
\begin{subfigure}[b]{.49\textwidth}
\includegraphics[scale=.64,clip=true,trim=2.3in 0in 2.8in 0in]{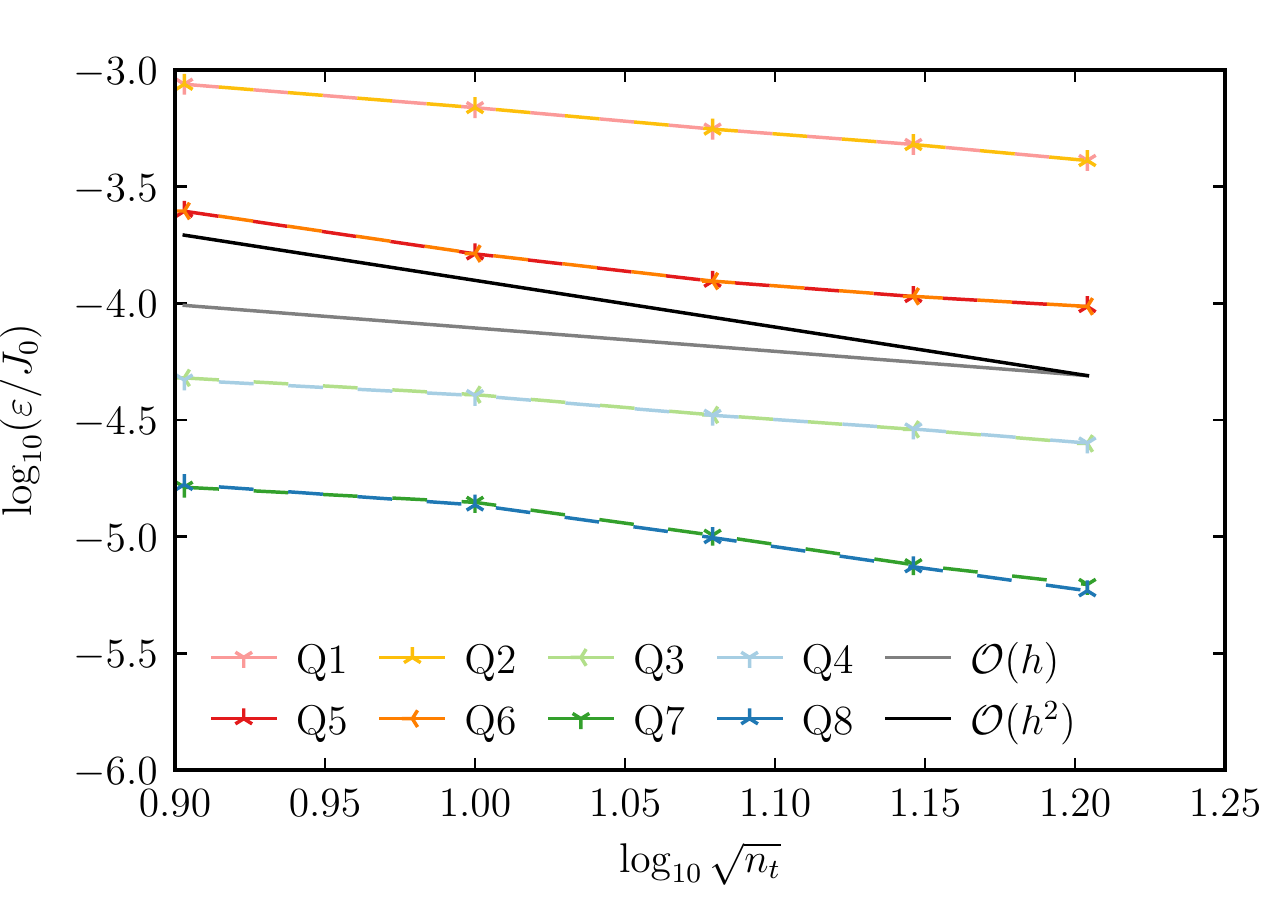}
\caption{Twisted mesh, $\theta=0^\circ$}
\end{subfigure}
\\
\begin{subfigure}[b]{.49\textwidth}
\includegraphics[scale=.64,clip=true,trim=2.3in 0in 2.8in 0in]{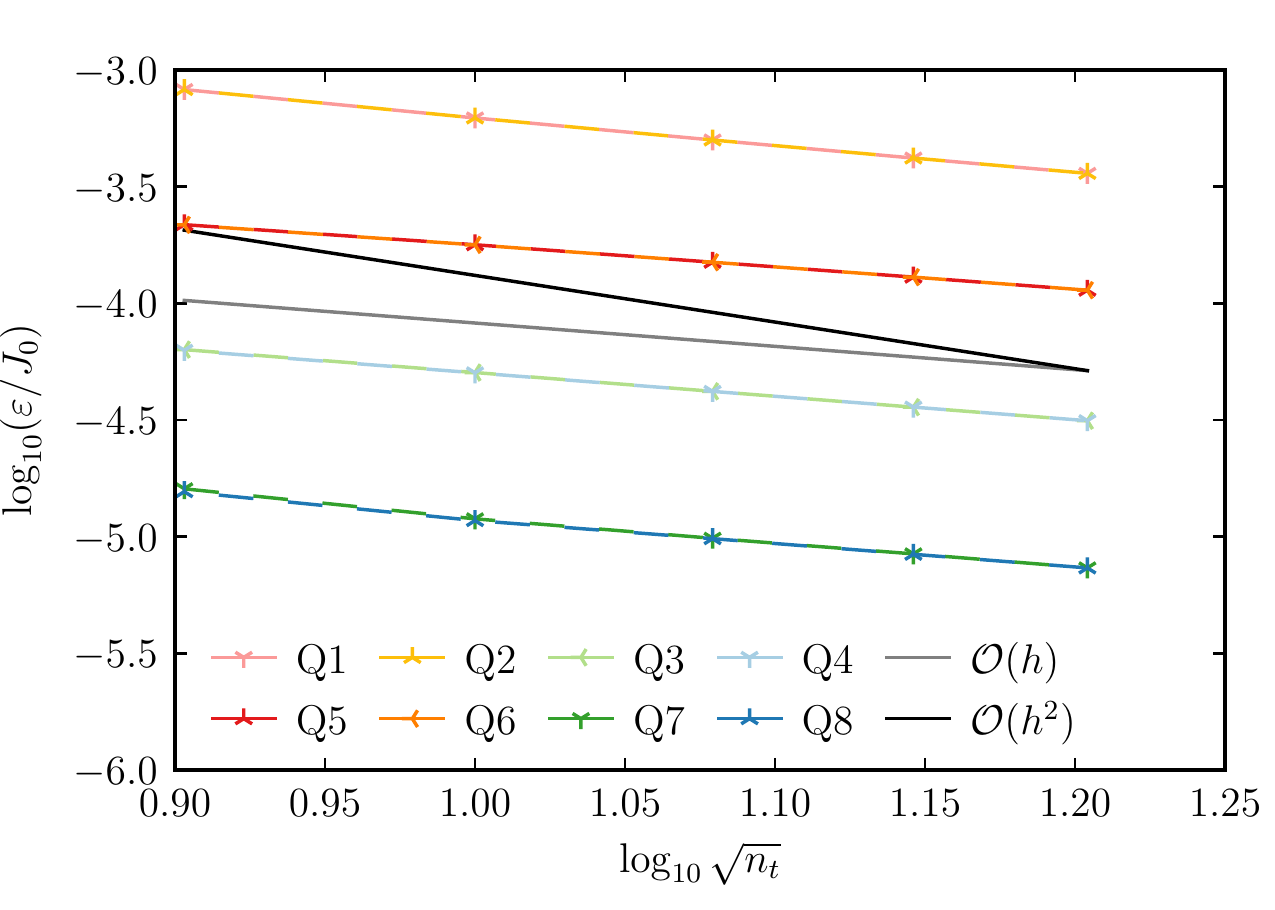}
\caption{Uniform mesh, $\theta=90^\circ$}
\end{subfigure}
\hspace{0.25em}
\begin{subfigure}[b]{.49\textwidth}
\includegraphics[scale=.64,clip=true,trim=2.3in 0in 2.8in 0in]{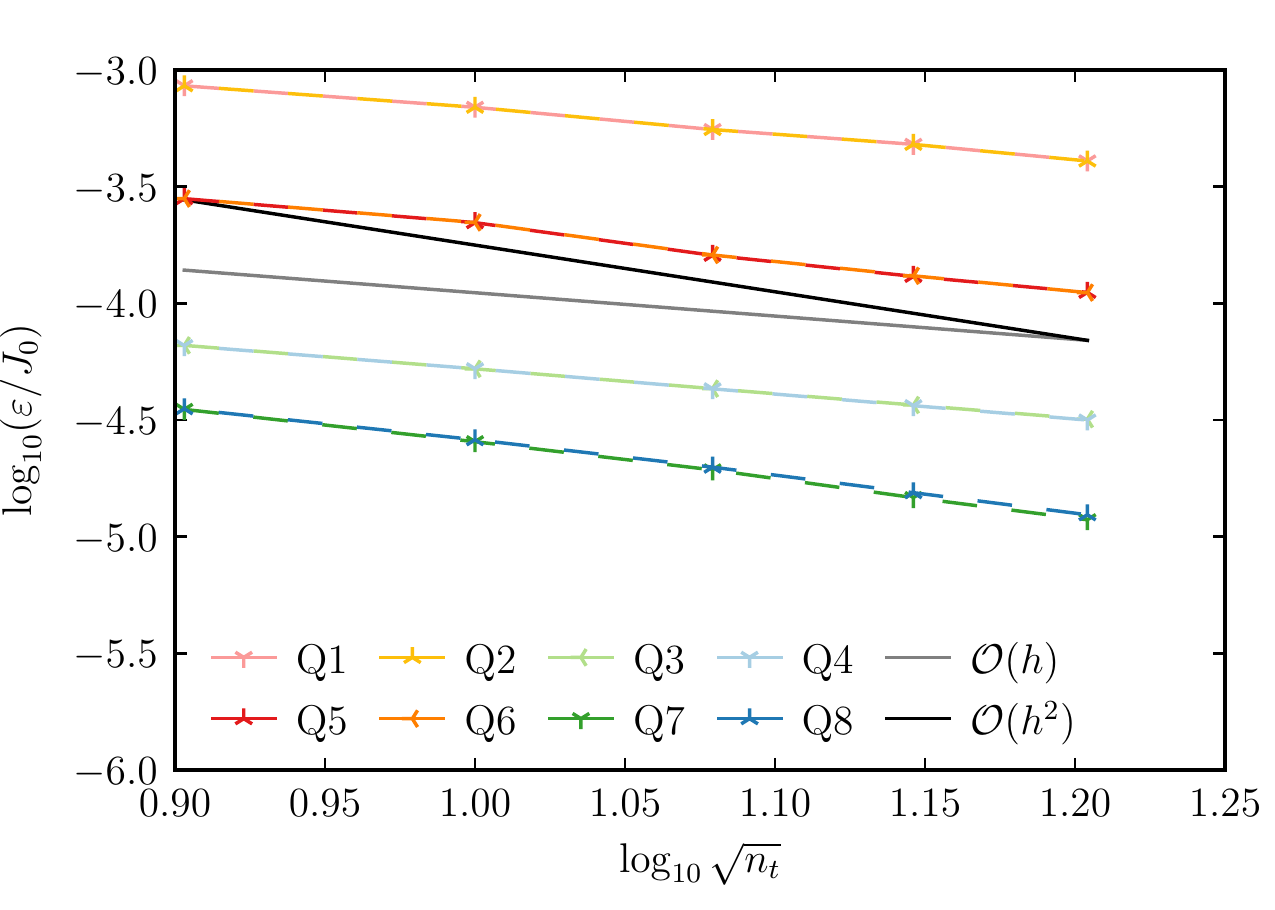}
\caption{Twisted mesh, $\theta=90^\circ$}
\end{subfigure}
\caption{Solution-discretization error cancellation: $\varepsilon=\left\|\mathbf{e}_n\right\|_\infty$ for $\mathbf{Z}^\Phi$, with $G_k$.}
\vskip-\dp\strutbox
\label{fig:part2g}
\end{figure}

\begin{figure}
\centering
\begin{subfigure}[b]{.49\textwidth}
\includegraphics[scale=.64,clip=true,trim=2.3in 0in 2.3in 0in]{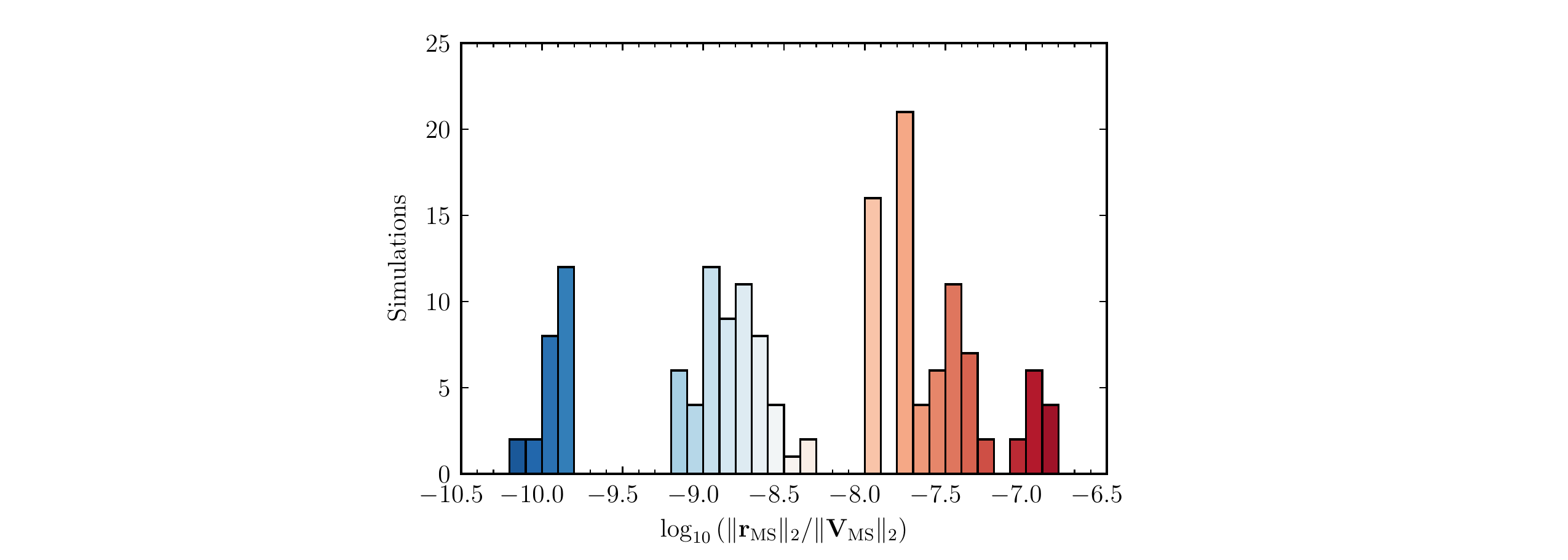}
\end{subfigure}
\caption{Solution-discretization error cancellation: $\mathbf{r}_\text{MS}=\mathbf{Z}\mathbf{J}^h - \mathbf{V}_{\text{MS}}(\mathbf{J}_{h_\text{MS}})$~\eqref{eq:zjv_mmsp} for the simulations in Figure~\ref{fig:part2g}.}
\vskip-\dp\strutbox
\label{fig:part2g_res}
\end{figure}

\begin{figure}[!t]
\centering
\begin{subfigure}[b]{.49\textwidth}
\includegraphics[scale=.64,clip=true,trim=2.3in 0in 2.8in 0in]{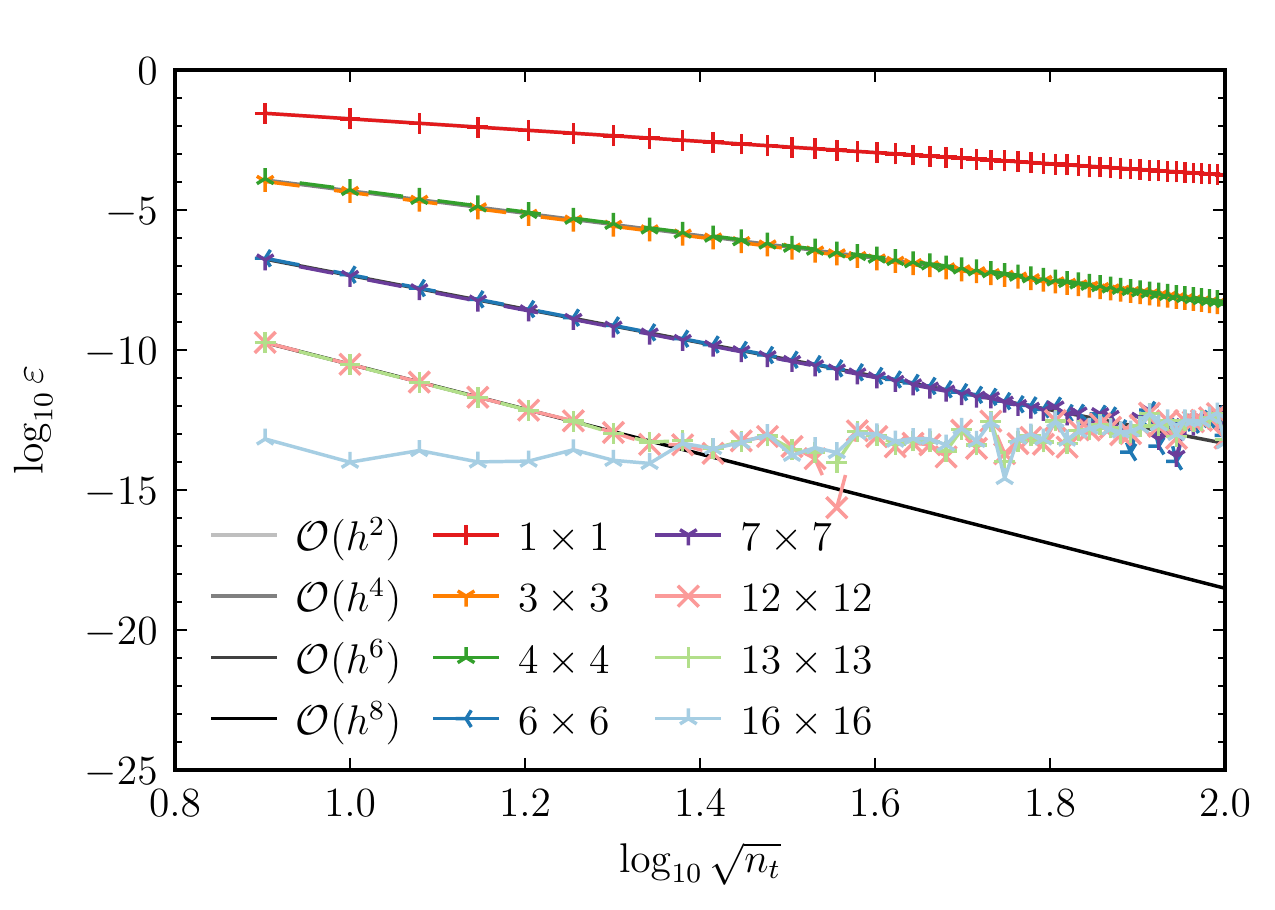}

\caption{Uniform mesh, $I^\mathbf{A}$}
\end{subfigure}
\hspace{0.25em}
\begin{subfigure}[b]{.49\textwidth}
\includegraphics[scale=.64,clip=true,trim=2.3in 0in 2.8in 0in]{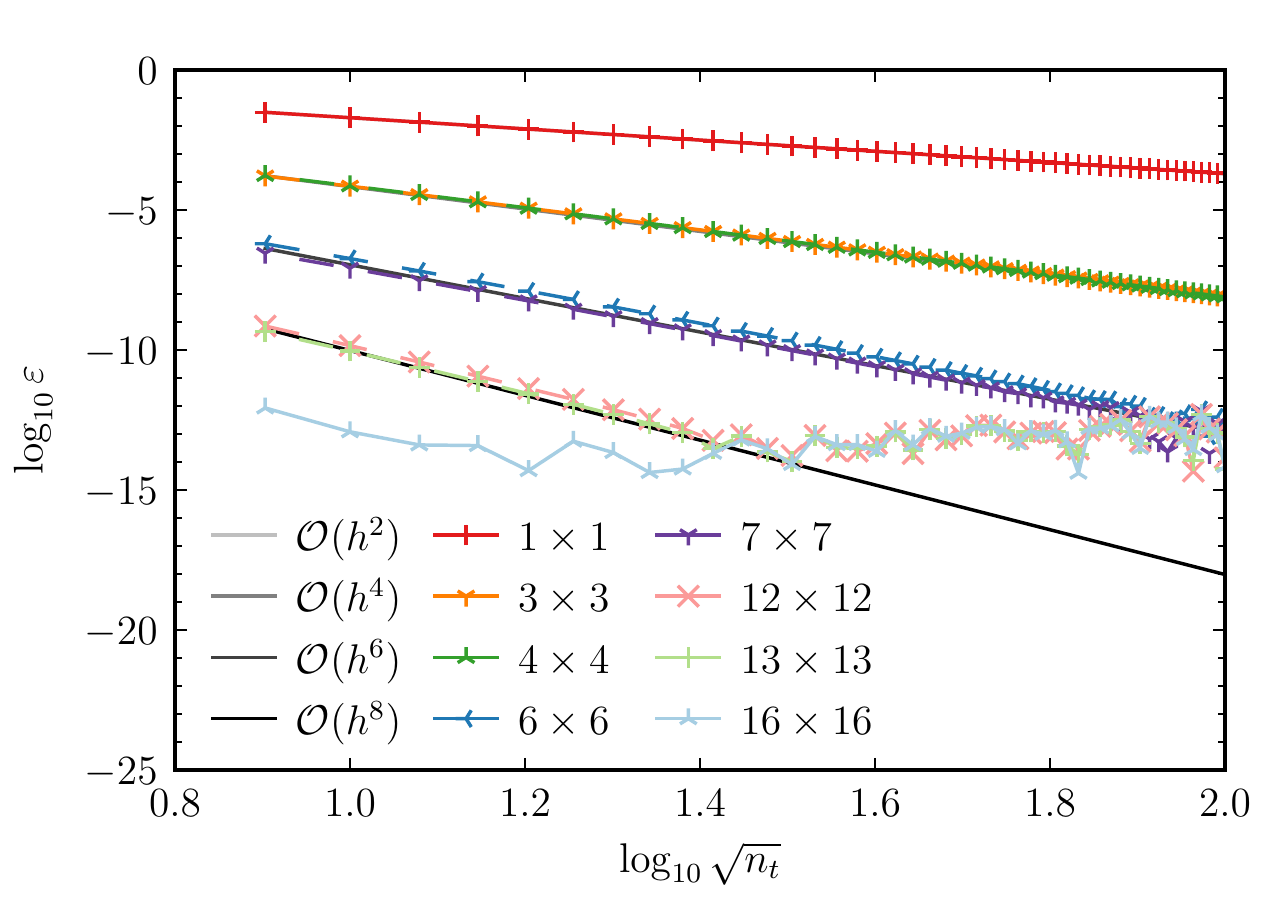}

\caption{Twisted mesh, $I^\mathbf{A}$}
\end{subfigure}
\\
\begin{subfigure}[b]{.49\textwidth}
\includegraphics[scale=.64,clip=true,trim=2.3in 0in 2.8in 0in]{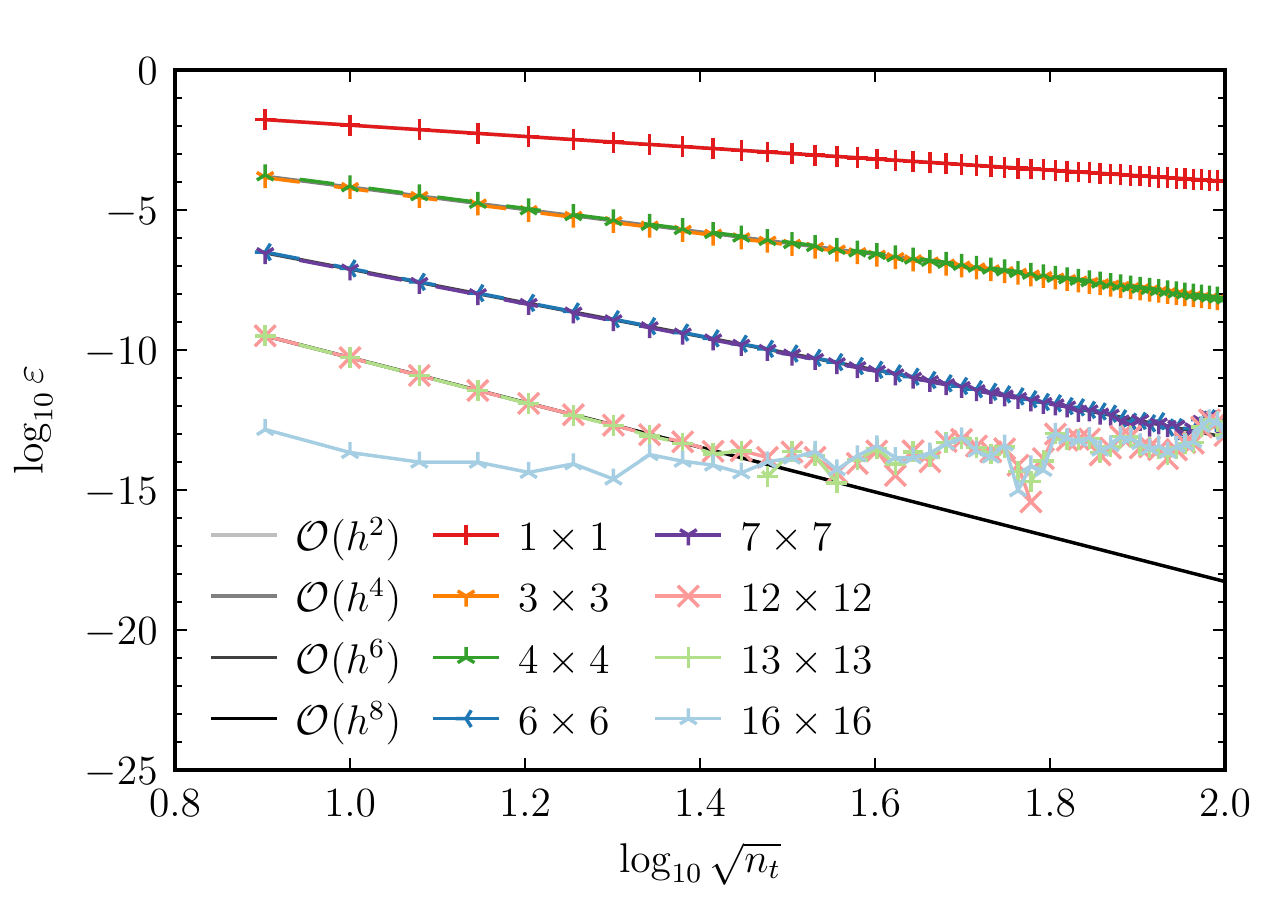}

\caption{Uniform mesh, $I^\Phi$}
\end{subfigure}
\hspace{0.25em}
\begin{subfigure}[b]{.49\textwidth}
\includegraphics[scale=.64,clip=true,trim=2.3in 0in 2.8in 0in]{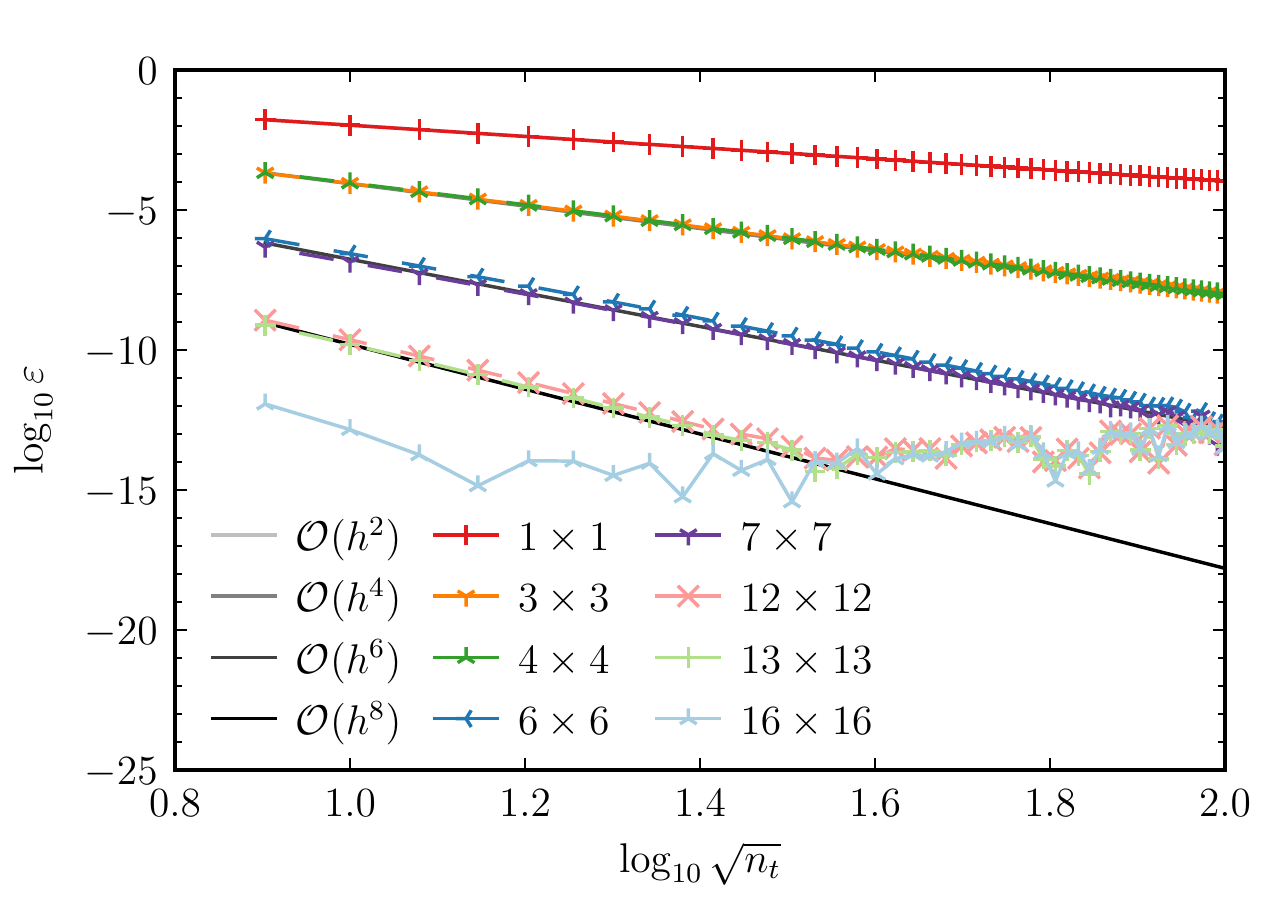}

\caption{Twisted mesh, $I^\Phi$}
\end{subfigure}
\caption{Solution-discretization error elimination: $\varepsilon=|I_h - I|/|I|$, with $G_\text{MS}$ and $\theta=0^\circ$.}
\vskip-\dp\strutbox
\label{fig:part3}
\end{figure}

\begin{figure}[!t]
\centering
\begin{subfigure}[b]{.49\textwidth}
\includegraphics[scale=.64,clip=true,trim=2.3in 0in 2.8in 0in]{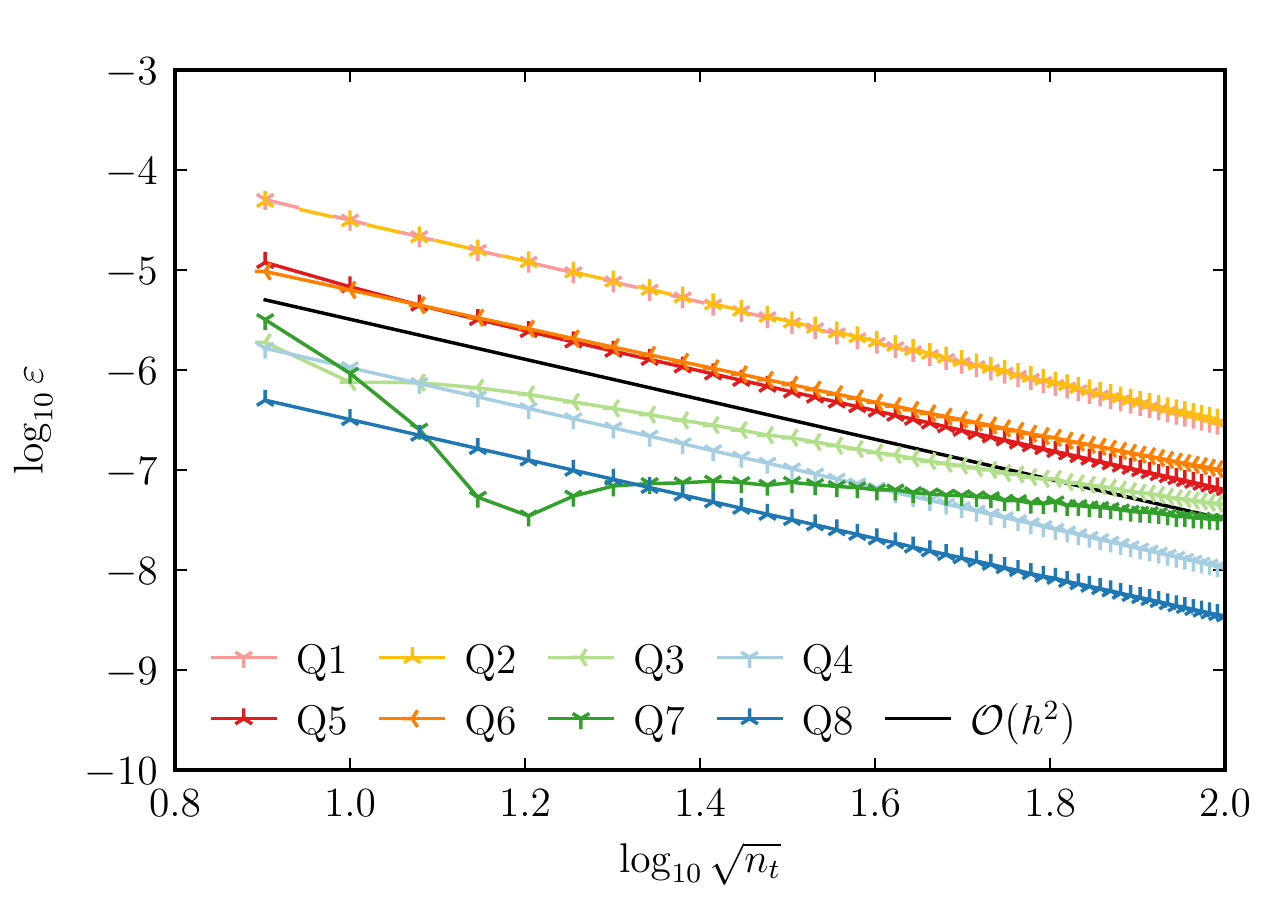}

\caption{Uniform mesh, $I^\mathbf{A}$}
\end{subfigure}
\hspace{0.25em}
\begin{subfigure}[b]{.49\textwidth}
\includegraphics[scale=.64,clip=true,trim=2.3in 0in 2.8in 0in]{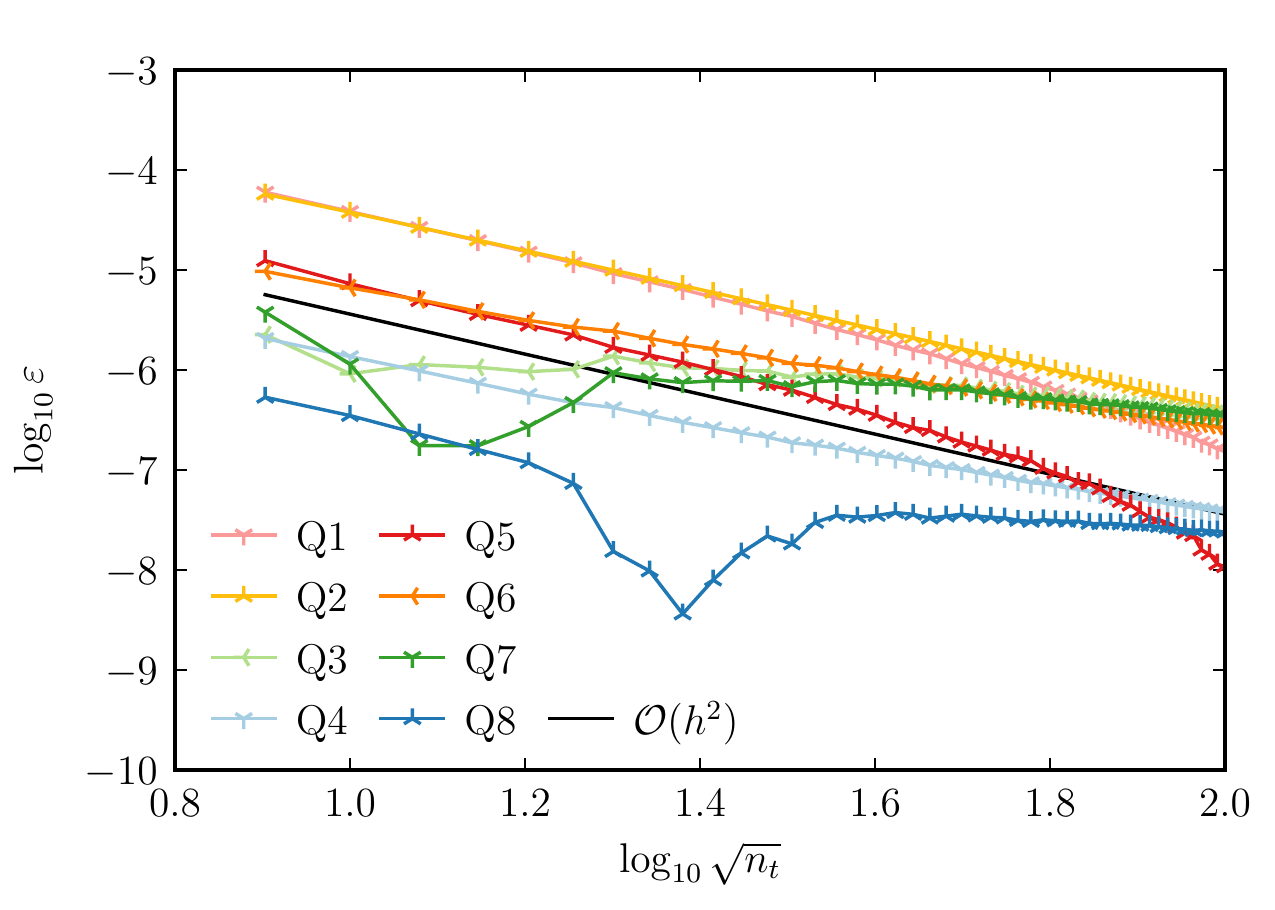}

\caption{Twisted mesh, $I^\mathbf{A}$}
\end{subfigure}
\\
\begin{subfigure}[b]{.49\textwidth}
\includegraphics[scale=.64,clip=true,trim=2.3in 0in 2.8in 0in]{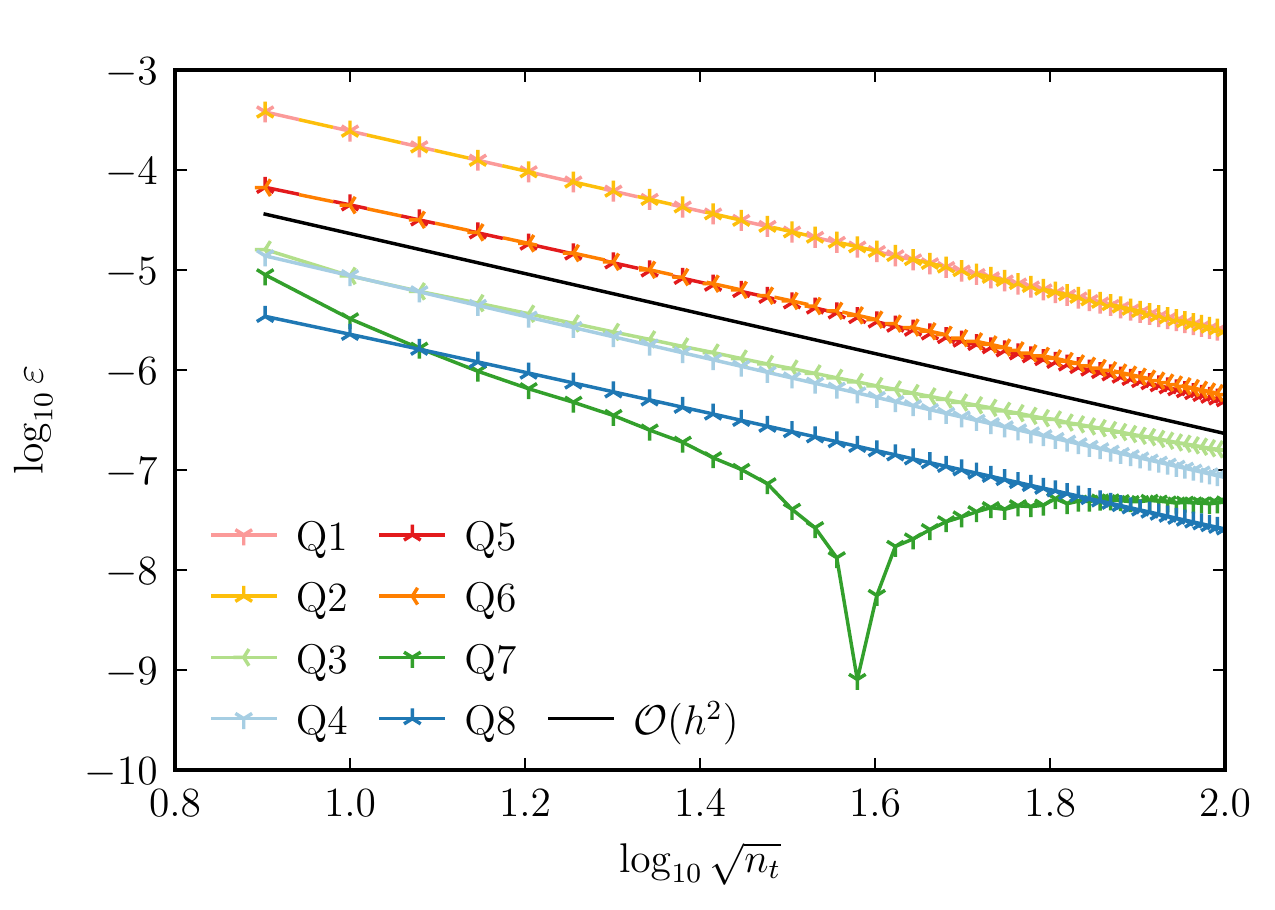}

\caption{Uniform mesh, $I^\Phi$}
\end{subfigure}
\hspace{0.25em}
\begin{subfigure}[b]{.49\textwidth}
\includegraphics[scale=.64,clip=true,trim=2.3in 0in 2.8in 0in]{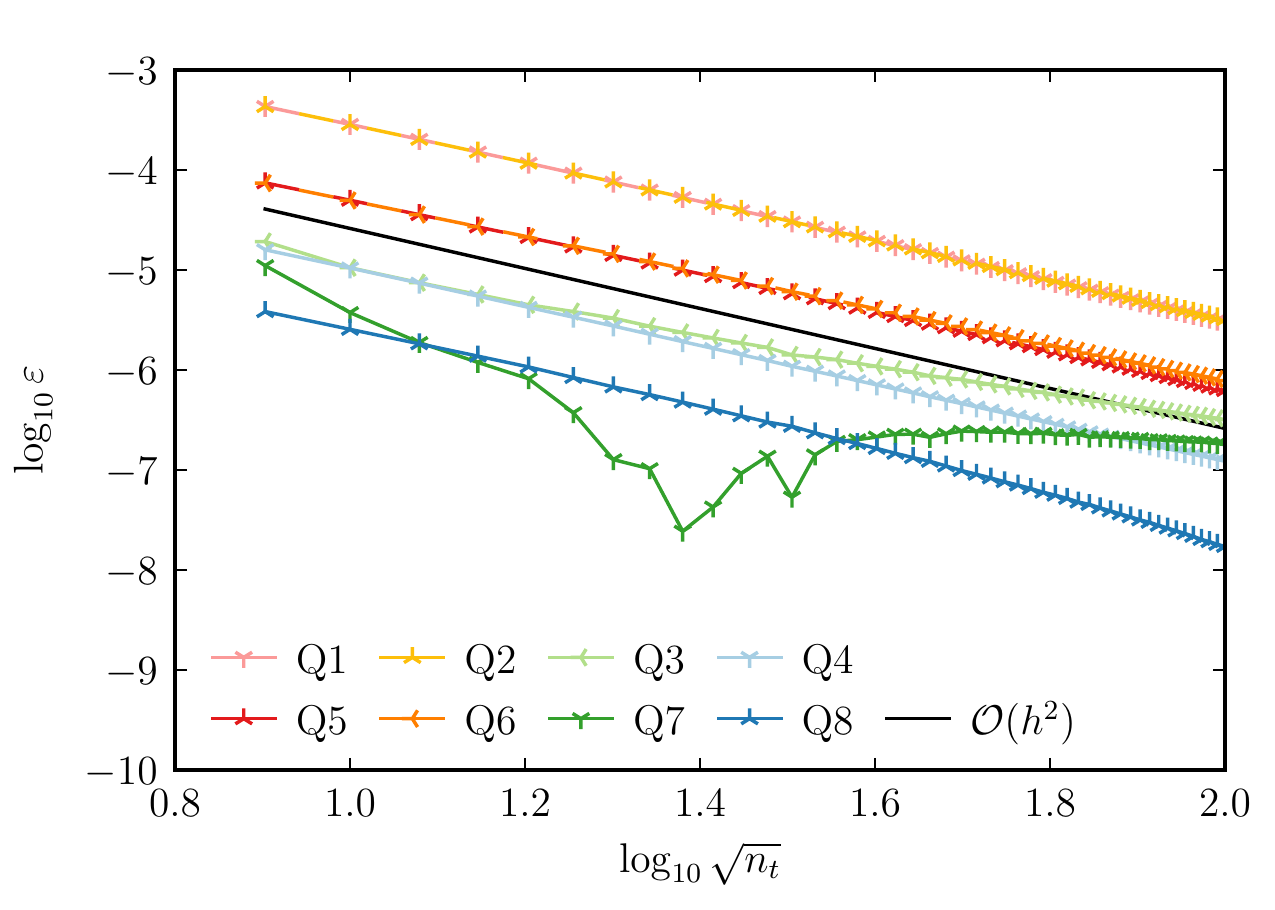}

\caption{Twisted mesh, $I^\Phi$}
\end{subfigure}
\caption{Solution-discretization error elimination: $\varepsilon=|I_h - I|/|I|$, with $G_k$ and $\theta=0^\circ$.}
\vskip-\dp\strutbox
\label{fig:part3_0}
\end{figure}


For a kernel with the regularity of $G_\text{MS}$, the solution-discretization error rate is expected to be $\mathcal{O}(h^2)$.  \reviewerOne{The errors in Figure~\ref{fig:part4} contain contributions from both the solution-discretization error and numerical-integration error.  In the absence of the solution-discretization error, we would expect the numerical-integration error to decrease as the number of integration points increases.  However, the contamination from the solution-discretization error inhibits this trend.  As a result,} the error from the inexact integrations not only decreases at the same rate as that from the exact integrations, but is also of a similar magnitude, complicating our ability to meaningfully compare quadrature options.  Therefore, to isolate the numerical-integration error, we undertake the studies of the following two subsections.

\subsubsection{Solution-Discretization Error Cancellation} 

Here, we consider the solution-discretization error cancellation approach of Section~\ref{sec:part2}.  For $G_\text{MS}$, $\theta=0^\circ$, and the uniform mesh, Figure~\ref{fig:part2} shows the error norm $\varepsilon=\left\|\mathbf{e}_n\right\|_\infty$ for different $n_\text{test} \times n_\text{source}$ polynomial triangle quadrature point combinations.  



Each of the errors plotted in Figure~\ref{fig:part2} decreases at least as fast as its expected rate in Table~\ref{tab:dunavant_properties}.  For $\mathbf{Z}$ and $\mathbf{Z}^\mathbf{A}$, the error from $3\times 4$ points decreases closer to a rate of $\mathcal{O}(h^6)$ than the expected rate of $\mathcal{O}(h^4)$, \reviewerBoth{and the error from $4\times 4$ points is the round-off error, which worsens as the mesh is refined}.  For $\mathbf{Z}^\Phi$, all quadrature point combinations exactly evaluate the integral, despite the expectation of $\mathcal{O}(h^2)$ for the one-point integrations.  \reviewerBoth{The errors here are due to round-off error.}

Figure~\ref{fig:part2_res} shows the residuals, which, though not as low as those in Figure~\ref{fig:part4_res}, are \reviewerTwo{less than $10^{-7}$}.

For $G_k$, Figure~\ref{fig:part2g} shows the error norm $\varepsilon=\left\|\mathbf{e}_n\right\|_\infty$ for the quadrature combinations in Table~\ref{tab:G_quad_combinations} for $\mathbf{Z}^\Phi$.  Because of the singular integrands, the errors in the quadrature integration, which are functions of the unbounded derivatives of the integrand, are unbounded, yielding less monotonic trends than what are obtained for $G_\text{MS}$ in Figure~\ref{fig:part2}.  

The relative convergence tolerance $\epsilon$ for the adaptive integration of $\mathbf{V}_{\text{MS}}(\mathbf{J}_{h_\text{MS}})$~\eqref{eq:part2_problem} \reviewerBoth{via the \texttt{CUBA} library} is initialized to $\epsilon=10^{-3}$ and reduced by a factor of ten until the relative change in $\left\|\mathbf{e}_n\right\|_\infty$~\eqref{eq:linf_norm} is less than $10^{-3}$.  Because $\mathbf{J}_{h_\text{MS}}$ is a piecewise function, computation of $\mathbf{V}_{\text{MS}}(\mathbf{J}_{h_\text{MS}})$ becomes prohibitively expensive beyond the coarsest meshes.  Nonetheless, Figure~\ref{fig:part2g} shows that Quadrature Combinations~Q4 and Q8 generally outperform Quadrature Combinations~Q1 and Q5, and Quadrature Combination~Q8 generally outperforms Quadrature Combination~Q4.   

Figure~\ref{fig:part2g_res} shows the residuals, which, though not as low as those in Figure~\ref{fig:part4_res}, are \reviewerTwo{less than $10^{-6}$}.

\subsubsection{Solution-Discretization Error Elimination} 

Finally, we consider the solution-discretization error elimination approach of Section~\ref{sec:part3}.  For $G_\text{MS}$, $\theta=0^\circ$, and both meshes, Figure~\ref{fig:part3} shows the relative error $\varepsilon=|I_h - I|/|I|$.  For each of the $n_\text{test} \times n_\text{source}$ polynomial triangle quadrature point combinations, the error rate matches the expected rate in Table~\ref{tab:dunavant_properties}, prior to \reviewerBoth{being dominated by the round-off error.}

\begin{figure}[!t]
\centering
\begin{subfigure}[b]{.49\textwidth}
\includegraphics[scale=.64,clip=true,trim=2.3in 0in 2.8in 0in]{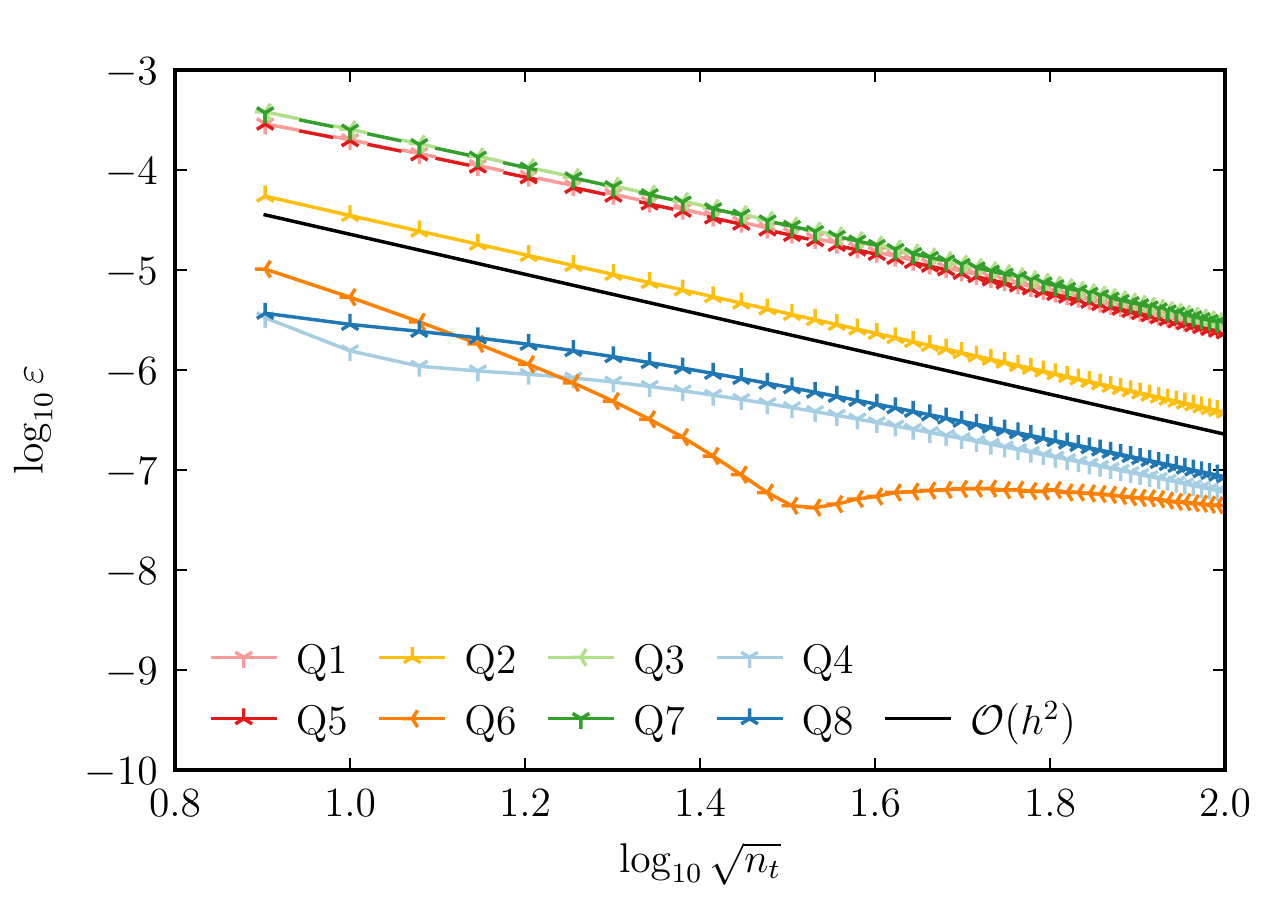}

\caption{Uniform mesh, $I^\mathbf{A}$}
\end{subfigure}
\hspace{0.25em}
\begin{subfigure}[b]{.49\textwidth}
\includegraphics[scale=.64,clip=true,trim=2.3in 0in 2.8in 0in]{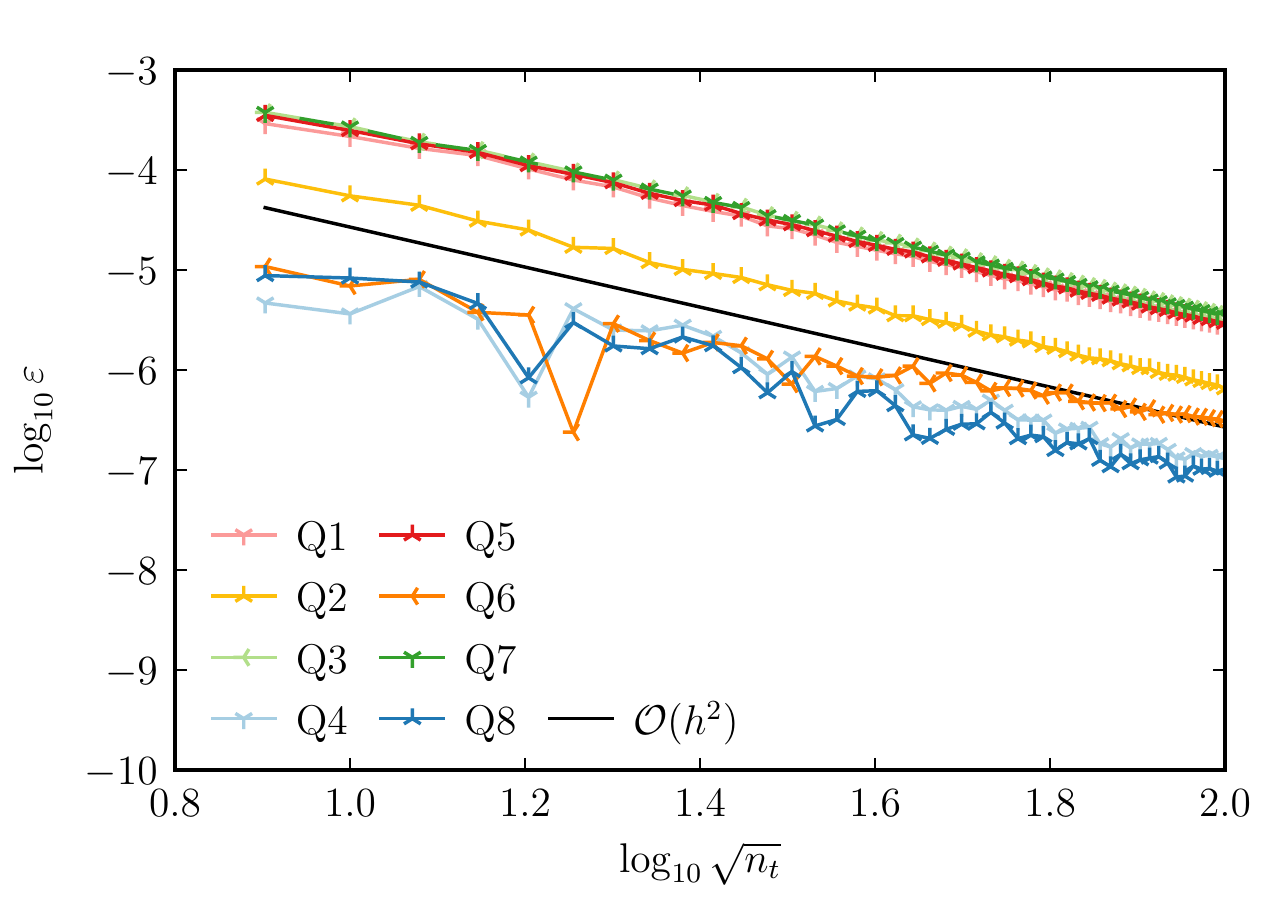}

\caption{Twisted mesh, $I^\mathbf{A}$}
\end{subfigure}
\\
\begin{subfigure}[b]{.49\textwidth}
\includegraphics[scale=.64,clip=true,trim=2.3in 0in 2.8in 0in]{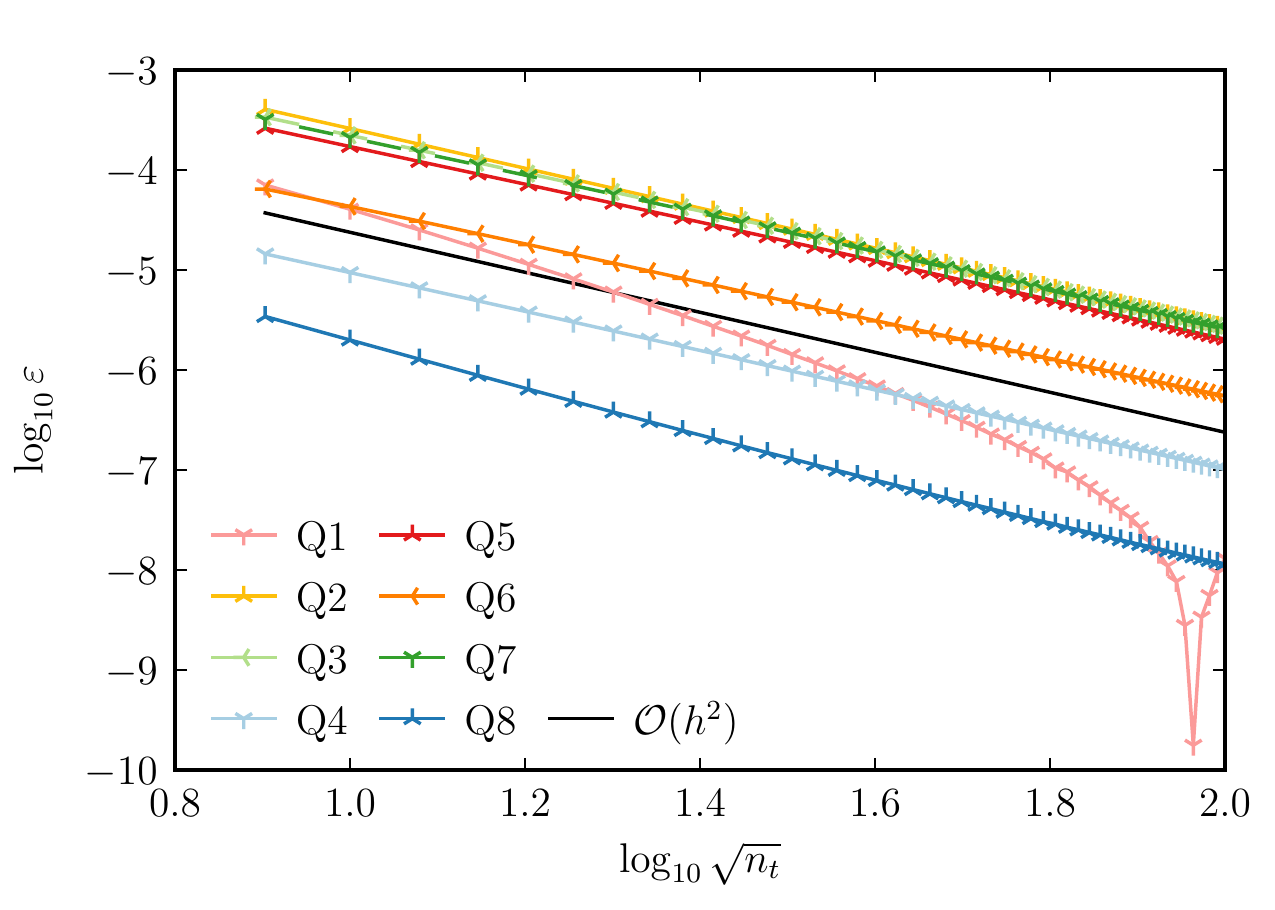}

\caption{Uniform mesh, $I^\Phi$}
\end{subfigure}
\hspace{0.25em}
\begin{subfigure}[b]{.49\textwidth}
\includegraphics[scale=.64,clip=true,trim=2.3in 0in 2.8in 0in]{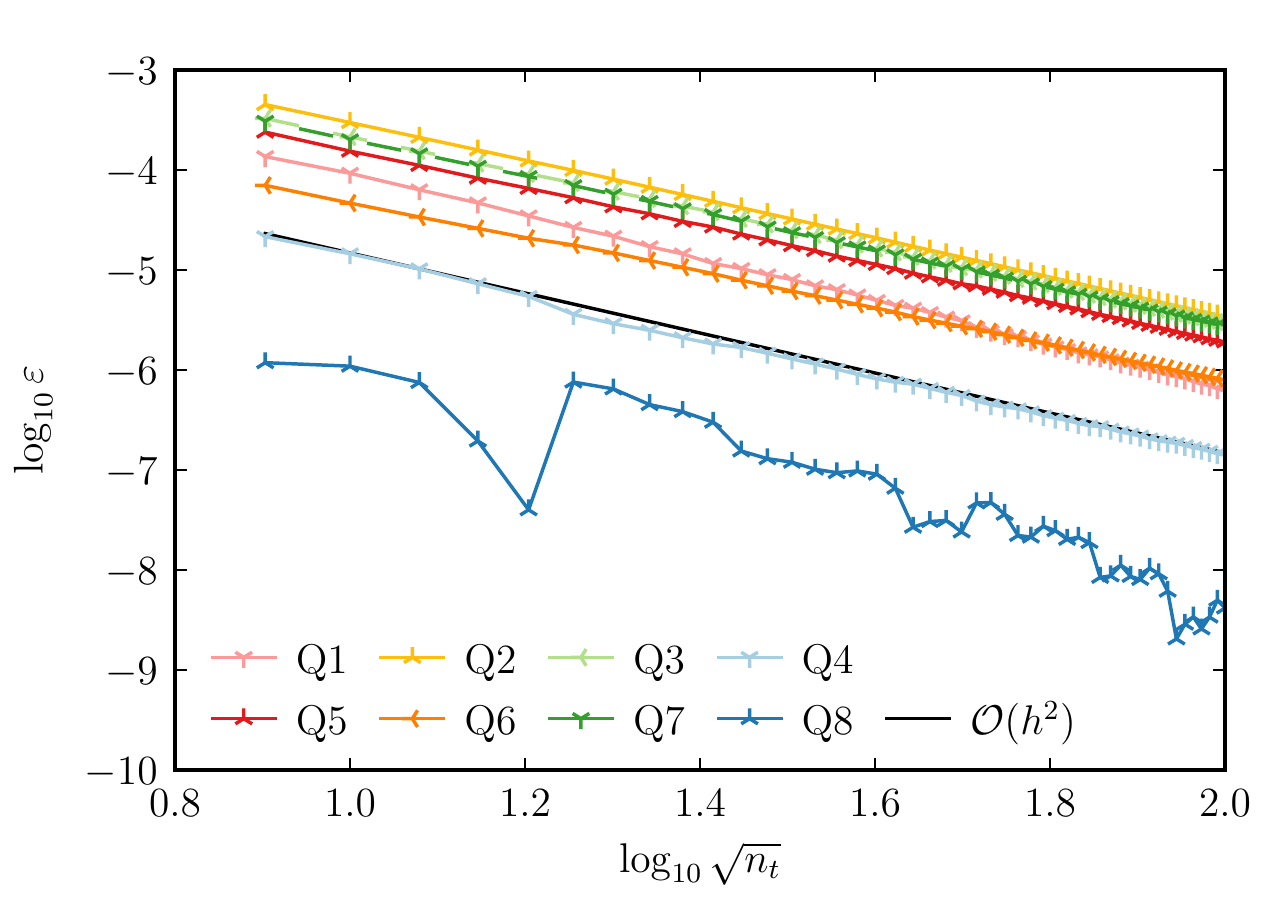}

\caption{Twisted mesh, $I^\Phi$}
\end{subfigure}
\caption{Solution-discretization error elimination: $\varepsilon=|I_h - I|/|I|$, with $G_k$ and $\theta=45^\circ$.}
\vskip-\dp\strutbox
\label{fig:part3_45}
\end{figure}

\begin{figure}[!t]
\centering
\begin{subfigure}[b]{.49\textwidth}
\includegraphics[scale=.64,clip=true,trim=2.3in 0in 2.8in 0in]{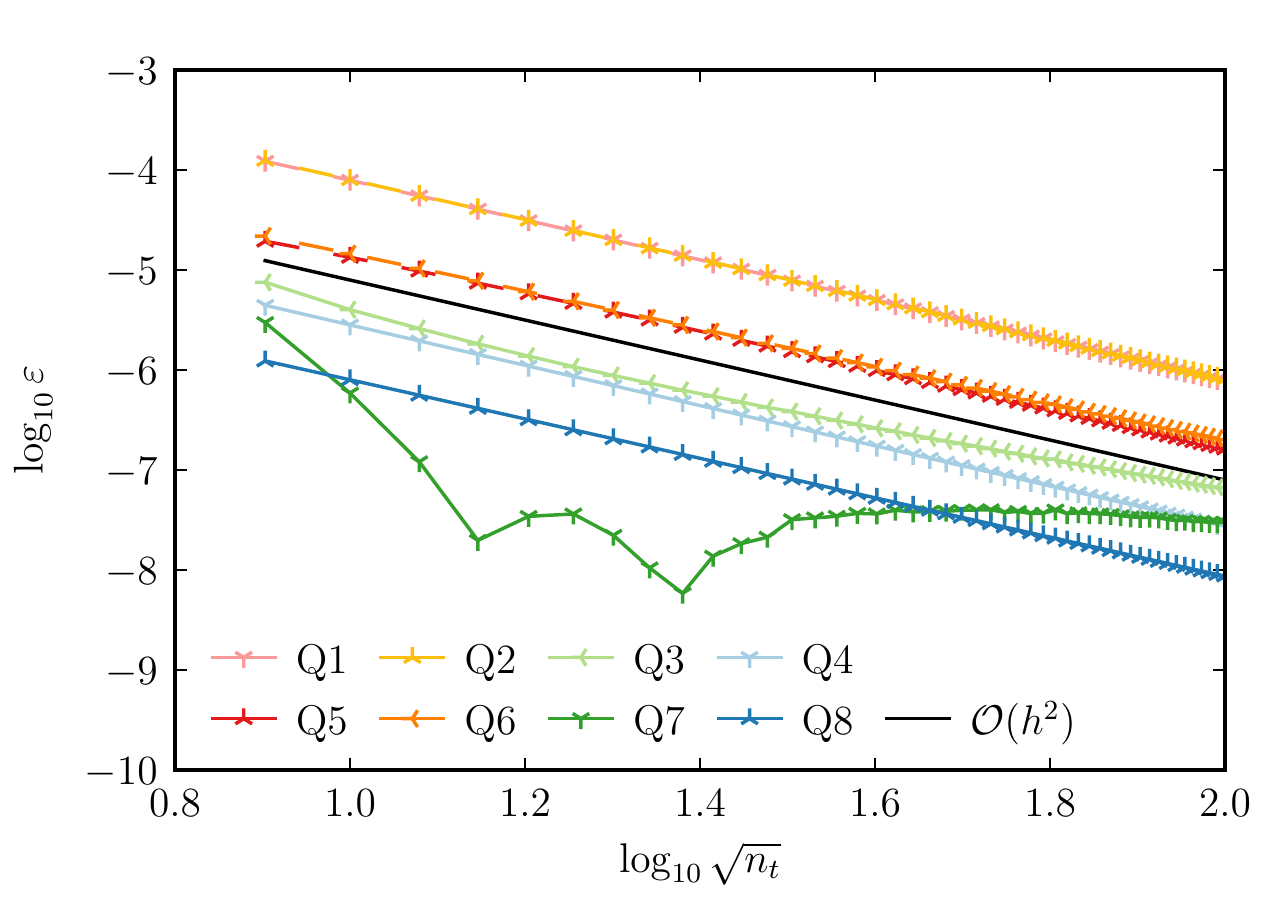}

\caption{Uniform mesh, $I^\mathbf{A}$}
\end{subfigure}
\hspace{0.25em}
\begin{subfigure}[b]{.49\textwidth}
\includegraphics[scale=.64,clip=true,trim=2.3in 0in 2.8in 0in]{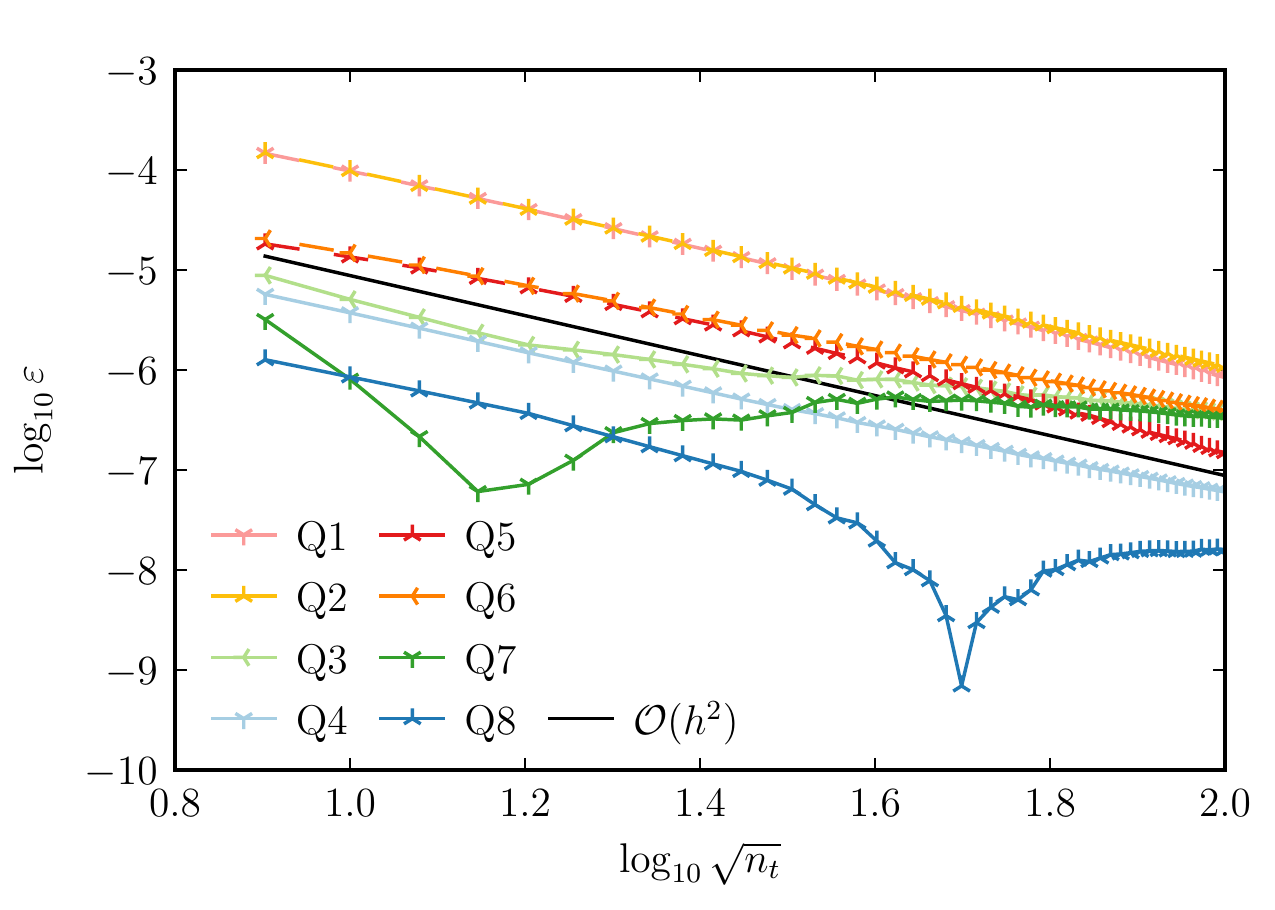}

\caption{Twisted mesh, $I^\mathbf{A}$}
\end{subfigure}
\\
\begin{subfigure}[b]{.49\textwidth}
\includegraphics[scale=.64,clip=true,trim=2.3in 0in 2.8in 0in]{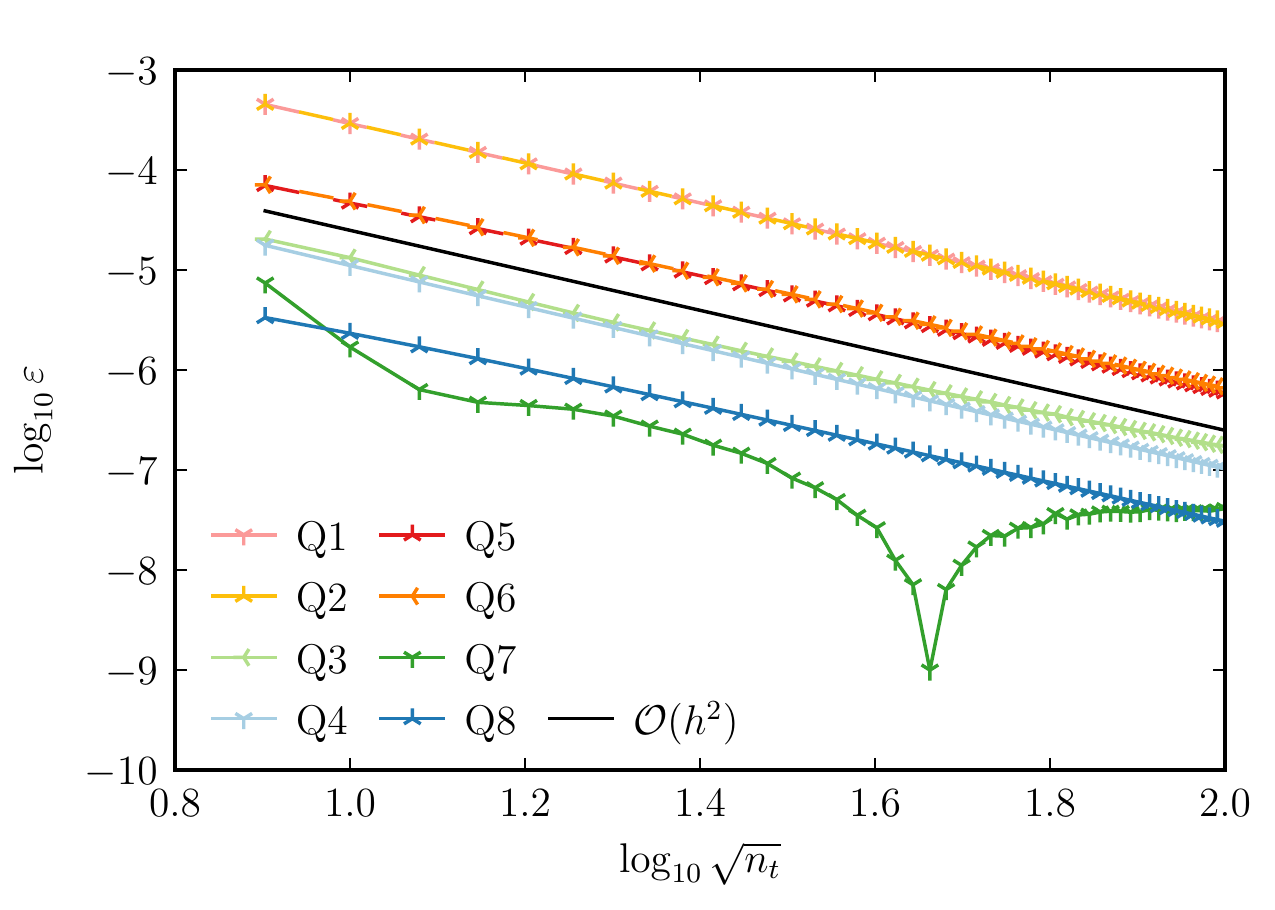}

\caption{Uniform mesh, $I^\Phi$}
\end{subfigure}
\hspace{0.25em}
\begin{subfigure}[b]{.49\textwidth}
\includegraphics[scale=.64,clip=true,trim=2.3in 0in 2.8in 0in]{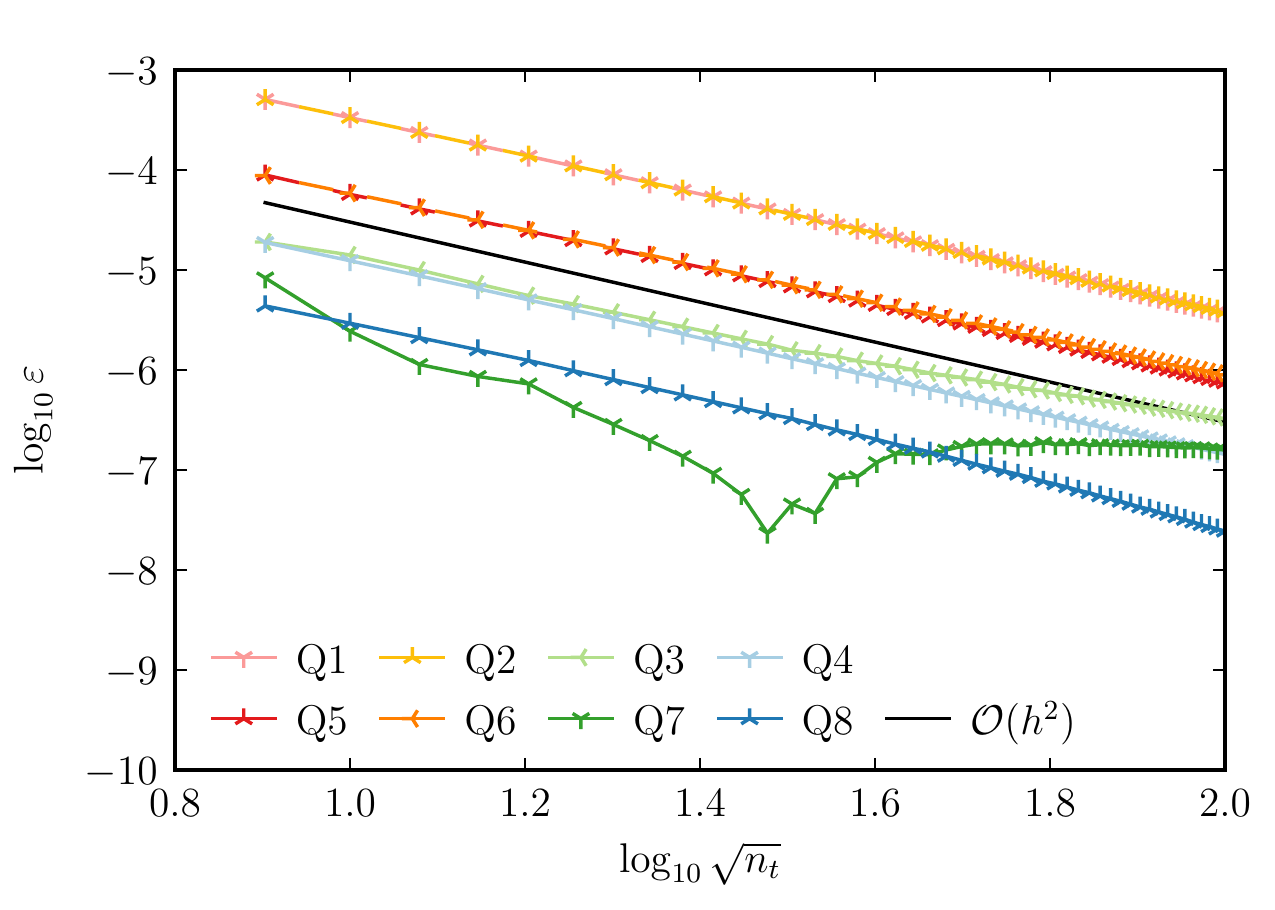}

\caption{Twisted mesh, $I^\Phi$}
\end{subfigure}
\caption{Solution-discretization error elimination: $\varepsilon=|I_h - I|/|I|$, with $G_k$ and $\theta=90^\circ$.}
\vskip-\dp\strutbox
\label{fig:part3_90}
\end{figure}

\begin{figure}[!t]
\centering
\begin{subfigure}[b]{.49\textwidth}
\includegraphics[scale=.64,clip=true,trim=2.3in 0in 2.8in 0in]{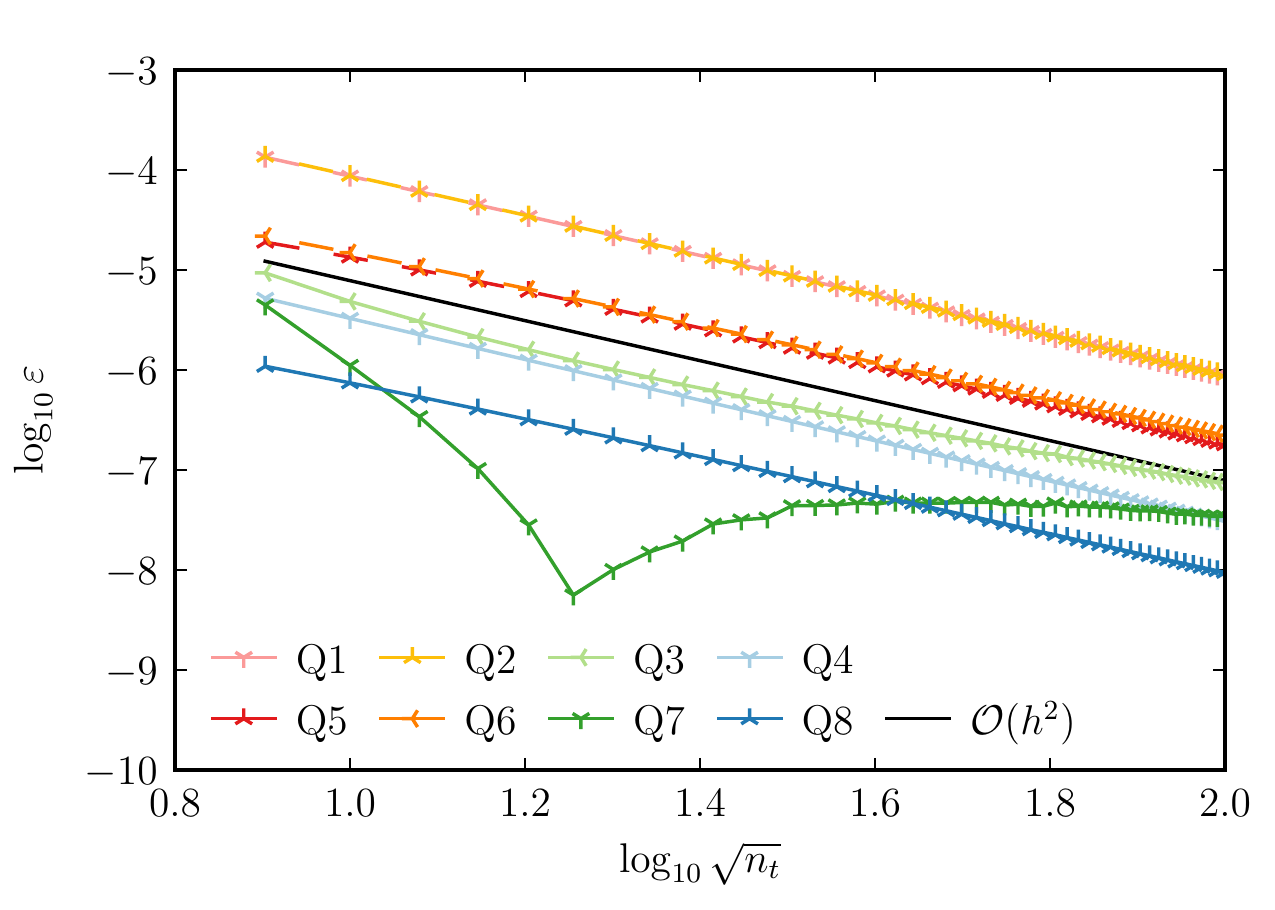}

\caption{Uniform mesh, $I^\mathbf{A}$}
\end{subfigure}
\hspace{0.25em}
\begin{subfigure}[b]{.49\textwidth}
\includegraphics[scale=.64,clip=true,trim=2.3in 0in 2.8in 0in]{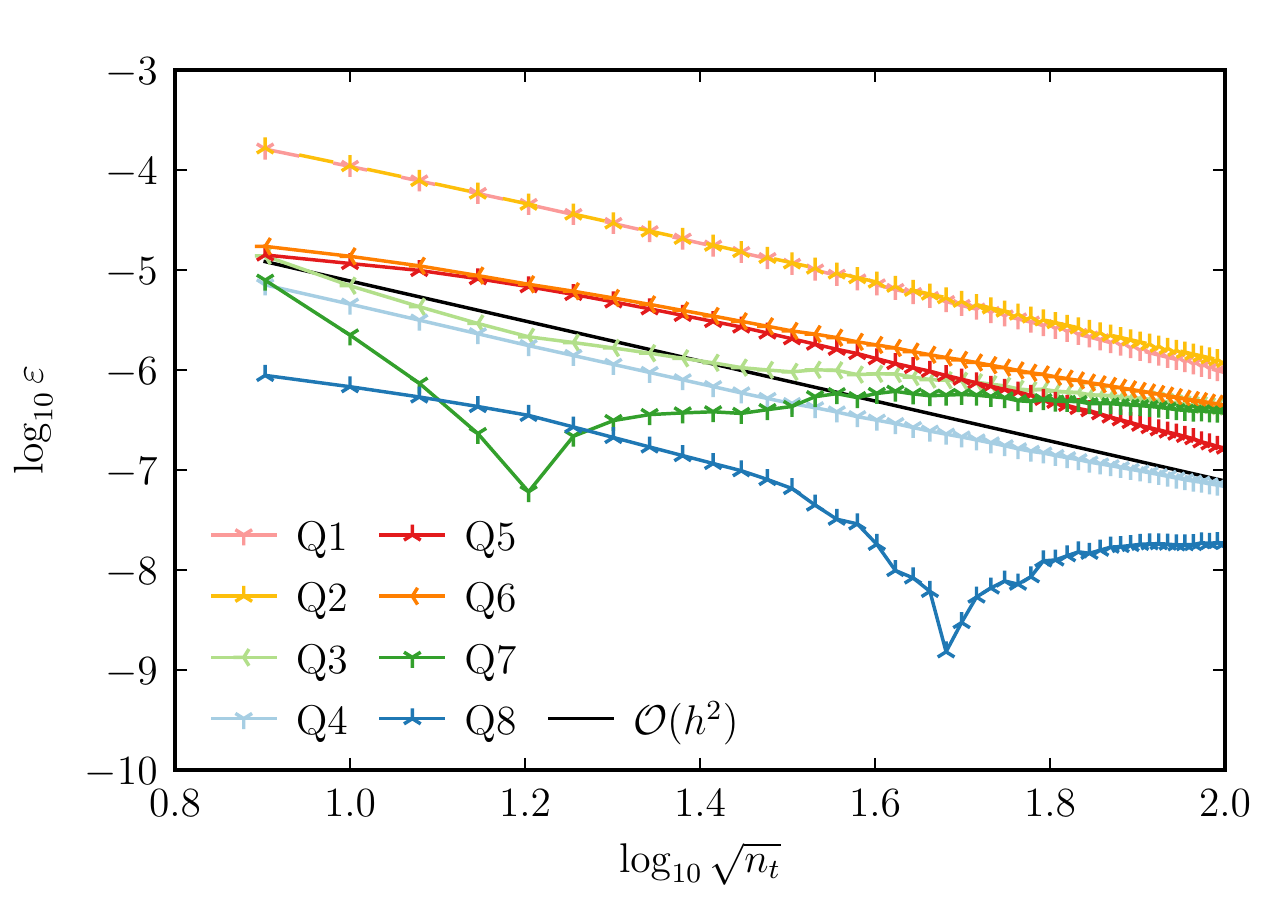}

\caption{Twisted mesh, $I^\mathbf{A}$}
\end{subfigure}
\\
\begin{subfigure}[b]{.49\textwidth}
\includegraphics[scale=.64,clip=true,trim=2.3in 0in 2.8in 0in]{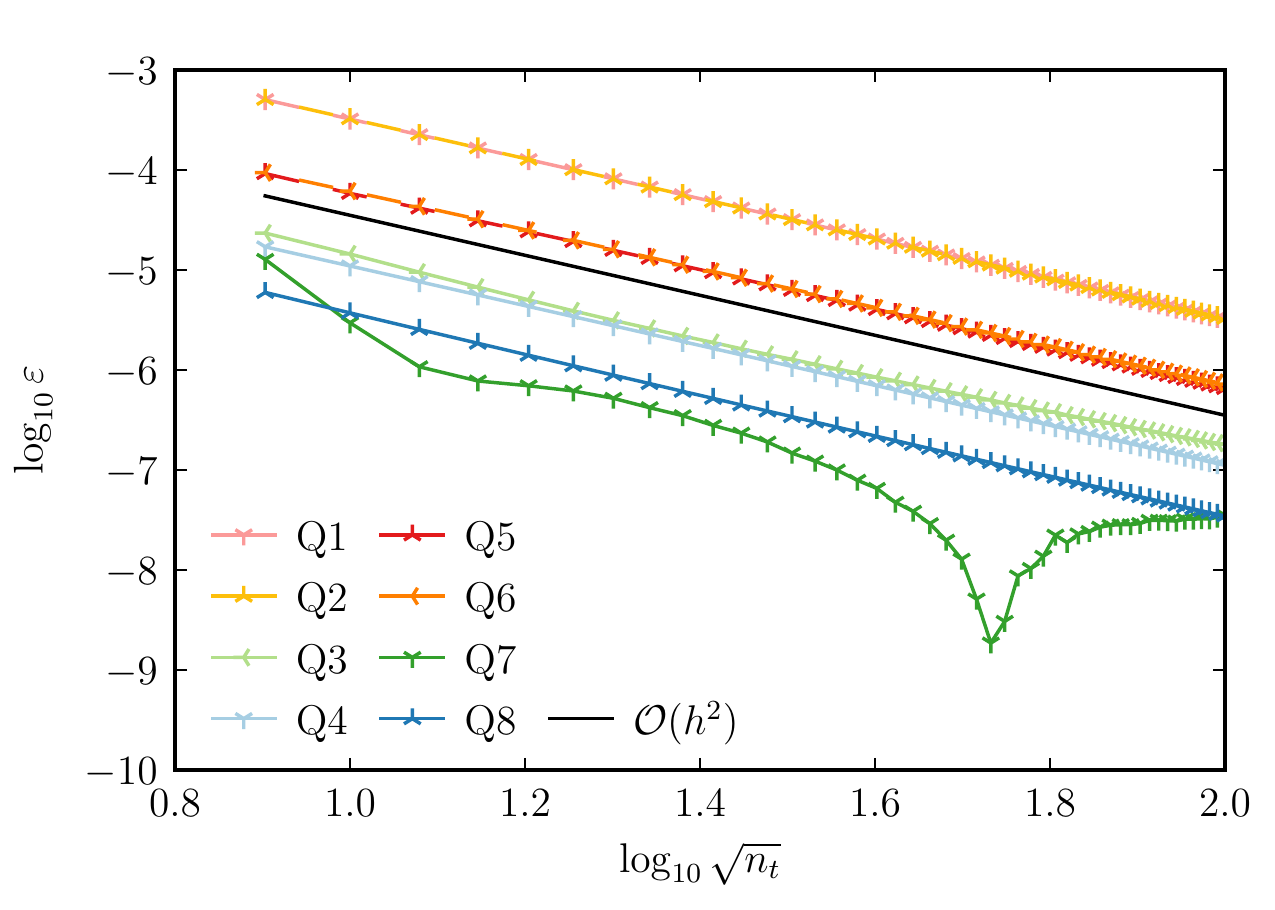}

\caption{Uniform mesh, $I^\Phi$}
\end{subfigure}
\hspace{0.25em}
\begin{subfigure}[b]{.49\textwidth}
\includegraphics[scale=.64,clip=true,trim=2.3in 0in 2.8in 0in]{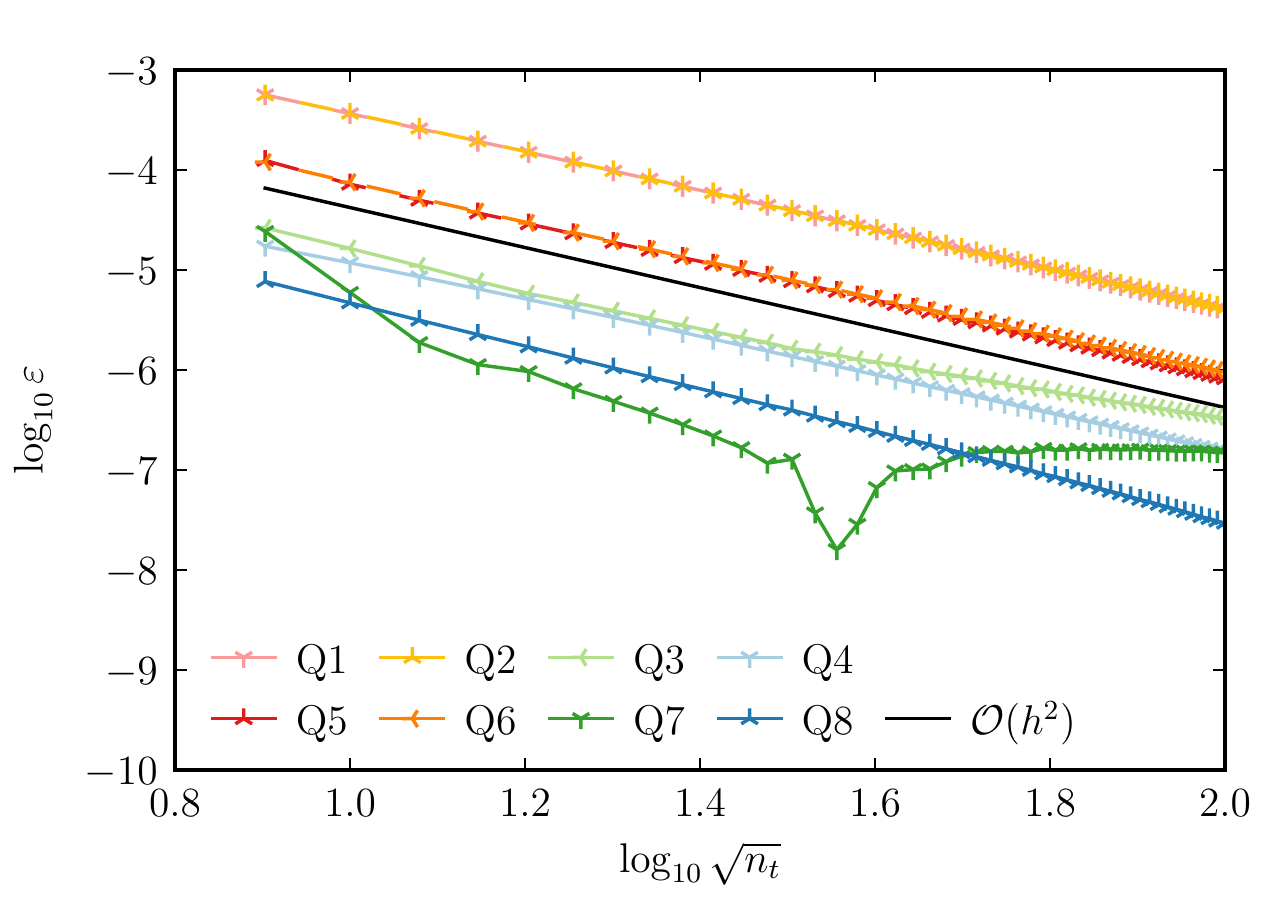}

\caption{Twisted mesh, $I^\Phi$}
\end{subfigure}
\caption{Solution-discretization error elimination: $\varepsilon=|I_h - I|/|I|$, with $G_k$ and $\theta=135^\circ$.}
\vskip-\dp\strutbox
\label{fig:part3_135}
\end{figure}

Figures~\ref{fig:part3_0}--\ref{fig:part3_135} show the relative error for the quadrature combinations in Table~\ref{tab:G_quad_combinations} for $G_k$ for $I^\mathbf{A}$ and $I^\Phi$ for the uniform and twisted meshes and four $\theta$ values.  Because of the singular integrands, the errors in the quadrature integration, which are functions of the unbounded derivatives of the integrand, are unbounded, yielding less monotonic trends than what are obtained for $G_\text{MS}$ in Figure~\ref{fig:part3}.  Nonetheless, Quadrature Combinations~Q4 and Q8 generally outperform Quadrature Combinations~Q1 and Q5, and Quadrature Combination~Q8 generally outperforms Quadrature Combination~Q4.

To compute $I$~\eqref{eq:I} for $G_k$, we select a relative \reviewerBoth{convergence} tolerance $\epsilon\ge|I'-I|/|I|$ when we compute its approximation $I'$ using the \texttt{CUBA} library.  Letting $\delta$ denote the relative difference between the actual $e=I_h - I$ and $e'=I_h - I'$, we see that
\begin{align*}
\delta = 
\left|\frac{e'-e}{e}\right| = 
\left|\frac{\left(I_h-I'\right)-\left(I_h-I\right)}{e}\right| =
\left|\frac{I-I'}{e}\right| \le
\epsilon\frac{|I|}{|e|} =
\frac{\epsilon}{\varepsilon}.
\end{align*}
Therefore, the ratio of the relative \reviewerBoth{convergence} tolerance $\epsilon$ to the relative error $\varepsilon$ provides an indication of how accurate the measured $\varepsilon'$ is.  For these results, we set $\epsilon=10^{-11}$, such that the measured $\varepsilon'$ is within $\delta=1\%$ if $\varepsilon<10^{-9}$, which is generally the case for Figures~\ref{fig:part3_0}--\ref{fig:part3_135}.

\section{Conclusions}
\label{sec:conclusions}


The method-of-moments implementation of the electric-field integral equation yields many code-verification challenges, which are largely due to the presence of singular integrals.
To mitigate these challenges, we assessed the solution-discretization and numerical-integration errors together and separately using the actual Green's function, as well as a manufactured Green's function.  In addition to considering the solution-discretization error, we isolated the numerical-integration error by 1) canceling the solution-discretization error by computing the MMS source term as a function of the basis-function representation of the manufactured surface current, and 2) eliminating the solution-discretization error by projecting the governing equations onto the manufactured surface current and avoiding a basis-function representation of the surface current.

Through these approaches, we were able to verify the code and compare the error from different quadrature options.  A comparison of the quadrature options alone would have been otherwise contaminated by the solution-discretization error.
\section*{Acknowledgments} 
\label{sec:acknowledgments}
The authors thank Robert Pfeiffer and Brian Zinser for their insightful feedback. 
This paper describes objective technical results and analysis. Any subjective views or opinions that might be expressed in the paper do not necessarily represent the views of the U.S. Department of Energy or the United States Government.
Sandia National Laboratories is a multimission laboratory managed and operated by National Technology and Engineering Solutions of Sandia, LLC, a wholly owned subsidiary of Honeywell International, Inc., for the U.S. Department of Energy's National Nuclear Security Administration under contract DE-NA-0003525.

\addcontentsline{toc}{section}{\refname}
\bibliographystyle{elsarticle-num}

\makeatletter
\interlinepenalty=10000
\bibliography{quadrature.bib}

\makeatother

\end{document}